\documentclass[letterpaper,twocolumn,10pt]{article}

\usepackage[twoside=true, head=13pt,
     paperwidth=8.5in, paperheight=11in,
     includeheadfoot=false, columnsep=2pc,
     top=1in, bottom=1in, inner=0.75in, outer=0.75in,
     marginparwidth=2pc, heightrounded]{geometry}

\usepackage[english]{babel}
\usepackage[T1]{fontenc}
\usepackage[tt=false, type1=true]{libertine}
\usepackage{mathptmx}
\usepackage[varqu]{zi4}
\usepackage{amssymb}
\usepackage{textcomp}
\usepackage{graphicx}
\usepackage{microtype}
\usepackage[font={footnotesize},labelfont={footnotesize,bf},textfont={footnotesize,it}]{caption}
\usepackage{subcaption}
\usepackage{booktabs}
\usepackage{longtable}
\usepackage{makecell}
\usepackage{float}
\usepackage{pdflscape}
\usepackage{siunitx}
\usepackage{listings}
\usepackage{footnote}
\usepackage{tablefootnote}
\usepackage{multirow}
\usepackage{soul}
\usepackage{xcolor}
\usepackage{pgf-pie}
\usepackage{tikz}
\usepackage{verbatim}
\usepackage{enumitem}
\usepackage{url}
\usepackage[hidelinks]{hyperref}

\hypersetup{
	pdftitle={Who Resolves Your DNS? Measuring Resolver Opacity and Improving DNS Observability},
	pdfauthor={Kedar Thiagarajan and Fabian E. Bustamante}
}

\usetikzlibrary{pie}
\usetikzlibrary{arrows.meta, calc, positioning, decorations.pathreplacing}

\graphicspath{{./}}

\makeatletter
\renewcommand\paragraph{%
	\@startsection{paragraph}{4}{\z@}%
	{0.6ex plus .2ex minus .2ex}%
	{-1em}%
	{\normalfont\normalsize\bfseries}%
}
\renewcommand\subparagraph{%
	\@startsection{subparagraph}{5}{\z@}%
	{0.5ex plus .2ex minus .2ex}%
	{-1em}%
	{\normalfont\normalsize\itshape}%
}
\makeatother

\newcommand{\Description}[1]{}

\title{Who Resolves Your DNS? Measuring Resolver Opacity and Improving DNS Observability}
\author{
	Kedar Thiagarajan\\
	Northwestern University\\
	\texttt{kedarthiagarajan2028@u.northwestern.edu}
	\and
	Fabi\'an E. Bustamante\\
	Northwestern University\\
	\texttt{fabianb@northwestern.edu}
}
\date{}

\begin{document}
\pagestyle{plain}

\maketitle

\begin{abstract}
	
	DNS resolution has no notion of a verifiable resolver path. When an ISP outsources resolution to a third-party provider, a user's queries can cross organizational and national boundaries without their awareness---and the client that issued them has no standardized, per-query mechanism to obtain the ordered resolver path. We argue that this opacity is an architectural gap rather than a deployment accident, and that a reported, verifiable resolver path should be a first-class goal of the resolution protocol.
	
	We quantify this opacity using RIPE Atlas measurements from 190 countries and then show that it is inexpensive to address. Even under a conservative treatment that assigns unattributable observations to the client's AS, 39.8\% of Campaign~B observations (6,622 of 16,636) cross the client--frontend AS boundary. One in four geolocatable anycast frontend pairs resolves outside the client's country, and Google Public DNS accounts for roughly two-thirds of those out-of-country cases. We present \textsc{Resolver-Path}, which enables participating resolvers to report their identity as they forward queries. We implement resolver-path disclosure in BIND~9, Unbound, and Knot Resolver with near-neutral throughput, latency, and CPU overhead. Because participation is voluntary, disclosure establishes the \emph{verifiable presence} of reported hops, not the \emph{absence} of hidden ones. An optional attestation extension authenticates received assertions, ordering, and freshness---not completeness. Together, disclosure and attestation provide bounded evidence about a query's reported resolver path, giving DNS a transparency mechanism it currently lacks.
	
\end{abstract}

\section{Introduction}
\label{sec:introduction}

Consider a DNS query issued by a client---the stub resolver on an end
host---in France. It is answered by an anycast public resolver whose
client-facing instance sits inside the country: by every signal available to
the client, resolution looks domestic. Yet the resolver that actually contacts the authoritative server sits in
Finland---the query has quietly left the country, and the client has no way to
know. That blindness is not unique to this query; it is the normal condition of
DNS resolution. ISPs increasingly outsource resolution to third-party
operators, forwarding queries through resolution structures that can span
organizations, Autonomous Systems, and national boundaries, and no standardized,
per-query mechanism lets a client obtain the ordered resolver path~\cite{schomp:dnsinfrastructure}. We argue
that this opacity is an architectural gap rather than a deployment accident,
and that a reported, verifiable resolver path should be a first-class goal of
the resolution protocol.

\begin{figure}[t]
	\centering
	\includegraphics[width=0.8\columnwidth]{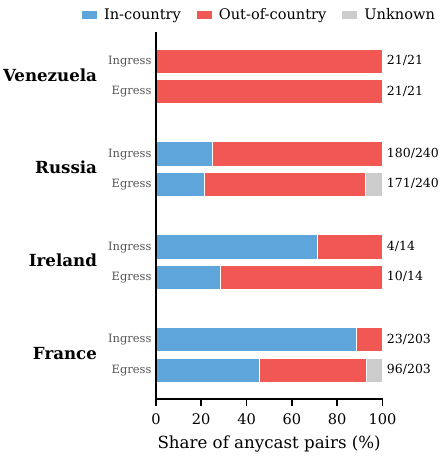}
	\caption{Ingress and egress cross-border rates for anycast resolution in selected countries. In Ireland and France, cross-border egress far exceeds cross-border ingress, a pattern consistent with hidden forwarding or provider-internal egress selection. Section~\ref{sec:Findings} presents the full global analysis.}
	\Description{Paired horizontal stacked bars for four countries (Venezuela, Russia, Ireland, France), each showing ingress and egress in-country versus out-of-country shares of anycast probe--resolver pairs. Ireland and France exhibit much higher egress than ingress cross-border rates.}
	\label{fig:motivating}
\end{figure}

The gap carries real consequences. A France-to-Belgium or Ireland-to-UK
crossing may remain within broadly comparable European legal
regimes~\cite{eu:gdpr,ec:adequacy-decisions}, but our
measurements also show domestic-looking anycast resolution from Colombia
egressing in the United States; in those cases, the query is exposed to a
different jurisdiction even though the client-facing resolver appeared local.
Such cross-border handling can subject DNS queries to different surveillance,
censorship, or data-retention regimes~\cite{dnscensor:venezuela,nytimes:article}.
Out-of-AS resolution is also associated with higher latency in our measurements.
Figure~\ref{fig:motivating}
previews the scale. Using RIPE Atlas across 190 countries, we find that one in four 
geolocatable anycast frontend pairs resolves outside
the client's country, and the two ends of the chain diverge: in Ireland and
France, frontend (ingress) resolvers appear mostly domestic, yet the egress
observed at our authoritative server crosses borders far more often---consistent
with hidden forwarding or provider-internal egress selection, though our vantage
observes two endpoints rather than a complete chain. A single operator, Google
Public DNS, dominates the out-of-country anycast subset, accounting for roughly
two-thirds of those cases. Resolver opacity is thus both common and concentrated.

\textbf{We argue that DNS resolution should be made visible---and show that it can be.} 
We present \textsc{Resolver-Path}, in which participating resolvers report their identity
as they forward a query. We implement this disclosure mechanism in three major resolver
implementations with near-neutral overhead; the same information can also be collected
through an alternative iterative query mechanism. Because non-participating resolvers may ignore or strip the metadata, deployment is incremental: modified resolvers contribute metadata while non-participants continue resolving as before. The resulting reported path, however, remains unauthenticated. An optional attestation extension authenticates each participating resolver's contribution by binding every received assertion to the selected DNS response and a fresh client nonce. Our
dataset, code, and analysis scripts will be released for replication.
We make three contributions:

\begin{itemize}[leftmargin=*]
	\item \textbf{Resolver opacity is widespread.} Under the conservative attribution convention, 39.8\% of Campaign~B observations (6,622 of 16,636) cross the client--frontend AS boundary. This establishes AS separation, not an intervening hop; separate endpoint comparisons expose pervasive address and prefix divergence without identifying the mechanism between them (\S\ref{sec:Findings}).
	\item \textbf{Resolver-path disclosure is practical.} We design and implement an EDNS-based disclosure mechanism in BIND 9, Unbound, and Knot Resolver and show that it incurs near-neutral overhead (\S\ref{sec:tracingdnspath}).
	\item \textbf{Attestation binds resolver assertions to responses.} We specify and evaluate a response-bound hash-chain construction using published conformance vectors across two codebases and signed load tests (\S\ref{sec:attestation}).
\end{itemize}


\section{Measuring Resolver Opacity}
\label{sec:measuring-opacity}

\begin{figure}[t]
	\centering
	\includegraphics[width=\columnwidth]{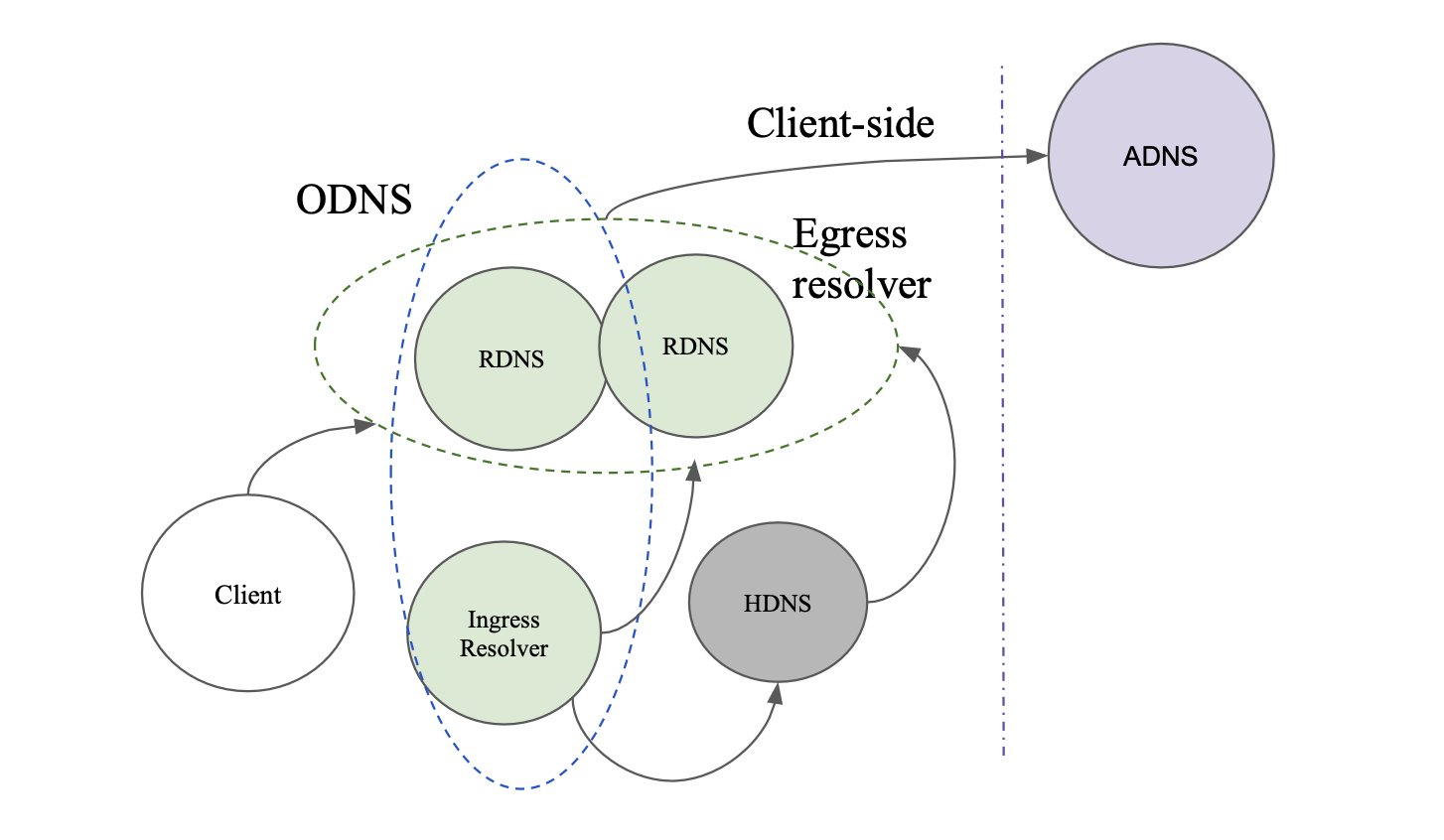}
	\caption{DNS resolution often spans multiple intermediaries, organizations, and ASes~\cite{schomp:dnsinfrastructure}.}
	\Description{A schematic DNS resolution path from client to ingress and egress resolvers, showing that different intermediaries may belong to different organizations and autonomous systems.}
	\label{fig:dnsinfrastructure}
\end{figure}

The conceptual gap motivates a direct measurement question: how often does resolution leave the client's organization and country, and how concentrated are the responsible providers?
The measurement campaign infers the observed client--ingress--egress resolver chain (Fig.~\ref{fig:dnsinfrastructure}) using controlled authoritative domains and metadata provided by the RIPE Atlas platform.
This observed chain reveals only the client-facing ingress resolver and the authoritative-facing egress resolver; it does not expose intermediate forwarders, forwarding depth, policy-dependent fan-out, or cache-dependent path changes in the full hidden forwarding path.
The analysis is organized around three research questions:

\begin{description}
\item[RQ1] How often do observed client--ingress--egress resolver chains cross AS, organizational, and national boundaries, and how has this changed over time?
\item[RQ2] Which providers and destination countries dominate out-of-country resolution in observed chains, and what cross-border handling is relevant to further legal or governance review?
\item[RQ3] Given the opacity documented by RQ1--RQ2, can lightweight protocol extensions expose more of the hidden forwarding path by reporting a resolver path to clients without degrading resolver performance?
\end{description}

\noindent Sections~\ref{sec:methodology}--\ref{sec:Findings} address RQ1 and RQ2 by measuring the problem; Section~\ref{sec:tracingdnspath} addresses RQ3 by showing that closing this visibility gap is technically inexpensive.

\subsection{Methodology and Data Collection}
\label{sec:methodology}

The methodology identifies mismatches between client-facing and authoritative-facing DNS resolvers by mapping publicly routable addresses to ASes and organizations and applying an explicitly conservative convention to private ingress.

\paragraph{Measurement design.}
All connected RIPE Atlas probes~\cite{ripeatlas} serve as DNS clients.
Each probe issues a query for a subdomain under a controlled authoritative domain.
Probes perform the resolution using their locally configured (\emph{on-probe}) resolvers. RIPE Atlas records which resolver IP receives the query and the resulting answer; that resolver IP is the \emph{ingress resolver}. If a probe is configured with multiple local resolvers or fails over between resolvers, different ingress IPs appear across measurements, but each individual query contributes only one observed ingress resolver.
All such queries reach the authoritative name server, which returns as its answer the source IP address of the packet that reached it---the \emph{egress resolver}. This egress endpoint is the authoritative-facing source address, not proof of the exact recursive process that performed every internal lookup behind NAT, forwarding, or provider-internal fan-out.
Combining the Atlas-reported ingress IP with the egress IP encoded in the authoritative response reconstructs an observed client--ingress--egress resolver chain for each query. The Atlas API also provides the probe's IP address, country, and ASN, which associates clients with their access networks.

\paragraph{Private-ingress convention.}
Private ingress supplies neither a public BGP origin ASN nor an independently
geolocatable address. Full-population analyses assign it to the client AS for AS
comparisons and the client country for country comparisons; the 83 Campaign~B
rows lacking a client and/or ingress ASN are also counted as same-AS. These
choices bias toward same-AS and domestic classifications. The measurement
exposes only the ingress and egress endpoints. Exposing hidden intermediate
resolvers requires a disclosure mechanism; we propose \textsc{Resolver-Path} to
provide it.

\subsection{Dataset and Scope}
\label{sec:dataset}

Four collection runs span three years: January~5--9, 2023; August~19--28,
2024; May~29--30, 2025 (Campaign~A); and October~1--2, 2025 (Campaign~B).
Across the three years, 10,082--11,918 unique probes returned results per year across
3,100--3,150 ISPs in 187--190 countries. Campaign-level ASN-completeness filtering retained
23,995, 9,899, 19,033, and 16,553 resolver chains, respectively, for 69,480
retained chains in total. Campaign~B supplies the 16,636 successful rows used
as the raw 2025 analytical coverage base; analyses then apply the
field-specific completeness filters reported in Appendix
Tables~\ref{tab:campaign-outcomes} and~\ref{tab:filter-accounting}. Appendix
Table~\ref{tab:unique_entities} summarizes the 2025 entities. The longitudinal
analysis instead uses a deduplicated union of Campaigns~A and~B for its 2025
year-level dataset.

Probes are broadly distributed across continents but are concentrated in fixed-broadband and enterprise access networks, with correspondingly less coverage of mobile and VPN users.
This repeated multi-year sampling allows us to observe temporal shifts in resolver behavior, including the growing adoption of cloud-based providers.

\subsubsection{Measurement Coverage}
\label{sec:measurement-coverage}

Latency was measured using three ICMP pings per probe--resolver pair and DNS resolution timing from successive cached lookups. The campaign geolocated 3,209 ingress and 7,563 egress resolvers with public IP addresses. Frontend RTT coverage is 44.6\% (56.1\% among known-AS targets); the primary coverage gap is that 51\% of ingress resolvers use private IP addresses rather than refusing ICMP. The 2025 dataset contains 6,713 private-ingress observations (40.4\% of 16,636 observations), reported separately to make the full-population denominator explicit.

RIPE Atlas probes are predominantly deployed in access networks (Appendix Figure~\ref{fig:asntype}), consistent with prior measurements showing that roughly 80\% operate within service-provider ASes~\cite{bajpai2017ripeatlas,ripe2017distribution}. This placement makes RIPE Atlas well suited to characterize DNS resolution from the perspective of end users.

\paragraph{Resolver attribution.}
Resolver attribution maps each client and each publicly routable ingress and egress resolver IP to its origin AS using RouteViews BGP dumps~\cite{routeviews} and to its organization using Borges~\cite{selmo:borges}. This mapping identifies observed AS and organizational boundary crossings. Private-ingress labels follow the convention above rather than an operator-ownership inference.

\paragraph{Latency measurement.}
Latency measurement sends three ICMP pings from each client to its ingress and egress resolvers and records the minimum RTT. The AS-stratified latency analysis includes only publicly attributable ingress and egress addresses with known client and resolver ASNs; private ingress is excluded from the frontend population. Within this population, the analysis compares resolvers in the client AS with resolvers in a different AS.

Because some resolvers block ICMP, DNS query timing provides a complementary estimate. Each probe performs two successive resolutions of a highly popular domain, maximizing the likelihood that the record is already present in the resolver cache. The second lookup therefore primarily reflects communication between the client and resolver rather than recursive fetch latency. Cache state and resolver behavior may still introduce noise.

For intuitive interpretation, the analysis translates measured RTTs into order-of-magnitude path lengths using a fiber propagation speed of approximately $c_f \approx 2/3 c$. These estimates are illustrative rather than literal: Internet paths rarely follow geodesic routes, and queuing, forwarding, and processing delays can all inflate RTT beyond pure propagation delay.

\paragraph{Geolocation.}
Resolver geolocation combines multiple complementary signals to identify cross-border resolution.
PTR records supply city-level hints extracted with \emph{The Aleph} framework~\cite{aqua:aleph}; the IPinfo database supplies the fallback for resolvers without embedded location hints~\cite{ipinfo:io}. RIPE Atlas RTT and traceroute constraints actively test every unicast resolver address whose country is supplied by IPinfo. When those measurements contradict the IPinfo location, the full-population analysis conservatively classifies the observation as in-country rather than as a cross-border mismatch; the stricter plausibility-consistent subset excludes it (Appendix~\ref{sec:appx:geo}).

For public anycast ingress resolvers, operator-provided serving-site identifiers supply the location signal: Google Public DNS NSID airport-code labels, Cloudflare \texttt{id.server} values, Quad9 NSIDs, and Cisco Umbrella/OpenDNS diagnostic identifiers~\cite{google-public-dns-nsid,cloudflare-1111-idserver,cloudflare-datacenter-iata,quad9-idserver-faq,cisco-umbrella-georouting}. These identifiers represent the operator-selected serving site, not the infrastructure's registration geography. RTT- and traceroute-based validation applies only to unicast addresses because repeated measurements to the same anycast IP may reach different points of presence. Table~\ref{tab:geolocation-signals} summarizes the available signals and their agreement with IPinfo. Operator-reported serving sites and IP registration databases capture different properties: the former identifies the serving location selected for a query, whereas the latter reflects registration geography~\cite{ipinfo:anycast}.

\begin{table}[t]
	\centering
	\scriptsize
	\setlength{\tabcolsep}{3pt}
	\begin{tabular}{lccc|ccc}
		\toprule
		& \multicolumn{3}{c|}{Frontend} & \multicolumn{3}{c}{Egress} \\
		\cmidrule(lr){2-4} \cmidrule(l){5-7}
		Source & \makecell{Count} & \makecell{\% of\\Total} & \makecell{\%\\Conflict} & \makecell{Count} & \makecell{\% of\\Total} & \makecell{\%\\Conflict} \\
		\midrule
		TXT records & 3,973 & 23.9\% & 86.6\% & 0 & 0.0\% & 0.0\% \\
		PTRs & 632 & 3.8\% & 3.3\% & 3,312 & 19.9\% & 8.2\% \\
		IPInfo & 5,318 & 32.0\% & 0.0\% & 13,309 & 80.0\% & 0.0\% \\
		Private ingress (near-side) & 6,713 & 40.4\% & -- & -- & -- & -- \\
		\midrule
		Total geolocated & 9,923 & 59.6\% & -- & 16,621 & 99.9\% & -- \\
		\bottomrule
	\end{tabular}
	\caption{Geolocation sources used for country attribution in the 2025 dataset ($N=16{,}636$ observed resolver chains per side). Methods are applied in priority order (TXT/NSID/CHAOS identity, PTR, then IPInfo), yielding the exclusive attribution shown. Private ingress is reported separately because it is not independently geolocated; complete-population country analyses follow the \S\ref{sec:methodology} convention.}
	\label{tab:geolocation_source_exclusive_coverage_2025}
\end{table}

Table~\ref{tab:geolocation_source_exclusive_coverage_2025} summarizes the country-attribution sources used throughout this section. Public anycast frontend location is derived primarily from operator-reported serving-site
identities, whereas egress location is derived primarily from actively validated PTR/IP geolocation. Private ingress is reported separately as a near-side category and is not independently geolocated.

For Cloudflare, location information is obtained from CHAOS TXT
\texttt{id.server} responses rather than NSID. Cloudflare confirmed in private
correspondence with the authors that, under normal operation,
\texttt{id.server} identifies the colo performing recursion. During overload,
provider-internal forwarding can instead cause that client-facing identity to
differ from the recursive egress. We therefore treat \texttt{id.server} as an
operator-confirmed ingress-serving-site indicator, not evidence of a complete
forwarding path.

When an ingress IP belongs to a known anycast resolver, ingress country is determined in the following order of precedence: (i) NSID or \texttt{id.server}, (ii) Aleph-derived location, and (iii) the probe country when neither source provides a location.

\paragraph{Anycast ingress analysis.}
Each probe--resolver pair constitutes one anycast-ingress observation and contributes to the geolocated subset only when the resolver exposes an identity string that maps to a documented serving site. Multiple probes may observe the same resolver and anycast site, so reported percentages are descriptive population shares rather than inferential estimates.

Egress geolocation excludes anycast IPs because a single anycast address may represent multiple serving locations. The remaining egress population is overwhelmingly unicast, allowing standard IP geolocation after the active consistency checks described above. Anycast egress addresses represent only about 1\% of unique egress IPs; excluding them removes 1,438 of 16,563 valid probe--frontend--egress country triples used in the observed-population analysis (Appendix~\ref{sec:appx:addl-geo}).

\subsection{Latency Results}
\label{sec:latency-results}
\graphicspath{{./}}

The latency analysis compares same-AS and out-of-AS resolvers using both ICMP RTT and cached DNS lookup time for the frontend and egress roles. Both comparisons are restricted to publicly attributable resolver addresses with known client and resolver ASNs; private ingress is excluded from the frontend latency population. Across both measurement methods and both resolver roles, the out-of-AS category consistently exhibits higher latency.

\begin{figure}[t]
    \centering
    \includegraphics[width=\columnwidth]{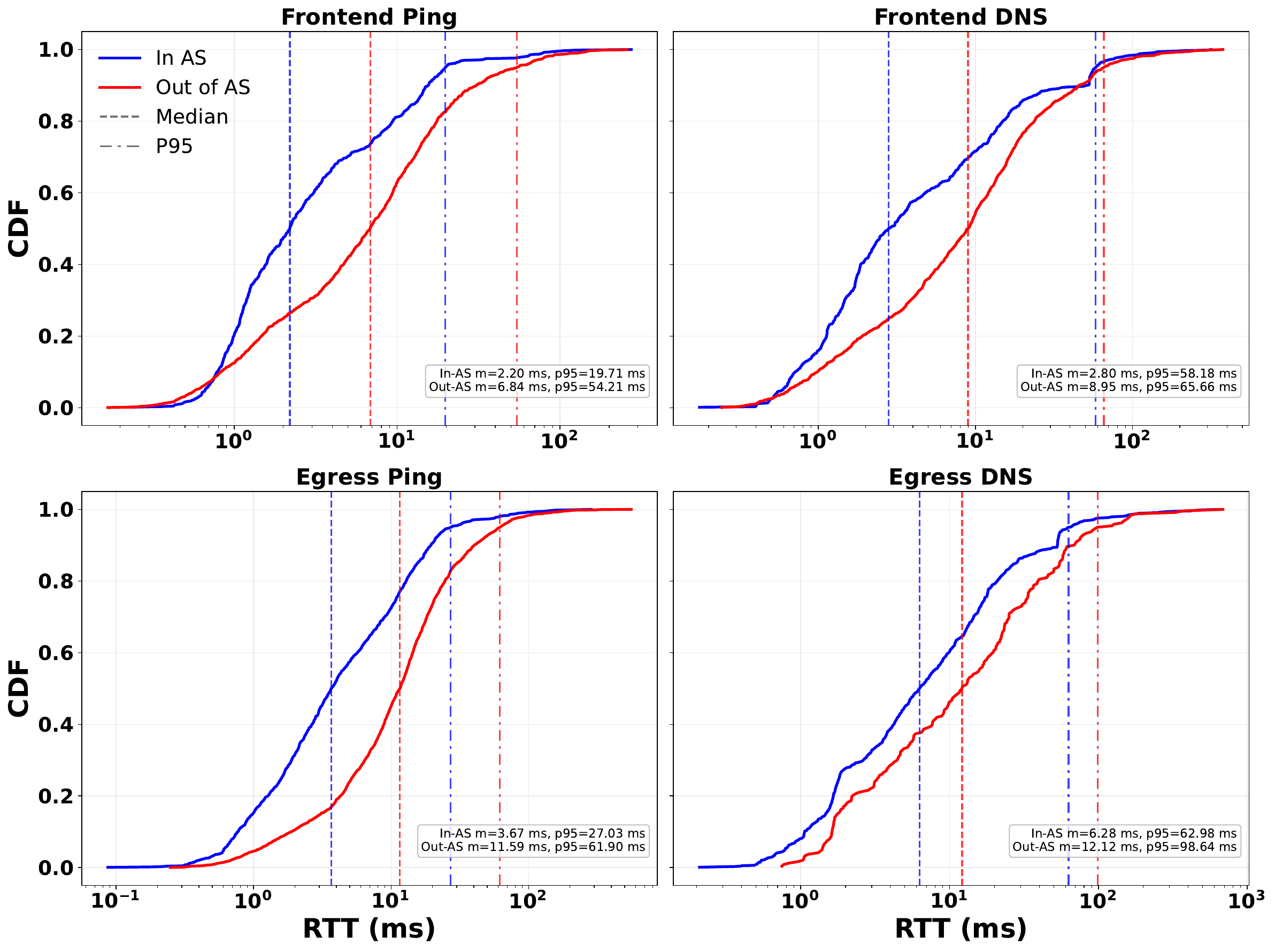}
    \caption{Latency to same-AS and out-of-AS resolvers. The four panels compare publicly attributable, AS-classified frontend and egress resolvers using ICMP RTT and cached DNS lookup time; private ingress is excluded from the frontend comparison. In every panel, the out-of-AS category exhibits higher medians and heavier tails.}
    \Description{A four-panel latency figure comparing same-AS and out-of-AS publicly attributable frontend and egress resolvers using ping and cached DNS lookup time. Private ingress is excluded from the frontend comparison. In all panels, the out-of-AS distribution is shifted upward, with higher medians and higher upper-percentile values.}
    \label{fig:ping_resolution_time}
\end{figure}

Figure~\ref{fig:ping_resolution_time} shows a remarkably consistent pattern. Regardless of whether latency is measured using ICMP RTT or cached DNS lookup time, and regardless of whether the resolver is the frontend or the egress, out-of-AS placement shifts the latency distribution toward higher values. Median latency approximately triples for frontend ICMP measurements (2.20 vs.\ 6.84 ms) and increases similarly for frontend DNS lookups (2.80 vs.\ 8.95 ms). Comparable shifts appear for egress resolvers (3.67 vs.\ 11.59 ms ICMP; 6.28 vs.\ 12.13 ms DNS), with consistently heavier latency tails.


The association is consistent across measurement methods and resolver roles, and holds when stratifying by probe continent (higher medians in 22/24 comparisons, higher p95 in 14/16 well-sampled strata). The agreement between ICMP RTT and cached DNS lookup time increases confidence that the observed latency differences reflect resolver placement rather than artifacts of either measurement method alone. AS-level mismatch is a proxy for, not equivalent to, geographic distance -- two ASes may peer at the same IXP, while one AS may span continents -- so the observed differences are interpreted as an association with AS-level resolver placement rather than a proven causal effect of geographic displacement or operator identity.

\smallskip\noindent\textbf{Takeaway (RQ1, latency):} Among publicly attributable, AS-classified frontend and egress resolvers, out-of-AS placement is consistently associated with higher latency: out-of-AS resolvers exhibit higher RTTs, higher cached DNS lookup times, and heavier latency tails.
\smallskip

\subsection{Resolver Boundary Crossings}
\label{sec:Findings}
\graphicspath{{./}}

Throughout this section, \emph{frontend} denotes the client-facing resolver (called \emph{ingress} in the methodology), while \emph{egress} denotes the resolver that contacts the authoritative server. We begin with country-level observations, then examine AS and organizational boundaries, identify the principal destination countries and providers, and finally study how these patterns evolve over time.

\subsubsection{Cross-border resolution is common}

We begin by asking how often observed resolver chains leave the client's country. We first consider public anycast frontend resolvers, which provide the strongest serving-site geolocation signals, and then compare them with the full
observed resolver population.

\begin{figure*}[t]
	\centering
	\includegraphics[width=\textwidth]{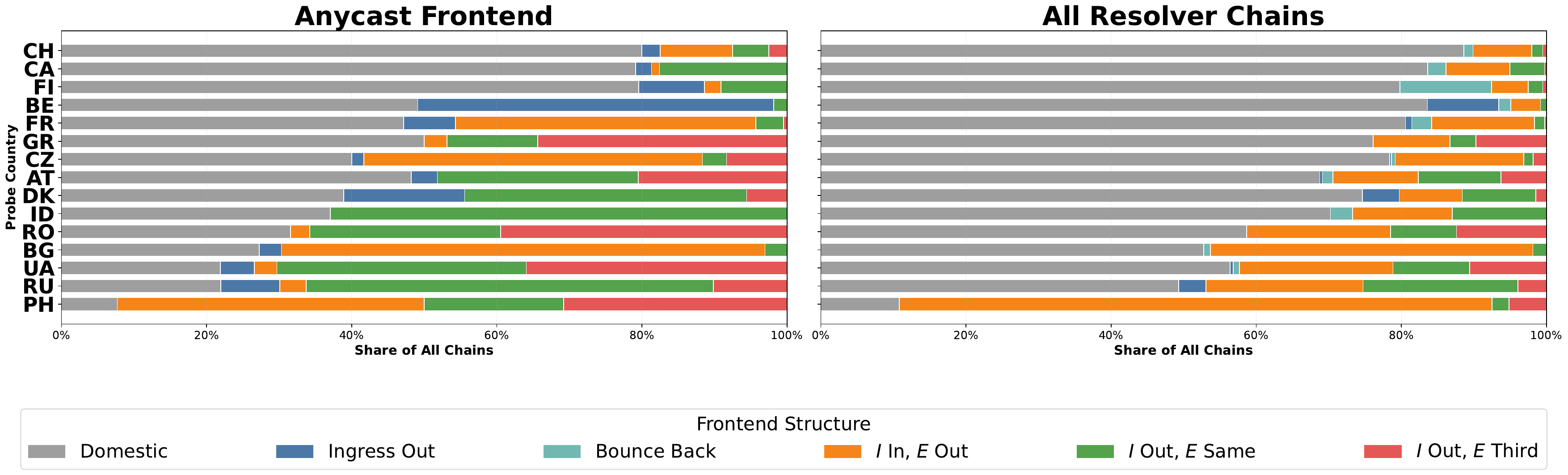}
	\caption{Frontend structure for the top 15 selected countries. Left: public Anycast Frontend. Right: All Observed Resolver Chains under the \S\ref{sec:methodology} convention. Bars are normalized by all observed chains per probe country and partitioned into Domestic, Ingress Out, Bounce Back, $I$ In/$E$ Out, $I$ Out/$E$ Same, and $I$ Out/$E$ Third. Key takeaway: public-anycast frontends exhibit substantially higher cross-border shares than the general observed-chain population.}
	\Description{Two-panel stacked-bar figure for the top-15 countries: public Anycast Frontend and All Observed Resolver Chains under the Section 2.1 convention. Each bar is normalized by all observed chains and partitioned into six structural categories: Domestic, Ingress Out, Bounce Back, I In E Out, I Out E Same, and I Out E Third.}
	\label{fig:frontend_structure_comparison}
\end{figure*}

Figure~\ref{fig:frontend_structure_comparison} compares public-anycast frontends and all observed resolver chains for the 15 probe countries with the highest mean cross-border frontend rate among the 29 countries meeting the minimum thresholds of 25 anycast chains, 100 observed resolver chains, and 15 unique probes.


Figure~\ref{fig:frontend_structure_comparison} shows that cross-border frontend resolution is common among public anycast resolvers. Across the full 2025 dataset, probe--frontend mismatch occurs in 1,430 of 16,563 chains (8.63\%). For public anycast, 989 of 5,498 scheduled probe--resolver pairs (18.0\%) geolocate outside the client country; among the 3,869 pairs with usable serving-site geolocation, the share rises to 25.6\%.

Even under the conservative assumption that every unresolved anycast pair remains in-country, nearly one in five frontend resolutions cross national boundaries. 

Egress mismatches remain common in the broader dataset: probe--egress mismatch occurs in 2,826 of 15,125 non-anycast-egress chains with known egress country (18.7\%), and the corresponding full observed-population rate is 17.06\% over all 16,563 valid probe--frontend--egress country triples, conservatively treating every IPinfo attribution contradicted by active checks as in-country.

\smallskip
\noindent\textbf{Takeaway (RQ1, country level):}
The cross-border share is 25.6\% among geolocatable public-anycast pairs and a
conservative 18.0\% when all unresolved pairs are counted as domestic.
\smallskip

\subsubsection{Resolver endpoints diverge and cross administrative and national boundaries}

\paragraph{Endpoint divergence.}
In Campaign~B, 91.70\% of public-ingress observations (9,103/9,927) expose
different ingress and egress IPs; 83.57\% of comparable IPv4 observations
(6,078/7,273) place them in different \texttt{/24}s. This establishes endpoint
and prefix divergence, not distinct operators or an intervening hop; NAT/SNAT,
multiple interfaces, egress selection, and forwarding remain indistinguishable.
Appendix
Tables~\ref{tab:endpoint-ip-divergence}--\ref{tab:endpoint-divergence-administration}
give complete campaign and population accounting.

\paragraph{AS-boundary crossing.}
Under the \S\ref{sec:methodology} convention, 6,622 of 16,636 Campaign~B
observations (39.8\%) cross the client--frontend AS boundary (Appendix
Table~\ref{tab:dnscategorization2025}). This is a lower bound, not an ownership
claim about unattributable observations, and it does not establish an
unobserved forwarding hop.

Restricting the analysis to public ingress removes the ambiguity introduced by private resolvers. Under this restriction, 6,480 of 9,877 public-ingress observations (65.6\%) have client ASN $\neq$ frontend ASN. At the organization level, 6,206 of 9,773 classified observations (63.5\%) belong to different client and frontend organizations.

Endpoint variability reinforces this opacity: 40.42\% of
$(\texttt{probe\_id},\texttt{frontend\_IP})$ chains map to multiple egress IPs
and 6.73\% to multiple egress ASNs. This is consistent with forwarding or
provider-internal selection without distinguishing them, motivating
standardized per-query path reporting.

\subsubsection{Cross-border resolution is concentrated}

\begin{figure*}[t]
	\centering
	\includegraphics[width=\textwidth]{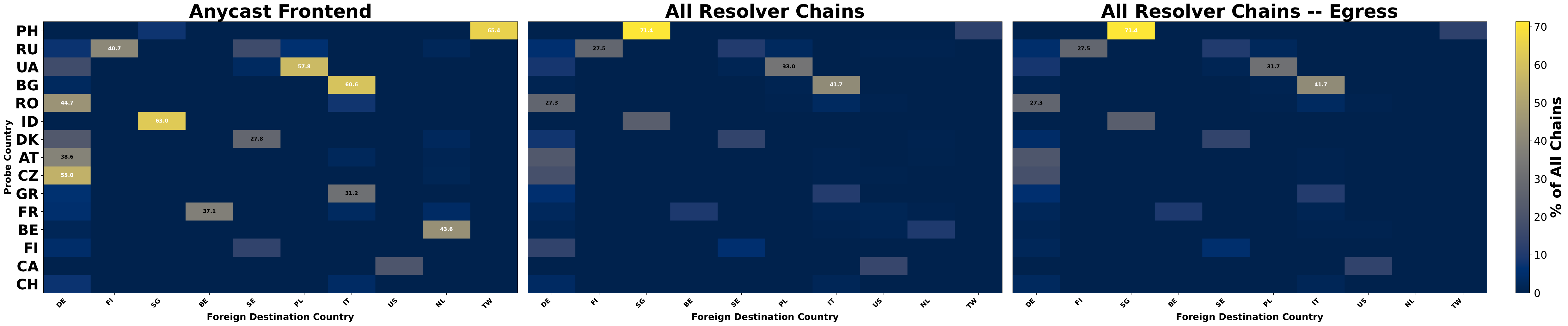}
	\caption{Foreign-destination heatmap for the countries in Figure~\ref{fig:frontend_structure_comparison}. Cells report the percentage of all chains for that probe country. Key takeaway: cross-border queries concentrate in a small number of regional hubs (notably DE, FI, SG), reflecting the geographic footprint of major public DNS providers.}
	\Description{Two-panel destination heatmap with the same row order as the structure figure. Cells show percent of all observed chains per probe country in cross-border frontend events by destination.}
	\label{fig:frontend_destination_heatmap}
\end{figure*}

\begin{figure*}[t]
	\centering
	\includegraphics[width=\textwidth]{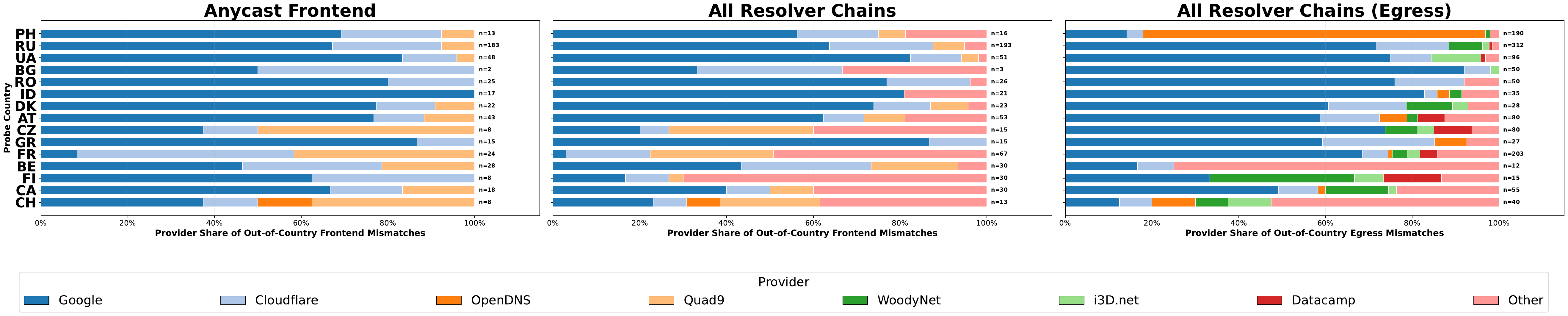}
	\caption{Provider mix for selected countries with shared row order: (left) public Anycast Frontend, (middle) All Observed Resolver Chains frontend mismatches, (right) All Observed Resolver Chains egress mismatches. Private ingress has no independently attributable provider identity. Key takeaway: Google Public DNS dominates the geolocatable out-of-country public-anycast frontend subset (67.3\%), while egress mismatches are more evenly distributed across providers.}
	\Description{Three-panel horizontal stacked-bar figure showing provider share for Anycast frontend mismatch, general frontend mismatch, and general egress mismatch.}
	\label{fig:frontend_provider_mix}
\end{figure*}

Having established that cross-border resolution is common, we next examine where cross-border queries terminate. Figure~\ref{fig:frontend_destination_heatmap} summarizes their destination countries. For third-country cases, attribution uses foreign egress country only.

Among the 989 out-of-country anycast frontend cases, 570 (57.6\%) keep egress in the same foreign country as ingress, 264 (26.7\%) continue to a third country, and 155 (15.7\%) remain in the residual Ingress Out category because the egress lacks usable geolocation.

The rightmost panel extends the analysis to all observed resolver chains, showing the distribution of foreign egress destinations for the same set of countries. In this selected set, 1,233 of 3,609 observed chains (34.16\%) have out-of-country egress. The largest egress destinations are DE (240, 19.46\%), SG (191, 15.49\%), FI (190, 15.41\%), BE (124, 10.06\%), and PL (121, 9.81\%). The ten plotted destination columns together cover 1,097 of 1,233 out-of-country egress cases (88.97\%), with the remaining 136 (11.03\%) aggregated in \textit{Other}.

We next ask which providers are responsible for these cross-border events. Figure~\ref{fig:frontend_provider_mix} summarizes the provider mix. Across all 3,869 geolocatable anycast frontend chains, Google Public DNS
accounts for 52.66\%, Cloudflare 32.75\%, Quad9 14.07\%, and OpenDNS
0.53\%. The out-of-country subset is substantially more concentrated:
Google Public DNS represents 67.3\% of geolocatable out-of-country
frontend cases, compared with 52.66\% overall, while Cloudflare and
Quad9 account for 20.6\% and 11.4\%, respectively. In contrast, provider
shares are more evenly distributed among out-of-country egress
mismatches in the full observed population.

\smallskip\noindent\textbf{Takeaway (RQ2):} Cross-border anycast frontend resolution is highly concentrated. Google Public DNS accounts for two-thirds of geolocatable out-of-country anycast frontend cases, and most foreign resolution terminates in a small set of regional hubs (DE, FI, SG). In most cases the egress endpoint remains in the same foreign country rather than returning to the client's jurisdiction.
\smallskip

\paragraph{Geographic grouping and legal interpretation.}
The 989 out-of-country anycast frontend pairs are also grouped by an
EU/EEA-and-adequacy scope in Appendix Figure~\ref{fig:appx_anycast_ingress_eu}.
This is a descriptive geographic grouping, not an ``EU-equivalent'' legal classification. Observed country pairs identify candidate cases for further legal or regulatory review, but the consequences of any crossing depend on the relevant actors, transfers, jurisdictions, safeguards, and applicable law. Neither an
operator-reported country label nor our geolocation pipeline establishes physical location, legal jurisdiction, or compliance. Intra-EU, EEA, and adequacy-jurisdiction crossings can still matter for privacy,
governance, data transfers, latency, and CDN selection~\cite{otto2012content}.

\smallskip\noindent\textbf{Takeaway (RQ2, continued):} The measurements expose
cross-border handling relevant to legal and governance review; they do not
decide the review's outcome or verify regulatory compliance.
\smallskip

\subsubsection{Longitudinal evolution and implications}

The preceding analyses provide a snapshot; we next ask whether these patterns evolve over time.
The longitudinal analysis uses a strict stable panel of 2,468 probes that
appear in each of the three year-level datasets and remain in the same client
AS and country. The 2025 year-level dataset is the deduplicated union of
Campaigns~A and~B.
All longitudinal results are reported under the \S\ref{sec:methodology}
convention; within each year-level dataset, each probe is given total weight 1, divided across its
observed resolver chains. Appendix Table~\ref{tab:longitudinal-stable-panel-appendix} reports the resulting shares. 

Even within this controlled panel, fully same-AS observed chains decline from 41.5\% in 2023 to 37.9\% in 2025, while different-AS ingress increases from 35.5\% to 42.1\%, different-AS egress from 39.6\% to 43.9\%, and chains exhibiting both from 16.5\% to 23.9\%. The directional result survives composition control under the \S\ref{sec:methodology} convention.

\smallskip\noindent\textbf{Takeaway (RQ1, longitudinal):} Controlling for probe composition preserves the observed shift away from fully same-AS resolver chains from 2023 through 2025. Under the \S\ref{sec:methodology} convention, the increase in different-AS ingress is a lower-bound trend.

Taken together, these results show that resolver behavior frequently crosses administrative and national boundaries and that these patterns are becoming more common over time. These findings have implications for ISPs, regulators, and resolver operators.

ISPs should audit whether their configured resolvers forward queries across
organizational and national boundaries: the public-anycast and egress
observations expose candidate cross-border handling, while the AS lower
bound shows pervasive boundary crossing under the \S\ref{sec:methodology}
convention.

Regulators and auditors can use authenticated operator assertions as one
input to compliance review, but the protocol neither proves physical or
legal jurisdiction nor verifies regulatory compliance.

Resolver operators already expose single-hop identities (e.g., NSID and id.server) for operational debugging. RESOLVER-PATH extends this practice to the participating resolver path, supporting transparency, auditing, and standardized disclosure with negligible measured overhead.


\section{Closing the Visibility Gap}
\label{sec:tracingdnspath}


Section~\ref{sec:measuring-opacity} showed that resolver endpoints frequently
cross AS, organizational, and national boundaries and that client-facing and
authoritative-facing endpoints widely diverge. These properties evolve over
time, yet endpoint observations cannot distinguish NAT, multiple interfaces,
egress selection, or forwarding; the logical path between them remains opaque.
This limitation is architectural: existing identity mechanisms---including
NSID~\cite{rfc5001} and CHAOS TXT records---are fundamentally single-hop,
allowing a client to identify only the resolver it directly contacts.

To address this limitation, we introduce \textsc{Resolver-Path}, a cooperative
protocol for disclosing resolver forwarding behavior. The protocol supports both
on-demand diagnostics and in-band disclosure during ordinary DNS resolution,
with an optional attestation extension. Throughout this section, a
\emph{reported} resolver path contains only participating resolvers.

\subsection{Resolver-Path Disclosure Mechanisms}

\begin{figure*}[t]
	\centering
	\begin{subfigure}[t]{0.48\textwidth}
		\centering
		\begin{tikzpicture}[
			entity/.style={rectangle, draw, fill=gray!12, minimum width=1.1cm, minimum height=0.55cm, font=\scriptsize\bfseries, rounded corners=1.5pt},
			msg/.style={-{Stealth[length=4pt]}, thick, font=\scriptsize},
			ret/.style={-{Stealth[length=4pt]}, thick, dashed, font=\scriptsize},
			life/.style={dashed, gray!60},
			annot/.style={font=\tiny\itshape, text=gray!70},
			]
			\node[entity] (C) at (0,0) {Client};
			\node[entity] (R1) at (2.2,0) {$R_1$};
			\node[entity] (R2) at (4.4,0) {$R_2$};
			\node[entity] (R3) at (6.6,0) {$R_3$};
			\node[annot, below=1pt] at (R1.south) {ingress};
			\node[annot, below=1pt] at (R2.south) {forwarder};
			\node[annot, below=1pt] at (R3.south) {egress};
			\draw[life] (C.south) ++(0,-0.35) -- ++(0,-5.0);
			\draw[life] (R1.south) ++(0,-0.35) -- ++(0,-5.0);
			\draw[life] (R2.south) ++(0,-0.35) -- ++(0,-5.0);
			\draw[life] (R3.south) ++(0,-0.35) -- ++(0,-5.0);
			\draw[msg] (0,-1.2) -- node[above] {\texttt{TYPE65400} to $R_1$} (2.2,-1.6);
			\draw[ret] (2.2,-2.0) -- node[above] {\texttt{next=$R_2$}, path[$R_1$]} (0,-2.4);
			\draw[msg] (0,-2.8) -- node[above] {\texttt{TYPE65400} to $R_2$} (4.4,-3.2);
			\draw[ret] (4.4,-3.6) -- node[above] {\texttt{next=$R_3$}, path[$R_1$,$R_2$]} (0,-4.0);
			\draw[msg] (0,-4.4) -- node[above] {\texttt{TYPE65400} to $R_3$} (6.6,-4.8);
			\draw[ret] (6.6,-5.2) -- node[above] {\texttt{next=$\varnothing$}, path[$R_1$,$R_2$,$R_3$]} (0,-5.6);
			\draw[decorate, decoration={brace, amplitude=4pt, mirror}] (-0.6,-1.0) -- (-0.6,-5.7) node[midway, left=5pt, font=\tiny, align=right] {iterative\\client\\probing};
		\end{tikzpicture}
		\caption{Technique~1: on-demand \texttt{TYPE65400} diagnostic. The client walks each resolver's reported next-hop topology, which need not equal an ordinary cache-hit path.}
		\label{fig:msc-technique1}
	\end{subfigure}
	\hfill
	\begin{subfigure}[t]{0.48\textwidth}
		\centering
		\begin{tikzpicture}[
			entity/.style={rectangle, draw, fill=gray!12, minimum width=1.1cm, minimum height=0.55cm, font=\scriptsize\bfseries, rounded corners=1.5pt},
			msg/.style={-{Stealth[length=4pt]}, thick, font=\scriptsize},
			ret/.style={-{Stealth[length=4pt]}, thick, dashed, font=\scriptsize},
			life/.style={dashed, gray!60},
			annot/.style={font=\tiny\itshape, text=gray!70},
			]
			\node[entity] (C) at (0,0) {Client};
			\node[entity] (R1) at (1.5,0) {$R_1$};
			\node[entity] (R2) at (3.0,0) {$R_2$};
			\node[entity] (R3) at (4.5,0) {$R_3$};
			\node[entity] (A) at (6.0,0) {Auth};
			\node[annot, below=1pt] at (R1.south) {ingress};
			\node[annot, below=1pt] at (R2.south) {fwd};
			\node[annot, below=1pt] at (R3.south) {egress};
			\draw[life] (C.south) ++(0,-0.35) -- ++(0,-5.0);
			\draw[life] (R1.south) ++(0,-0.35) -- ++(0,-5.0);
			\draw[life] (R2.south) ++(0,-0.35) -- ++(0,-5.0);
			\draw[life] (R3.south) ++(0,-0.35) -- ++(0,-5.0);
			\draw[life] (A.south) ++(0,-0.35) -- ++(0,-5.0);
			\draw[msg] (0,-1.2) -- node[above, font=\tiny] {Q + \texttt{opt[]}} (1.5,-1.5);
			\draw[msg] (1.5,-1.8) -- node[above, font=\tiny] {Q + \texttt{opt[$R_1$]}} (3.0,-2.1);
			\draw[msg] (3.0,-2.4) -- node[above, font=\tiny] {Q + \texttt{opt[$R_1$,$R_2$]}} (4.5,-2.7);
			\draw[msg] (4.5,-3.0) -- node[above, font=\tiny] {Q + \texttt{opt[$R_1$,$R_2$,$R_3$]}} (6.0,-3.3);
			\draw[ret] (6.0,-3.8) -- node[above, font=\tiny] {R + \texttt{path[$R_1$,$R_2$,$R_3$]}} (4.5,-4.1);
			\draw[ret] (4.5,-4.3) -- node[above, font=\tiny] {R + \texttt{path[$R_1$,$R_2$,$R_3$]}} (3.0,-4.6);
			\draw[ret] (3.0,-4.8) -- node[above, font=\tiny] {R + \texttt{path[$R_1$,$R_2$,$R_3$]}} (1.5,-5.1);
			\draw[ret] (1.5,-5.3) -- node[above, font=\tiny] {R + \texttt{path[$R_1$,$R_2$,$R_3$]}} (0,-5.6);
			\draw[decorate, decoration={brace, amplitude=4pt, mirror}] (-0.6,-1.0) -- (-0.6,-3.4) node[midway, left=5pt, font=\tiny, align=right] {path\\builds\\forward};
			\draw[decorate, decoration={brace, amplitude=4pt, mirror}] (-0.6,-3.7) -- (-0.6,-5.7) node[midway, left=5pt, font=\tiny, align=right] {reported\\path returns};
		\end{tikzpicture}
		\caption{Technique~2: unsigned EDNS Resolver Path Tracing (option
\texttt{65024}). Each resolver appends its identity before forwarding the
normal query. This panel shows disclosure only; response-bound attestation
signs after responses exist (Appendix
Figure~\ref{fig:response-bound-attestation}).}
		\label{fig:msc-technique2}
	\end{subfigure}
	\caption{Unsigned resolver disclosure. Technique~1 (left) walks reported
forwarding topology with dedicated probes. Technique~2 (right) accumulates the
participating path of an ordinary forwarded query. Dashed arrows indicate
responses; signatures are not shown here.}
	\Description{Two side-by-side message-sequence diagrams showing unsigned
resolver disclosure across three resolvers. Technique 1 uses iterative
TYPE65400 probes and client-side path construction. Technique 2 appends an EDNS
path option while forwarding an ordinary query. A separate figure explains
response-bound attestation.}
	\label{fig:msc-tracing}
\end{figure*}
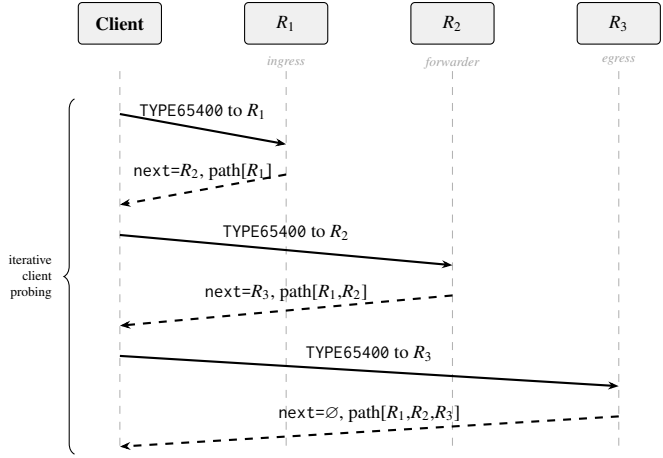
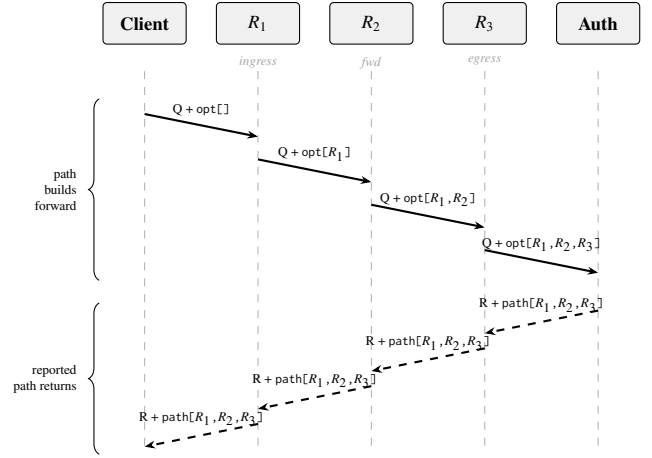

\textsc{Resolver-Path} provides two complementary disclosure mechanisms. Technique~1 is an on-demand diagnostic that allows clients to inspect reported forwarding behavior explicitly. Technique~2 integrates resolver-path disclosure into ordinary DNS resolution through an EDNS option, enabling continuous reporting during normal query processing. Figure~\ref{fig:msc-tracing} illustrates both mechanisms.

We define a \emph{response path} as the ordered participating resolvers whose
processing contributed to one DNS response. A query may produce several
upstream attempts or responses; each response retains its own path, and the
client ordinarily observes only the path carried by the selected response.
Attempts that produce no response, discarded losing responses, and
non-participating resolvers are outside this construct. An ingress cache hit has
path length one, while a hit at the $k$th participating resolver terminates that
response path at length $k$. A cache answer constructs fresh current-response
metadata for the client's nonce and never reuses the path or signatures from
the transaction that populated the cached RRset.

\paragraph{Technique 1: Recursive DNS Query Tracing}

This technique provides an on-demand diagnostic of the reported forwarding topology. A client walks the resolver chain by querying each participating resolver for its reported next hop using a new recursive query type (\texttt{TYPE65400}). Because this diagnostic is independent of an ordinary lookup, it may reveal a configured next hop that a cache-hit query would not traverse, and conditional forwarding or failover policy may select a different downstream resolver. \texttt{TYPE65400} uses an experimental/private-use codepoint rather than a proposed permanent assignment.

\paragraph{Technique 2: EDNS Resolver-Path Disclosure}

Technique~2 uses a new EDNS option (code~65024) to report the participating path
carried by an ordinary DNS response. The unsigned disclosure implementation
appends identity before forwarding and returns the accumulated list with the
response; signed mode instead binds a return-path chain to the selected response
(\S\ref{sec:attestation}). Because EDNS is hop-by-hop under RFC~6891, recursive
resolvers are not required to preserve unknown options. Consequently, Technique~2 reports
only participating resolvers and cannot rely on arbitrary legacy deployments to
propagate the option. Code~65024 is a prototype assignment; deployment would
require DNSOP discussion and IANA Expert Review.

Techniques~1 and 2 serve complementary operational purposes.
Technique~1 supports explicit diagnostics by walking the reported forwarding
topology through dedicated queries, whereas Technique~2 continuously reports the
participating resolver path during ordinary query processing without additional
client probes. Technique~2 therefore provides routine transparency during
ordinary resolution.

\subsection{Prototype}

We implement both disclosure mechanisms in BIND~9, Unbound, and Knot Resolver. The implementations preserve each resolver's existing forwarding and caching behavior while adding support for reporting resolver identity and propagating resolver-path metadata.

The Technique~1 prototype adds support for TYPE65400 queries, allowing participating resolvers to return their identity and reported next hop. The prototype requires 427 lines of code across six BIND files. 

Technique~2's prototype adds EDNS option~65024 processing, allowing participating resolvers to accumulate resolver-path metadata during ordinary query forwarding. The BIND implementation requires 153 lines of code across two files.

Equivalent functionality is implemented in Unbound and Knot Resolver.

\paragraph{Measurement Instrumentation.}

An EDNS-aware client records Technique~2 wire-level metadata, including the returned hop list, response size, truncation, and TCP fallback. As an experimental control, we also evaluate a modified-but-disabled configuration in
which all implementation changes are present but resolver-path disclosure is disabled. This allows us to separate the cost of the additional code paths from the cost of carrying resolver-path metadata.

\subsection{Evaluation}

We evaluate \textsc{Resolver-Path}'s (i) throughput and latency overhead,
(ii) scalability across resolver depths and heterogeneous deployments, and
(iii) correctness of reported-path recovery.

\paragraph{Experimental Setup} All experiments use the same replay of a university DNS trace from a peak 10-minute interval at 2~p.m. The trace contains 2,616,894 queries to 106,102 unique hostnames, of which 24,413 (23\%) are locally managed. Before replay, query-source identifiers and qnames were anonymized. The authoritative side is configured to answer every replay name locally, so the benchmark does not send those names to external authoritative services.

Experiments span cloud deployments, controlled local testbeds, and heterogeneous resolver chains. Cloud evaluation compares four BIND configurations: (i) vanilla BIND,
(ii) a modified-but-disabled baseline that isolates implementation overhead,
(iii) Recursive DNS Query Tracing (Technique~1), and
(iv) EDNS Resolver-Path Disclosure (Technique~2). Because Technique~1 operates
as an explicit diagnostic rather than during ordinary query processing, the
performance analysis focuses on Technique~2; Technique~1 is evaluated
separately. For Technique~2, an EDNS-aware client records wire-level metadata (hop lists, response sizes, truncation, and TCP fallback). Each experiment isolates a different aspect of the design: cloud deployments measure performance under realistic workloads, local testbeds evaluate scalability with resolver depth and cache behavior, and heterogeneous deployments verify interoperability across resolver implementations.
Unless noted otherwise, the results below apply to these configurations,
workloads, and participating resolvers; they do not estimate Internet-wide
performance, path-length, or deployment distributions.

\paragraph{Response-path correctness.}
Across 185 controlled runs spanning 23 scenarios and 32 topologies, validators
compared each disclosure with linked per-resolver events. Every fully
participating path matched: 63,042 individual and 47,042 selected-response
paths. The P1 condition marked all 1,000 partially participating responses
incomplete, and no sibling- or cross-response merges occurred. Appendix
Table~\ref{tab:response-path-correctness} gives the condition-level accounting.

\paragraph{Cloud Performance}

We measure the performance impact of the EDNS Resolver-Path disclosure
mechanism (Technique~2) using BIND on AWS EC2 \texttt{c7i.large} instances
(2~vCPUs, 4~GiB RAM) across Ireland, London, and Germany. The experiments span
five path behaviors, four BIND configurations (vanilla,
modified-tracing-disabled, Technique~1, and Technique~2), offered loads from
2{,}000--20{,}000~QPS, and five repetitions per operating point.

Table~\ref{tab:capacity-summary} summarizes cloud-capacity results. 
At 20{,}000~QPS, vanilla BIND completes 5,075--18,085~QPS across behaviors,
with p95 latency between 13.3 and 60.4~ms. The modified-but-disabled baseline
remains statistically close to vanilla (median throughput delta $-1.32\%$,
median p95 delta $0.00\%$). Relative to that baseline, Technique~2 incurs a
median throughput change of only $-0.90\%$ and a median p95 latency increase of
$2.08\%$. Resolver CPU increases by up to 17\% in the
\emph{Ingress Out/Egress Third} scenario. No median packet loss is observed;
saturation manifests only as reduced throughput.


\smallskip\noindent\textbf{Takeaway (RQ3, cloud performance):}
Technique~2 adds negligible throughput and tail-latency overhead; one forwarding
scenario increases CPU utilization by 17\%.
\smallskip

\begin{table}[t]
	\centering
	\scriptsize
	\caption{Cloud-capacity summary at a 20{,}000~QPS offered load on 2-vCPU \texttt{c7i.large} BIND testbeds (medians over five repetitions). Technique~1 is omitted because it operates as an explicit diagnostic rather than
		during ordinary query processing. Full load-sweep curves appear in Appendix Figures~\ref{fig:appx-capacity-latency} and~\ref{fig:appx-capacity-throughput}; per-behavior detail in Appendix Table~\ref{tab:cloud-20k-breakdown}.}
	\label{tab:capacity-summary}
	\begin{tabular}{lrrrrrr}
		\toprule
		& \multicolumn{3}{c}{p95 latency (ms)} & \multicolumn{3}{c}{CPU (\si{\micro\second/q})} \\
		\cmidrule(lr){2-4} \cmidrule(lr){5-7}
		Behavior & Van.\ & Off & T2 & Van.\ & Off & T2 \\
		\midrule
		Domestic          & 13.3 & 12.3 & 12.5 & 216 & 239 & 207 \\
		Egr.\ Out         & 44.0 & 44.0 & 44.0 & 313 & 246 & 352 \\
		Ingr.\ Out/Same   & 14.6 & 15.4 & 14.6 & 359 & 378 & 389 \\
		Ingr.\ Out/Third  & 60.4 & 60.4 & 62.5 & 454 & 452 & 531 \\
		Bounce Back       & 51.2 & 53.2 & 55.3 & 447 & 428 & 469 \\
		\bottomrule
	\end{tabular}
\end{table}

\paragraph{Resource Costs.}

Technique~2 adds little resource overhead. Relative to the vanilla baseline,
CPU utilization ranges from $-4\%$ to $+17\%$
(Table~\ref{tab:capacity-summary}, right), memory remains within 2\,MB, and
wire-level response sizes stay between 94.5 and 98.2\,B with zero truncation or
TCP fallback.

\paragraph{Scalability}

We evaluate scalability in a local Docker testbed by varying resolver-chain
depth (2--15 resolvers) at a 20{,}000~QPS offered load and cache state
(warm/cold) at a 10{,}000~QPS offered load, with five repetitions per
configuration.

Technique~2 scales cleanly with resolver-chain depth and cache behavior.
Throughput changes by only $-0.10$--$0.40\%$ as chain depth increases
(Appendix Figure~\ref{fig:appx-depth-scaling}), while p95 latency rises
modestly from 0.077 to 0.097\,ms at depth~15. Reported-path recovery remains
100\% at each forwarded depth, with zero truncation or TCP fallback.

Controlled cache hits tested termination and historical reuse: all 17,000
selected-response paths matched current-query events, and none reused the path
that populated the cached RRset.

\smallskip\noindent\textbf{Takeaway (RQ3, scalability):}
Technique~2 preserves path recovery with little overhead as resolver depth
increases.
\smallskip

\paragraph{Completed local workloads.}
To measure warm-cache overhead, we ran all 20 mode/load conditions with five
repetitions each (100 runs). At a
20,000-QPS offered load, signed mode completed 19,993.933~QPS versus
19,999.633~QPS for baseline, while p95 latency rose from 0.017 to 0.127~ms,
chain CPU from 13.124 to 76.135\,\si{\micro\second/query}, and wire volume from
127 to 309 bytes/query. Median client verification cost was
226.01\,\si{\micro\second}. Separate 20-condition experiments tested cold-cache
and conditional-forwarding behavior; their signed-mode median completion rates
at 20,000 offered QPS were 19,999.899 and 17,633.980~QPS, respectively.
Appendix Tables~\ref{tab:completed-local-throughput}
and~\ref{tab:warm-performance-complete} report the completed workload results.

\paragraph{Heterogeneous Deployments}

We evaluate interoperability across resolver implementations using depth-3
chains composed of BIND, Unbound, and Knot Resolver. Four topologies are
considered: homogeneous BIND (B-B-B), B$\rightarrow$U$\rightarrow$B,
B$\rightarrow$K$\rightarrow$B, and B$\rightarrow$U$\rightarrow$K. Each resolver
runs native \textsc{Resolver-Path} support rather than simply forwarding an
unknown EDNS option.

Table~\ref{tab:hetero-depth-summary} shows that all topologies maintain
sub-millisecond p95 latency while achieving 100\% reported-path recovery.
Under both warm- and cold-cache conditions, packet loss remains
$\leq$0.01\% (Appendix Figure~\ref{fig:appx-heterogeneous-performance}).

\smallskip\noindent\textbf{Takeaway (RQ3, interoperability):}
Technique~2 maintains correct resolver-path accumulation and negligible
performance overhead across heterogeneous resolver implementations.
\smallskip

\begin{table}[t]
	\centering
	\scriptsize
	\caption{Heterogeneous depth-3 summary at a 20{,}000~QPS offered load in the local Docker testbed (medians over five repetitions). CPU is the summed resolver-chain cost in \si{\micro\second/query}, excluding the authoritative server. The final column shows Technique~2 wire-level recovery as positive-path rate / conditional path length.}
	\label{tab:hetero-depth-summary}
	\begin{tabular}{lrrrrrrr}
		\toprule
		\multicolumn{1}{c}{} & \multicolumn{3}{c}{p95 latency (ms)} & \multicolumn{3}{c}{CPU (\si{\micro\second/q})} & T2 path \\
		\cmidrule(lr){2-4} \cmidrule(lr){5-7}
		Topology & Off & T1 & T2 & Off & T1 & T2 & rate / len \\
		\midrule
		B-B-B & 0.024 & 0.019 & 0.026 & 15.2 & 11.5 & 16.2 & 100\% / 3.0 \\
		B-U-B & 0.083 & 0.019 & 0.077 & 18.8 & 11.6 & 18.0 & 100\% / 3.0 \\
		B-K-B & 0.024 & 0.022 & 0.028 & 14.6 & 12.7 & 18.2 & 100\% / 3.0 \\
		B-U-K & 0.095 & 0.022 & 0.091 & 20.2 & 12.8 & 19.9 & 100\% / 3.0 \\
		\bottomrule
	\end{tabular}
\end{table}

\paragraph{Recursive DNS Query Tracing (Technique~1).}

We evaluate Recursive DNS Query Tracing as an on-demand diagnostic rather than as part of
ordinary query processing. Across all cloud topologies and Docker depth
configurations, Technique~1 successfully completed every diagnostic walk
(200/200 in each configuration). Its latency reflects the dedicated sequence of
diagnostic queries and is therefore not directly comparable to ordinary-query
performance. Complete per-hop and depth results appear in Appendix
Tables~\ref{tab:t1-cloud-decomposition}
and~\ref{tab:t1-depth-scaling}.
Diagnostic and ordinary-response paths differed in 4,000/7,000 paired trials
(57.14\%); we therefore report them separately.

\smallskip\noindent\textbf{Takeaway (RQ3):}
Technique~2 reports the selected-response path with near-neutral throughput;
Technique~1 separately diagnoses reported forwarding topology.
\smallskip

\subsection{Path Attestation}
\label{sec:attestation}

Disclosure reveals what participating resolvers report, but not whether those reports 
are authentic. Attestation turns unsigned disclosure into \emph{authenticated cooperative
resolver disclosure}: participating resolvers sign after observing the response
they return. The hash chain binds each received assertion to the client nonce,
canonical question, response digest, and response-instance identifier.
The client can verify the signatures on received entries, intact hash-chain
order, and satisfaction of the nonce and timestamp checks.
Table~\ref{tab:attestation-guarantees} is the canonical detailed threat and
guarantee boundary.

\begin{table*}[t]
\centering
\scriptsize
\setlength{\tabcolsep}{4pt}
\begin{tabular}{p{0.18\textwidth}p{0.13\textwidth}p{0.61\textwidth}}
\toprule
Property & Guarantee & Boundary \\
\midrule
Authenticated assertions & Yes & A valid signature authenticates the assertion made under a reported participating operator's key; it does not establish that the asserted identity or country is true. \\
Signed-entry modification/reordering & Detectable & Verification fails when received signed bytes or intact hash links are changed. \\
Freshness & Bounded & Query/nonce/timestamp binding rejects the tested stale or mismatched replays; nonce-reuse detection requires verifier state. \\
Omitted entries or non-participating hops & No & A resolver can decline to append, and hidden hops need not preserve the option. \\
Stripping or arbitrary removal & No & Internal removal before a later intact entry can break a hash link, but suffix, whole-chain, or replacement stripping is not generally detectable. \\
False country, execution location, or physical traversal & No & The signature authenticates the operator's assertion, not the assertion's external truth. \\
Collusion & No & Colluding operators can coordinate assertions under keys they control. \\
Path completeness / regulatory compliance & No & The mechanism supplies audit evidence; it neither proves completeness nor decides legal compliance. \\
\bottomrule
\end{tabular}
\caption{Threat and guarantee summary for authenticated cooperative resolver disclosure.}
\label{tab:attestation-guarantees}
\end{table*}

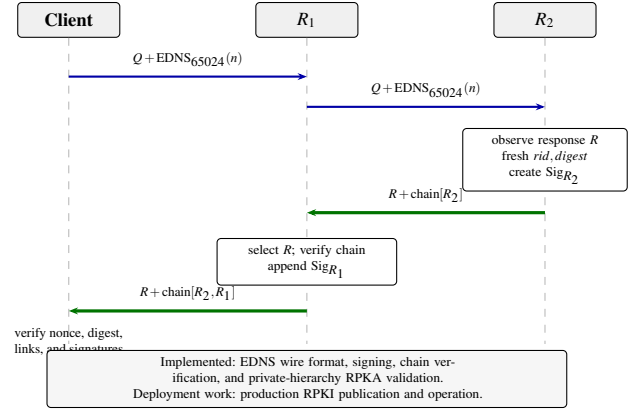
\begin{figure}[t]
\centering
\begin{tikzpicture}[
  x=1cm,
  y=1cm,
  entity/.style={rectangle, draw, fill=gray!12, minimum width=1.35cm,
    minimum height=0.46cm, font=\scriptsize\bfseries, rounded corners=1.5pt},
  query/.style={-{Stealth[length=3.5pt]}, thick, draw=blue!65!black},
  response/.style={-{Stealth[length=3.5pt]}, very thick, draw=green!45!black},
  life/.style={dashed, gray!55},
  action/.style={rectangle, draw, fill=white, rounded corners=1.5pt,
    font=\tiny, align=center, inner sep=2.5pt},
  status/.style={rectangle, draw, fill=gray!7, rounded corners=1.5pt,
    font=\tiny, align=center, text width=6.7cm, inner sep=2.5pt},
]
  \node[entity] (C) at (0,0) {Client};
  \node[entity] (R1) at (3.15,0) {$R_1$};
  \node[entity] (R2) at (6.3,0) {$R_2$};

  \draw[life] (C.south) -- ++(0,-4.05);
  \draw[life] (R1.south) -- ++(0,-4.05);
  \draw[life] (R2.south) -- ++(0,-4.05);

  \draw[query] (0,-0.75) -- node[above, font=\tiny]
    {$Q+\mathrm{EDNS}_{65024}(n)$} (3.15,-0.75);
  \draw[query] (3.15,-1.15) -- node[above, font=\tiny]
    {$Q+\mathrm{EDNS}_{65024}(n)$} (6.3,-1.15);

  \node[action, text width=2.0cm] at (6.3,-1.85)
    {observe response $R$\\fresh $rid,digest$\\create $\mathrm{Sig}_{R_2}$};

  \draw[response] (6.3,-2.55) -- node[above, font=\tiny]
    {$R+\mathrm{chain}[R_2]$} (3.15,-2.55);
  \node[action, text width=2.2cm] at (3.15,-3.2)
    {select $R$; verify chain\\append $\mathrm{Sig}_{R_1}$};

  \draw[response] (3.15,-3.85) -- node[above, font=\tiny]
    {$R+\mathrm{chain}[R_2,R_1]$} (0,-3.85);
  \node[font=\tiny, align=center, anchor=north] at (0,-3.95)
    {verify nonce, digest,\\links, and signatures};

  \node[status] at (3.15,-4.75)
    {Implemented: EDNS wire format, signing, chain verification, and private-hierarchy RPKA validation.\\
     Deployment work: production RPKI publication and operation.};
\end{tikzpicture}
\caption{Signed EDNS return path for Technique~2 (option~65024). The client
supplies a fresh nonce; after $R_2$ observes response $R$, it creates a fresh
response identifier and digest and signs its entry. $R_1$ selects that response,
verifies the downstream chain, and extends it before the client verifies the
received assertions.}
\Description{A compact message-sequence diagram for signed EDNS option 65024.
The client sends a query and fresh nonce through R1 to R2. After observing the
response, R2 creates a response identifier and digest and signs. R1 selects the
response, verifies and extends its chain, and returns it for client verification.
The diagram distinguishes implemented signing and verification from the
deployment work of publishing RPKI key-authorization objects.}
\label{fig:edns-signed-return-path}
\end{figure}

\paragraph{Design.}
The response producer creates a fresh identifier, canonicalizes the question
and selected response, and signs its digest only after observing the response.
Each upstream participant verifies the selected response's downstream chain and
appends its claimed ASN and country, timestamp, and prior-entry hash. Ed25519
covers both assertion fields; RPKI authorizes only the ASN-to-key binding. Appendix
~\ref{sec:appx:attestation-wire} specifies the bytes, and Appendix
Figure~\ref{fig:response-bound-attestation} expands the fan-out, retry, sibling,
and cache rules summarized in Figure~\ref{fig:edns-signed-return-path}.

Attestation follows the response-path semantics defined above. A participant
receiving an unattested response may start a gap-marked partial chain, and the
client distinguishes valid, invalid, unattested, and unknown/revoked-key
evidence.

The implementation signs every entry's ASN and country and validates ASN-to-key
authorization using an RPKI object under a private test trust anchor and the
standard CMS profile~\cite{rfc6488}. Validation covers the trust-anchor chain,
current manifest and CRL, certificate resource coverage, validity interval, and
key identifier. Production deployment requires standardizing the RPKA object
type and publishing it through production repositories (Appendix
~\ref{sec:appx:attestation-wire}).

\paragraph{Evaluation.}
Wire and transport validation covered zero through six entries over IPv4 and
IPv6, comparing serializer lengths with packet captures. Appendix
Table~\ref{tab:attestation-size-validation} reports exact sizes, budgets, and
verification work; Appendix Table~\ref{tab:response-bound-conformance} reports
cross-implementation conformance.

Two standalone, separately implemented codebases consumed the 18-case
published specification vector suite for five repetitions: Go using
standard-library Ed25519 and
Python using \texttt{cryptography} Ed25519. Neither imports the production
implementation or vector generator. Across 180 implementation decisions, all
18/18 vectors were evidence-ready; the codebases agreed on 90/90 paired
verdicts and 85/85 signer-applicable canonical signatures, and recorded
180/180 agreements with the published expected state, parser/serializer
result, and canonical hash. Every decision links to a client event, a separate
implementation log, and one of 180 packets in 5 PCAPs. Canonical re-signing
matched 130/170 signer-applicable presented payloads; the 40 expected
non-matches are the removed-entry, reordered-entry, modified-entry, and
copied-signature mutation
vectors. This establishes agreement between the two codebases for the published
vectors, not provenance from independent human teams or authors.

Across 240 runs covering 48 conditions, all 1,840 packet captures and 1,600
linked event sets matched. At a 1,232-byte UDP payload, signed depth-eight and
depth-15 cases used TCP in 10/55 and 45/125 cases per IP family, respectively;
Appendix~\ref{sec:appx:wire-matrix} gives the condition-level accounting.

Validation confirmed a 36-byte header and 144-byte entries, with option sizes
$36+144h$ bytes ($40+144h$ including the EDNS option header) for
$h=0,\ldots,6$. IPv4 and IPv6 had identical DNS-layer sizes. At six entries,
the option contains 900 bytes of data and is 904 bytes in full, bounding work at
six Ed25519 verifications and RPKA lookups. Captures confirmed TC/TCP fallback
for responses exceeding the UDP budget while preserving the complete chain
(Appendix Table~\ref{tab:attestation-size-validation}).

Per-entry Ed25519 signing was 36.8--37.1\,\si{\micro\second} p50 and verification
93.4--95.2\,\si{\micro\second} p50. Direct depth-two verification took
149\,\si{\micro\second} p50 and 158\,\si{\micro\second} p95 over ten iterations.

To measure throughput as load and reported-path depth increased, we ran five
300-second repetitions for every combination of depths 2, 3, and 5,
baseline/unsigned/signed modes, and ten offered loads from
2,000 to 20,000~QPS: 450 runs and 37.5 measured hours. At 20,000
offered QPS, median signed throughput was 19.7k, 19.5k, and 19.4k~QPS at
depths 2, 3, and 5, respectively. Across all loads, signed loss remained below
0.001\%; the maximum was 1,595 losses across 163.4 million queries at depth~5
(Appendix Table~\ref{tab:attestation-depth-load}). A
separate CPU/latency experiment ran at 2,000~QPS; completed heterogeneous runs
reported verification rate 1.0 and no observed truncation or TCP fallback.

\smallskip\noindent\textbf{Takeaway (RQ3, attestation):}
\textsc{Resolver-Path}'s wire behavior matched the validated model, signed mode
sustained 19.4k--19.7k median QPS at 20,000 offered QPS with all-load loss below
0.001\%, and the standalone Go and Python codebases agreed throughout the
published-vector experiment. These results do not establish path completeness.
\smallskip

\paragraph{Privacy considerations.}
The design adds no extra queries and defaults to AS and country identifiers to avoid exposing per-resolver topology. Appendix~\ref{sec:appx:protocol-security} gives additional detail on amplification, ECS coexistence, and infrastructure exposure.


\section{Related Work}
\label{sec:relwork}

X-Trace propagated EDNS0 metadata through recursive DNS operations and reconstructed cross-layer task trees from out-of-band reports~\cite{fonseca:xtrace}; \textsc{Resolver-Path} instead returns with the DNS response an ordered sequence of participating recursive resolvers, with optional response-bound attestation.

\paragraph{Measuring hidden DNS dependencies.}
Schomp et~al. established the ingress, forwarding, hidden, and egress taxonomy of client-side DNS infrastructure and inferred resolver pools from client and authoritative vantage points~\cite{schomp:dnsinfrastructure}. Their methodology cannot identify resolvers hidden between observed ingress and egress; the measurement in this paper inherits that boundary. Liu et~al. compared the resolver a client intended to contact with response and authoritative-side signals to detect on-path DNS interception~\cite{liu:who-answering-queries}. The measurement here likewise uses controlled queries and authoritative observations, but studies outsourcing and cross-border dependencies rather than treating this two-vantage construction as new. DNSRoute++ exploits transparent forwarding to extend TTL-limited probes beyond a forwarder and reveal the \emph{network-router path} toward its recursive resolver~\cite{nawrocki:forwarders}; it neither enumerates a logical chain of recursive resolvers nor provides authenticated resolver identities. These approaches reveal opacity from outside. \textsc{Resolver-Path} instead asks participating resolvers to disclose the logical resolver path used by a query.

Follow-up measurements cover open resolvers~\cite{al-dalky_schomp_2018,alzoubi:ldns}, cellular DNS~\cite{rula:behind}, egress dependencies~\cite{luo:who-querying-for-me}, and open-DNS clustering~\cite{wu:odns-clustering}. Public-DNS dominance is also well documented~\cite{globaldns:callejo,doan:eval_publicdns,huston:centralization,gmoura:apnic-blog20}. The Google concentration result is therefore motivational and partly confirmatory; the measurement contribution is its cross-border and jurisdictional analysis, stable-panel longitudinal control, and identity-first handling of anycast.

\paragraph{Information for selection versus evidence of a chain.}
RESINFO is the closest selection-oriented DNS standard: a contacted resolver self-publishes capabilities, such as QNAME minimization and supported extended errors, for clients to use in resolver selection~\cite{rfc9606}. RESINFO describes that resolver, not the downstream resolvers used for a particular query. CHAIN also returns additional material in one DNS exchange, but its ``chain'' is the DNSKEY, DS, and related RRsets needed to validate an authoritative DNSSEC delegation, not a chain of recursive operators~\cite{rfc7901}. NSID~\cite{rfc5001} and CHAOS TXT records~\cite{isc_bind_chaos} identify a single answering instance. In contrast, \textsc{Resolver-Path} carries disclosures about the identities, jurisdictional assertions, and order of reported participating resolver hops; attestation binds those assertions to the selected response and client nonce. Appendix Table~\ref{tab:comparison} makes these boundaries explicit.

\paragraph{From opaque steering to resolver-path accountability.}
Prior work characterizes CDN steering strategies and their implications for performance, resilience, and delivery geography~\cite{rkumar:steering}. That work infers which steering strategy a CDN deploys; it does not reveal the downstream resolvers handling a particular query. Resolver forwarding creates a related visibility gap when queries traverse multiple organizations. This paper's novelty is cooperative disclosure of the logical resolver path carried by a selected response and cryptographic attestation of its reported resolver assertions---not the discovery that resolver chains, centralization, or cross-border DNS behavior exist.

\paragraph{Encrypted DNS and path attestation.}
DoT~\cite{dickinson:ietf}, DoH~\cite{rfc8484}, and ODoH~\cite{rfc9230} protect queries from observers~\cite{lu:dns-encryption,doh:imc19,hounsel:dnsperformance,10.1145/3359989.3365429,doan:dot}; \textsc{Resolver-Path} is complementary because it exposes reported path evidence to the client without providing query confidentiality. BGPsec~\cite{rfc8205}, Certificate Transparency~\cite{rfc9162}, and SCION~\cite{scion:book} provide accountability analogies, while DNSSEC signs DNS data rather than resolver participation~\cite{goingwild:marc,corrupteddns:dagon,dnscensor:venezuela,aryan:iran}.

\section{Limitations and Ethical Considerations}
\label{sec:limitations-ethics}

\paragraph{Scope of findings.}
RIPE Atlas provides broad global coverage but overrepresents fixed-broadband networks, particularly in Europe; mobile, enterprise, IoT, and VPN-mediated paths are comparatively underrepresented. Probes in the same AS share resolver configuration, so observations are not independent and the reported rates describe the measured population rather than statistical estimates for the Internet as a whole (\S\ref{sec:Findings}). Our measurements characterize resolver behavior during the 2023--2025 campaigns; because resolver deployments evolve over time, the reported percentages should be interpreted as representative of that period rather than immutable Internet-wide constants.

\paragraph{Conservative AS attribution.}
Private ingress provides no publicly attributable BGP-origin ASN, so the
full-population comparison assigns each private ingress to its probe's client
AS. It also retains the 83 Campaign~B observations missing a client and/or
ingress ASN and counts them as same-AS. These choices place unattributable cases
in the non-mismatch category: they retain all 16,636 successful observations in
the denominator while 6,622 are classified as different-AS. The resulting
39.8\% client--frontend AS-mismatch rate is therefore a conservative lower bound
on observed AS-boundary crossing, not an ownership claim about private or
ASN-incomplete ingress (\S\ref{sec:methodology}).

\paragraph{Generalizability of tracing mechanisms.}
The \textsc{Resolver-Path} evaluation demonstrates feasibility under controlled
deployments, including cache hits, multiple upstream responses, and intentional
partial participation across 23 scenario and 32 topology conditions, rather
than production-scale resolver ecosystems. Production deployments involve
other stacks, middleboxes, more varied fan-out, and higher throughput than the
20{,}000~QPS offered-load experiments on 2-vCPU \texttt{c7i.large} BIND
testbeds. Its EDNS disclosure depends on intermediate
resolvers implementing and preserving the option; implementations that ignore
or strip unrecognized options~\cite{rfc6891} truncate the accumulated path. The
controlled cache matrix validates 17,000 selected cache-hit paths with no
historical-path reuse, but does not characterize the distribution of
cache-terminated path lengths in operational deployments.

\paragraph{Deployment, attestation, and privacy.}
Deployment depends on operator adoption and preservation of the EDNS option.
\textsc{Resolver-Path} generates and validates its RPKI ASN-to-key authorization
under a private test trust anchor and implements the response-bound wire format,
but production RPKA publication and rollover, production-scale nonce state, and
operational key-revocation behavior remain unevaluated. Two standalone,
separately implemented Go and Python codebases agree on the published
conformance vectors, but this does not establish independent human-team or
authorship provenance, production interoperability, or assertion truth.
Table~\ref{tab:attestation-guarantees} consolidates the remaining trust and
adversary boundaries; AS and country disclosure reduces topology exposure.

\paragraph{Ethical measurement practices.}
Atlas queries target controlled authoritative domains, and results are reported only in aggregate.
For performance tests, the replay trace was collected by the university IT
department during routine network operations. The authors received only an
anonymized version of the trace, in which client identifiers and queried names
had been removed or anonymized prior to release. The organization that owns and
operates the measured network approved secondary research use and release of
the trace in transformed form; a local authoritative service in the Docker
testbed answered every replay name.

\section{Conclusions}
\label{sec:conclusion}

We showed that observed resolver endpoints frequently cross organizational and
national boundaries, with measurable implications for performance,
jurisdiction, and infrastructure concentration. Existing DNS mechanisms expose
only the endpoints of resolver chains; widespread ingress--egress address and
prefix divergence leaves the connecting interval unresolved, but does not
identify its mechanism or prove an additional forwarding hop. We introduced
\textsc{Resolver-Path}, a lightweight cooperative disclosure mechanism whose
base mode exposes the reported resolver path with near-neutral completion
throughput in the completed controlled workloads. Its authenticated extension
binds received cooperative assertions to the selected response while leaving
non-participating hops outside the evidence.
By expanding the observable state available to Internet measurement,
\textsc{Resolver-Path} enables a new class of studies of resolver behavior,
infrastructure dependencies, and their evolution over time that are not
possible with today's endpoint-only observations.

\clearpage
\bibliographystyle{plain}
\bibliography{base,reference}
\clearpage
\appendix

\section{Appendix Overview}
\label{sec:appx:overview}

This appendix records reproducibility and filtering accounting, protocol and
attestation detail, supplementary measurement results, geolocation methodology,
and ethics limitations. It also distinguishes the evidence supporting the
submitted claims from validation that remains future work.

\begin{table*}[t]
\centering
\scriptsize
\caption{Closest prior work and the boundary of the novelty claim. ``Multihop IDs'' means identities of logical resolvers rather than IP-router hops; ``response path'' means an ordered resolver sequence carried by the selected response; attestation applies only to reported participating hops.}
\label{tab:comparison}
\begin{tabular}{@{}lp{7.2cm}ccc@{}}
\toprule
Work & Object exposed & Multihop IDs & Response path & Response-bound attestation \\
\midrule
Schomp et~al.~\cite{schomp:dnsinfrastructure} & Externally inferred ingress, egress, and resolver pools; intermediate resolvers remain hidden & partial & \texttimes & \texttimes \\
Liu et~al.~\cite{liu:who-answering-queries} & On-path interception of traffic intended for a chosen resolver & \texttimes & \texttimes & \texttimes \\
DNSRoute++~\cite{nawrocki:forwarders} & IP-router path through a transparent forwarder toward its recursive resolver & \texttimes & \texttimes & \texttimes \\
RESINFO~\cite{rfc9606} & Self-published capabilities of the contacted resolver & \texttimes & \texttimes & \texttimes \\
CHAIN~\cite{rfc7901} & DNSSEC validation RRsets along the authoritative delegation chain & \texttimes & \texttimes & \texttimes \\
Kumar et~al.~\cite{rkumar:steering} & CDN replica-selection strategy inferred from latency & \texttimes & n/a & \texttimes \\
\midrule
This work & Identities, jurisdictional assertions, and order carried by the selected response & \checkmark & \checkmark & \checkmark \\
\bottomrule
\end{tabular}
\end{table*}

\section{Protocol Security and Privacy Details}
\label{sec:appx:protocol-security}

\paragraph{Amplification and flooding.}
Technique~2 introduces no extra ordinary-query round trips, while Technique~1
is an explicit diagnostic walk. The unsigned Technique~2 experiments measured
94.5--98.2\,B responses and no truncation or TCP fallback in the tested cases.
\textsc{Resolver-Path} validation confirmed its 36-byte header, fixed 144-byte
entries, six-entry limit, and maximum 900-byte option data and 904-byte complete
option (Table~\ref{tab:attestation-size-validation}). The advertised EDNS UDP
payload bounds the DNS message, not the enclosing IP/UDP packet, so IPv4 and
IPv6 have the same hop budget. Ordinary IPv4/UDP adds 28 on-wire bytes beyond
the DNS message; IPv6/UDP adds 48 bytes absent extension headers. Packet
captures at the tested EDNS limits recorded TC signaling and TCP fallback for
over-budget selected responses while preserving every signed entry.

\paragraph{Coexistence with EDNS Client Subnet.}
EDNS Client Subnet (ECS, RFC~7871~\cite{rfc7871}) shares the EDNS option space
with \textsc{Resolver-Path}. Implementations may preserve recognized options
such as ECS while stripping unknown ones, truncating disclosure without
affecting normal resolution. ECS and \textsc{Resolver-Path} share the
requestor's advertised payload budget; an implementation applies the same
validated TC/TCP behavior rather than exceeding that budget or silently
deleting entries.

\paragraph{Infrastructure exposure and encrypted DNS.}
The EDNS path option reveals resolver identifiers to the client and, in the
raw-IP implementation, to the authoritative server. DoT/DoH protects metadata
in transit but not from endpoints. AS+country granularity reduces topology
exposure. Both fields are covered by the entry signature, but RPKI authorizes
only the ASN-to-key binding; the country remains an operator assertion rather
than independent location or jurisdiction proof. Finer-grained identifiers
add explicit topology exposure.

\paragraph{Attestation integrity guarantees.}
Assuming authenticated key distribution, signatures authenticate assertions
under reported operators' keys and expose modification or reordering of entries
that remain linked in the received chain. They do not validate the assertions'
truth. A resolver can omit itself, strip a suffix or the whole option, replace a
chain with a new partial chain, or collude under keys it controls. Only removal
that leaves a later intact hash-linked entry is necessarily exposed. The
mechanism therefore provides authenticated cooperative assertions, not proof of
physical traversal, complete participation, or regulatory compliance.

\subsection{Attestation Design and Evaluation Details}
\label{sec:appx:attestation-details}

\subsubsection{Response-Bound \textsc{Resolver-Path} Wire Specification}
\label{sec:appx:attestation-wire}

\paragraph{Option header and byte order.}
EDNS option code 65024 carries one 36-byte header followed immediately by
\emph{entry count} fixed-width entries. All integers use unsigned network byte
order. Octet strings have the widths in Table~\ref{tab:attestation-wire}; there
is no padding. A query has message kind 0, zero flags, zero entries, a 16-byte
client nonce, and an all-zero response identifier. A response has message kind
1, a nonzero 16-byte response identifier, and one to six entries. Both the
nonce and response identifier are generated independently with a
cryptographically secure random-number generator.

\begin{table*}[t]
\centering
\scriptsize
\begin{tabular}{@{}llrp{8.2cm}@{}}
\toprule
Structure & Field, in wire order & Bytes & Encoding and rule \\
\midrule
Header & protocol version & 1 & Value 1 \\
 & message kind & 1 & 0=query, 1=response \\
 & flags & 1 & All bits zero at protocol version 1 \\
 & entry count & 1 & 0 in a query; 1--6 in a response \\
 & client nonce & 16 & Uniform random octets supplied by the client \\
 & response identifier & 16 & Zero in a query; fresh uniform random octets per produced response \\
\midrule
Entry & identity namespace & 1 & Value 1 denotes an RPKI ASN \\
 & ASN & 4 & Unsigned 32-bit autonomous-system number \\
 & country & 2 & Signed uppercase US-ASCII ISO~3166-1 alpha-2 operator assertion; not RPKI-authorized \\
 & entry flags & 1 & Bit 0 GAP\_BELOW; bit 1 CACHE\_ANSWER; other bits zero \\
 & key identifier & 32 & SHA-256 of the key's DER SubjectPublicKeyInfo \\
 & timestamp & 8 & Unix seconds, UTC \\
 & prior-entry hash & 32 & SHA-256 of the complete preceding entry, or 32 zero octets for entry 0 \\
 & signature & 64 & Ed25519 signature defined below \\
\bottomrule
\end{tabular}
\caption{Normative \textsc{Resolver-Path} serialization. The header is 36 bytes
and each entry is exactly 144 bytes.}
\label{tab:attestation-wire}
\end{table*}

Entries are ordered in response direction: entry 0 is the deepest reported
participant that produced or first received the response, and later entries
move toward the client. A user interface may reverse this order for display but
never changes the signed bytes. The \texttt{GAP\_BELOW} bit states only that the
signer received the selected response without usable downstream attestation;
\texttt{CACHE\_ANSWER} states that the signer produced the response from local
cache. Neither bit asserts completeness.

\paragraph{Canonical question and response.}
The canonical question is QNAME in lowercase, uncompressed DNS wire form,
including its terminal zero octet, followed by QTYPE and QCLASS as two-byte
integers. Exactly one question is permitted. To compute the canonical response
digest, first encode each RR as canonical lowercase, uncompressed owner name;
two-byte TYPE; two-byte CLASS; two-byte RDATA length; and RDATA in the DNSSEC
canonical form of RFC~4034, Section~6.2. Embedded domain names are therefore
lowercase and uncompressed; for an unknown type, the RDATA wire octets are
unchanged. TTL is deliberately omitted so forwarding-time TTL decrement does
not break the chain. OPT, TSIG, and SIG(0) records are excluded, including the
\textsc{Resolver-Path} option itself. Within each answer, authority, and
additional section, retain duplicates and sort the encoded RRs
lexicographically by unsigned octet value.

The response serialization is the ASCII string
\texttt{Resolver-Path response v1} followed by one zero octet; the 16-byte
response identifier; a two-byte canonical-question length and the canonical
question; a two-byte status whose high four bits are OPCODE and low 12 bits are
the effective RCODE; one semantic-flags byte whose bits 0, 1, and 2 are AA, TC,
and AD (other bits zero); then answer, authority, and additional in that order.
Each section is encoded as its one-byte identifier (1, 2, or 3), a two-byte RR
count, and for each sorted RR a four-byte length followed by its bytes. The
canonical response digest is SHA-256 of this entire serialization. Thus
transaction ID, RD, RA, CD, TTL, record order, and EDNS options are
intentionally outside the digest; answer data, response status, and the unique
response instance are inside it.

\paragraph{Signed bytes and chaining.}
For each entry, the Ed25519 input is exactly the following concatenation: the
ASCII string \texttt{Resolver-Path signature v1} followed by one zero octet;
the one-byte protocol version; the 16-byte client nonce; a two-byte canonical
question length and the canonical question; the 32-byte canonical response
digest; the one-byte identity namespace; four-byte ASN; two-byte country;
one-byte entry flags; 32-byte key identifier; eight-byte timestamp; and
32-byte prior-entry hash. The signature is not part of its own input. Entry 0
uses 32 zero octets as its prior hash; entry $i>0$ uses SHA-256 over all 144
wire bytes of entry $i-1$, including that entry's signature. Consequently, a
forwarding resolver verifies the complete downstream chain for its selected
response before signing the hash of its last entry.

\paragraph{Identity and key authorization.}
\textsc{Resolver-Path} uses the RPKI ASN namespace. A \textsc{Resolver-Path} Key
Authorization (RPKA) is an RPKI signed object using the standard CMS
profile~\cite{rfc6488} and a DER-encoded payload containing, in order, the ASN, Ed25519
SubjectPublicKeyInfo, 32-byte key identifier, not-before time, not-after time,
and serial number. The key identifier must equal SHA-256 of the exact DER
SubjectPublicKeyInfo. Verification requires a valid CMS chain to an accepted
RPKI trust anchor, a current manifest and CRL, an end-entity certificate whose
AS resources cover the payload ASN, a current validity interval, and a matching
key identifier and Ed25519 signature. Failure of the authorization chain or
entry signature is \emph{invalid}; an unavailable object, unknown key, expired
interval, or revocation is \emph{unknown/revoked key}. The implemented entry
serialization includes the country in the Ed25519 input, so modification of
either the ASN or country invalidates the signature. RPKI authorizes only the
ASN-to-key binding and does not certify the signed country's geographic truth.
The prototype generates and validates the RPKA in a private RPKI hierarchy,
including its trust-anchor chain, manifest and CRL, end-entity certificate
resource coverage, validity interval, and key identifier. Allocating and
standardizing the RPKA eContentType and operating production publication
infrastructure remain deployment work; the RPKA payload and verifier algorithm
are specified here rather than left as alternatives.

\paragraph{Freshness.}
The client generates a uniform 128-bit nonce for each outstanding query and
never has two outstanding queries with the same nonce. Timestamps are unsigned
64-bit Unix seconds in network byte order. A verifier accepts an entry only
when its nonce matches the outstanding query and its timestamp is within
300 seconds of the verifier's clock. The client marks the nonce consumed when
it selects a response, rejects a second selected response for that nonce, and
retains consumed nonce--response-identifier pairs for 300 seconds. A response
producer, including a cache-answering resolver, generates a fresh response
identifier and chain for every response. Reuse of a historical cache chain,
an expired timestamp, a consumed nonce, or a response identifier transplanted
from a sibling response is invalid.

\paragraph{Partial deployment and response selection.}
The transition rules are deterministic. A valid selected downstream response
is preserved and extended. A response without the option causes the first
participating upstream resolver to create a fresh response identifier and entry
with \texttt{GAP\_BELOW}; option absence at the client is \emph{unattested}.
A cache or locally produced answer creates a fresh identifier and chain for the
current nonce and sets \texttt{CACHE\_ANSWER}. Malformed or cryptographically
invalid downstream data is relayed as invalid and is not converted into a new
valid partial chain; an unverifiable key similarly retains the
unknown/revoked-key outcome. With retries or fan-out, every produced response
has an independent identifier and chain; a resolver verifies and extends only
the response it selects, discards losing chains, and never merges entries. On
receiving six entries, an upstream participant forwards the chain without
appending and the verifier marks the result resource-limited.

\begin{figure*}[t]
\centering
\begin{tikzpicture}[
  x=1cm,
  y=1cm,
  >=Latex,
  entity/.style={draw, rounded corners=1.5pt, fill=gray!10,
    minimum width=2.35cm, minimum height=0.58cm,
    align=center, font=\footnotesize\bfseries},
  life/.style={gray!55, densely dashed},
  query/.style={-{Latex[length=2.2mm]}, semithick, blue!70!black},
  selectedcandidate/.style={-{Latex[length=2.2mm]}, very thick,
    green!45!black, densely dashed},
  discardedcandidate/.style={-{Latex[length=2.2mm]}, thick,
    gray!65, densely dashed},
  selectedresponse/.style={-{Latex[length=2.4mm]}, ultra thick,
    green!40!black},
  event/.style={draw, rounded corners=1.5pt, fill=blue!3,
    text width=3.55cm, align=center, inner sep=3pt, font=\scriptsize},
  selectbox/.style={draw=green!45!black, very thick, rounded corners=2pt,
    fill=green!6, text width=5.75cm, align=center, inner sep=4pt,
    font=\scriptsize},
  rejectbox/.style={draw=red!65!black, rounded corners=2pt, fill=red!4,
    text width=8.35cm, align=left, inner sep=4pt, font=\scriptsize},
  cachebox/.style={draw=blue!55!black, rounded corners=2pt, fill=blue!3,
    text width=6.05cm, align=left, inner sep=4pt, font=\scriptsize},
  boundary/.style={draw=black!65, rounded corners=2pt, fill=gray!8,
    text width=16.35cm, align=center, inner sep=4pt, font=\scriptsize}
]

\node[entity] (C)  at (0.9,0) {Client};
\node[entity] (R1) at (5.0,0) {Forwarder $R_1$};
\node[entity] (RA) at (10.1,0) {Downstream branch A};
\node[entity] (RB) at (15.2,0) {Downstream branch B};

\draw[life] (C.south)  -- ++(0,-6.55);
\draw[life] (R1.south) -- ++(0,-6.55);
\draw[life] (RA.south) -- ++(0,-5.00);
\draw[life] (RB.south) -- ++(0,-5.00);

\draw[query] (C.south) ++(0,-0.72) --
  node[above, font=\scriptsize] {one query $Q$ with fresh client nonce $n$}
  (R1.south |- 0,-1.30);
\draw[query] (R1.south |- 0,-1.72) --
  node[above, font=\scriptsize] {fan-out: $Q(n)$}
  (RA.south |- 0,-2.05);
\draw[query] (R1.south |- 0,-2.30) --
  node[above, font=\scriptsize] {fan-out / retry: $Q(n)$}
  (RB.south |- 0,-2.63);

\node[event] (EA) at (10.1,-3.18) {observe response A\\
  create \texttt{rid\_A}, \texttt{digest\_A}\\
  sign/extend \texttt{chain\_A} on return};
\node[event] (EB) at (15.2,-3.18) {observe response B\\
  create \texttt{rid\_B}, \texttt{digest\_B}\\
  sign/extend \texttt{chain\_B} on return};

\draw[selectedcandidate] (RA.south |- 0,-4.12) --
  node[above=2pt, align=center, font=\scriptsize]
    {candidate response A: \texttt{rid\_A}, \texttt{digest\_A}, \texttt{chain\_A}}
  (R1.south |- 0,-4.48);
\draw[discardedcandidate] (RB.south |- 0,-4.63) --
  node[above=2pt, align=center, font=\scriptsize, text=black!75]
    {candidate response B: \texttt{rid\_B}, \texttt{digest\_B}, \texttt{chain\_B}}
  (R1.south |- 0,-5.02);

\node[selectbox] (SEL) at (5.0,-5.65) {\textbf{selected: A}\\
  verify \texttt{chain\_A}, then append
  $\mathrm{Sig}_{R_1}(\mathrm{digest}_A,H(\mathrm{chain}_A))$\\
  after $R_1$ has observed response A};
\node[font=\scriptsize\bfseries, text=black!70, align=center]
  at (11.8,-5.40) {B discarded;\\never merged};

\draw[selectedresponse] (R1.south |- 0,-6.36) --
  node[below=2pt, align=center, font=\scriptsize\bfseries]
    {selected response A with extended \texttt{chain\_A}}
  (C.south |- 0,-6.66);

\node[rejectbox, anchor=north west] (REJ) at (0.15,-7.05)
  {\textbf{Sibling transplant:} placing \texttt{chain\_B} beneath the intact
   selecting-resolver entry for A is \textbf{INVALID}:
   \texttt{digest\_B}$\neq$\texttt{digest\_A} and
   $H(\texttt{chain\_B})\neq H(\texttt{chain\_A})$.};
\node[cachebox, anchor=north east] (CACHE) at (16.95,-7.05)
  {\textbf{Cache hit (current nonce $n$):} create a fresh
   \texttt{rid\_C}, \texttt{digest\_C}, and \texttt{chain\_C}; never reuse
   historical attestation.};

\node[boundary, anchor=north] at (8.55,-8.42)
  {\textbf{Guarantee boundary:} valid attestation covers only received
   cooperative assertions. It does not prove completeness or physical
   traversal, and arbitrary whole-option stripping or replacement remains
   outside the guarantee.};

\draw[query] (0.25,0.65) -- (1.25,0.65)
  node[right, font=\scriptsize] {query};
\draw[selectedcandidate] (2.35,0.65) -- (3.35,0.65)
  node[right, font=\scriptsize] {candidate response};
\draw[selectedresponse] (5.55,0.65) -- (6.55,0.65)
  node[right, font=\scriptsize] {selected response};

\end{tikzpicture}
\caption{Response-bound attestation under fan-out and retries. Solid blue
arrows carry one query and its fresh client nonce; dashed arrows carry candidate
responses, and the thick green arrow marks the selected response. Each
downstream branch first observes its response and then creates a distinct
response identifier, canonical response digest, and independent chain. $R_1$
selects A, verifies
\texttt{chain\_A}, and only then appends its signature; B is discarded, and
sibling chains are never merged. Transplanting \texttt{chain\_B} beneath the
intact selecting-resolver entry for A fails the response-digest/prior-hash
checks. Valid attestation covers only received cooperative assertions, not
completeness or physical traversal; arbitrary whole-option stripping or
replacement remains outside the guarantee.}
\Description{A message-sequence diagram in which a client sends one query with
a fresh nonce to forwarding resolver R1. R1 fans the query out to downstream
branches A and B. Each branch observes a different response, creates a distinct
response identifier and canonical digest, and signs an independent return-path
chain. R1 receives both candidates, selects A, verifies chain A, appends its
signature only to A, returns A to the client, and discards B without merging
chains. Callouts show rejection of a transplanted sibling chain, creation of a
fresh chain for a cache hit, and the non-completeness guarantee boundary.}
\label{fig:response-bound-attestation}
\end{figure*}
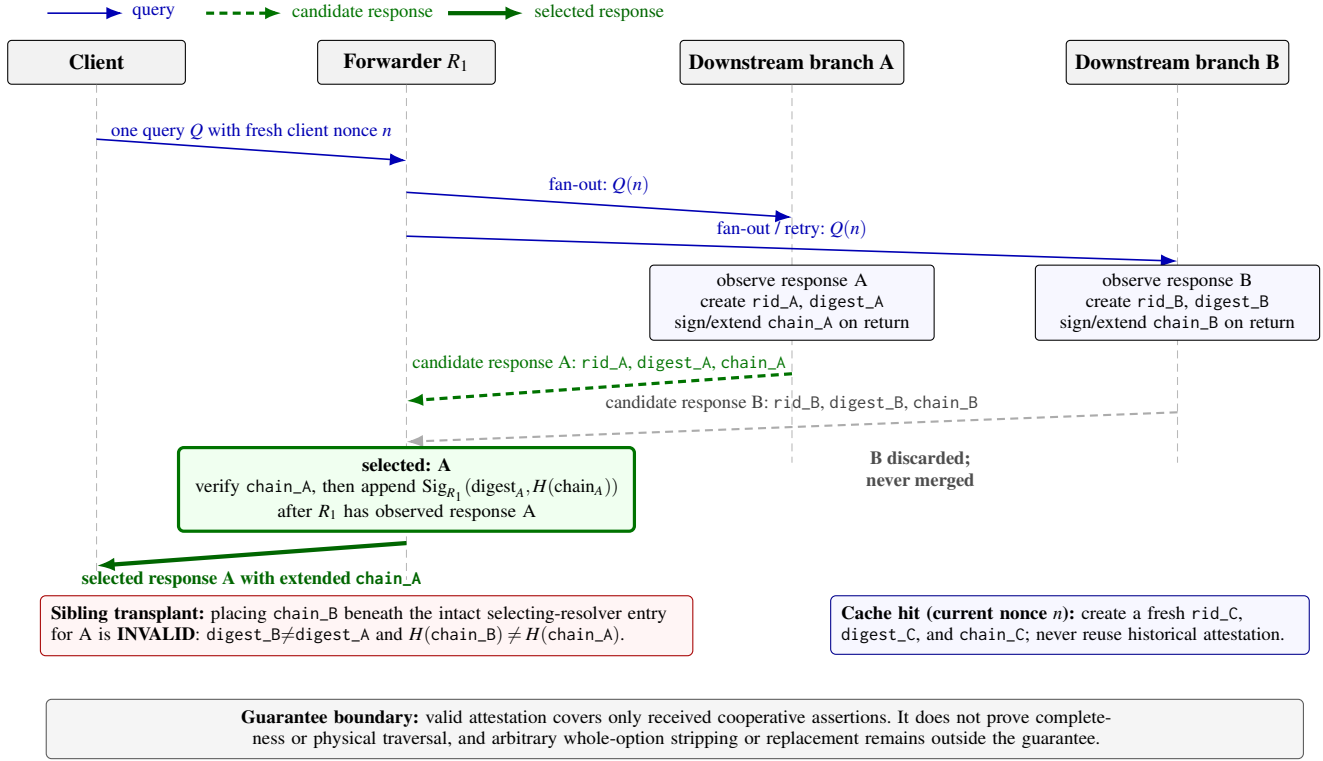

\begin{table*}[t]
\centering
\scriptsize
\begin{tabular}{@{}lp{3.4cm}p{8.7cm}@{}}
\toprule
Verifier state & Trigger & Interpretation \\
\midrule
VALID\_RECEIVED & Every received entry, link, digest, nonce, time, and key authorization verifies & Only the integrity, order, and freshness of these received cooperative assertions; never completeness \\
INVALID & Bad layout, reserved bit, count/length, nonce, time, digest, response identifier, link, authorization chain, or signature & Received attestation failed verification \\
UNATTESTED & Selected response contains no option & No attestation evidence; stripping and non-participation are indistinguishable \\
KEY\_UNAVAILABLE & A referenced authorization is unknown, expired, revoked, or unavailable & Assertions are not accepted as valid until the key state is resolved \\
\bottomrule
\end{tabular}
\caption{\textsc{Resolver-Path} verifier states. VALID\_RECEIVED may carry GAP\_BELOW and/or
RESOURCE\_LIMITED modifiers; neither modifier nor an unmodified valid state is
a claim that the path is complete.}
\label{tab:attestation-verdicts}
\end{table*}

Unsupported protocol-version values are unattested rather than parsed as
\textsc{Resolver-Path}. For an option with protocol version 1, a question count
other than one, header/length mismatch,
response identifier of zero, response entry count outside 1--6, nonzero
reserved bit, noncanonical country, or trailing octet is malformed and
therefore invalid. Verification stops after six
signature checks and six RPKA lookups. If the complete message exceeds the
advertised UDP payload, the resolver preserves the chain and uses the validated
TC/TCP behavior specified above rather than changing or merging the chain.

\paragraph{Response-bound cross-implementation conformance.}
Two standalone, separately implemented codebases consumed the published
specification vectors: Go using standard-library Ed25519 and Python using
\texttt{cryptography} Ed25519. Neither codebase imports the production
implementation or vector generator. The suite contains 18 canonical cases,
executed for five repetitions by both codebases, yielding 180 implementation
decisions. This establishes separate code paths, not provenance from
independent human teams or authors.

\begin{table*}[t]
\centering
\scriptsize
\setlength{\tabcolsep}{4pt}
\begin{tabular}{@{}p{4.2cm}p{3.0cm}p{2.2cm}p{6.4cm}@{}}
\toprule
Check & Population & Result & Interpretation \\
\midrule
Evidence-ready vectors & 18 canonical cases & 18/18 & Every vector has the required decision, log, event, and packet evidence. \\
Implementation decisions & $18\times5\times2$ & 180 & One decision per codebase, case, and repetition. \\
Cross-implementation verdicts & $18\times5$ paired trials & 90/90 & Go and Python returned the same verifier state. \\
Canonical signing & 17 signer-applicable cases $\times5$ & 85/85 & The codebases emitted identical canonical signatures. \\
Expected verifier state & 180 decisions & 180/180 & Each decision matched the vector's expected state. \\
Parser/serializer & 180 decisions & 180/180 & Each codebase matched the canonical parsed and serialized bytes. \\
Canonical response hash & 180 decisions & 180/180 & Each codebase matched the published canonical hash. \\
Evidence linkage & 180 decisions & 180/180 & Decisions link to client events, separate implementation logs, and 180 packets in 5 PCAPs. \\
Presented-payload re-signing & 170 signer-applicable payloads & 130/170 & The 40 expected non-matches are removed-entry, reordered-entry, modified-entry, and copied-signature mutation vectors. \\
\bottomrule
\end{tabular}
\caption{Published-vector conformance for two standalone, separately
implemented codebases. Go uses standard-library Ed25519; Python uses
\texttt{cryptography} Ed25519. Neither imports the production implementation or
vector generator. The experiment establishes cross-implementation agreement
for these vectors, not independent human-team or authorship provenance.}
\label{tab:response-bound-conformance}
\end{table*}

Coverage includes valid and fresh-cache responses, historical reuse, nonce
replay, selected-answer/status substitution, sibling transplant, partial
participation with \texttt{GAP\_BELOW}, stripping, unknown and revoked keys, and
six-entry overflow. Canonical re-signing matched 130/170 signer-applicable
presented payloads. The 40 non-matches are expected: those four mutation-vector
families deliberately present a removed, reordered, or modified entry, or a
copied signature, rather than the canonical payload the signer would produce.
All 180 decisions link to client events and separate per-codebase logs, and the
packet evidence comprises 180 packets in 5 PCAPs. These results test the
published vector suite; they do not establish assertion truth, path
completeness, physical traversal, or production interoperability.

\paragraph{Identity granularity and key distribution.}
\textsc{Resolver-Path} signs an AS+country assertion instead of a resolver IP.
Across 16,141 validated observed chains, 3,166 cross borders; 1,091 of those
3,166 cases (34.46\%) retain the same ASN across different ingress and egress
country labels. This motivates including country as a useful operator assertion,
not treating it as independently verified truth. Examples include Google
AS15169 chains from Russian ingress to Finnish or Swedish egress labels and
Orange AS3215 chains from French Guiana to France. Verification requires an
RPKI RPKA object as specified above~\cite{rfc6480}. Production publication,
rollover, revocation behavior under operational failure, and production-scale
nonce state remain unimplemented deployment work.

\paragraph{Wire-format and transport validation.}
The validation exercised entry counts zero through six over both IPv4 and IPv6.
Serializer-reported lengths were compared with independently parsed packet
captures while exercising the 1,232- and 4,096-byte EDNS payload limits. The
captures recorded TC signaling and TCP fallback for over-budget responses, and
\textsc{Resolver-Path} preserved the complete chain rather than silently deleting
entries. Every observed length matched the analytical values in
Table~\ref{tab:attestation-size-validation}.

\paragraph{Full-scope wire matrix.}
\label{sec:appx:wire-matrix}
The paper matrix completed 48/48 cells and 240/240 runs. Its strict artifact
checks accepted 1,840/1,840 packet captures and reconciled 1,600/1,600 linked
client/resolver event sets. At the 1,232-byte UDP payload, signed depth-eight
cases used TCP in 10/55 cases for each IP family; signed depth-15 cases used TCP
in 45/125 cases for each IP family. These counts characterize the tested
transport matrix and do not imply that a resource-limited disclosed path is
complete.

Wire-format validation confirmed the \textsc{Resolver-Path} size model across
both IP families. Every serialized entry was 144 bytes, and option data matched
$36+144h$ bytes for $h=0,\ldots,6$; including the EDNS option-code/length header
yielded $40+144h$ bytes. IPv4 and IPv6 produced identical DNS-layer sizes.
Packet captures independently confirmed the complete DNS-message sizes and the
specified TC/TCP behavior at the tested EDNS payload limits. The advertised
payload limits the DNS message rather than its enclosing packet: ordinary
IPv4/UDP adds 28 bytes on wire, while IPv6/UDP adds 48 bytes absent IPv6
extension headers.

If $B$ is the DNS-message size excluding only the \textsc{Resolver-Path} option
and $U$ is the advertised EDNS payload limit, the supported entry count is
\[
h_{\max}=\min\left(6,
\left\lfloor\frac{U-B-40}{144}\right\rfloor\right).
\]
This expression applies when $U-B\geq40$; otherwise the complete option does
not fit in the advertised UDP budget.

\begin{table}[t]
\centering
\scriptsize
\resizebox{\columnwidth}{!}{%
\begin{tabular}{rrrrrr}
\toprule
Entries $h$ & Option data & Full option &
Residual at 1232 & Residual at 4096 & Verifications \\
\midrule
0 & 36  & 40  & 1192 & 4056 & 0 \\
1 & 180 & 184 & 1048 & 3912 & 1 \\
2 & 324 & 328 & 904  & 3768 & 2 \\
3 & 468 & 472 & 760  & 3624 & 3 \\
4 & 612 & 616 & 616  & 3480 & 4 \\
5 & 756 & 760 & 472  & 3336 & 5 \\
6 & 900 & 904 & 328  & 3192 & 6 \\
\bottomrule
\end{tabular}%
}
\caption{Validated \textsc{Resolver-Path} wire sizes and verification bounds.
Sizes are bytes. ``Option data'' is the EDNS option payload, ``Full option''
includes the four-byte option-code/length header, and each residual is the
DNS-message budget remaining for the response and other EDNS data. IPv4 and
IPv6 produce identical values because IP headers are outside the advertised
EDNS payload.}
\label{tab:attestation-size-validation}
\end{table}

\paragraph{Cryptographic cost.}
Per-entry Ed25519 signing was 36.8--37.1\,\si{\micro\second} p50 and verification
93.4--95.2\,\si{\micro\second} p50. Direct depth-two verification took
149\,\si{\micro\second} p50 and 158\,\si{\micro\second} p95 over ten iterations.

\paragraph{Depth-load throughput and loss.}
The depth-load matrix crossed three depths, three modes, ten offered loads from
2,000 to 20,000~QPS, and five 300-second repetitions per cell. The resulting 450
runs covered 135,000.268 measured seconds.
Table~\ref{tab:attestation-depth-load} reports median achieved throughput at the
20,000-QPS load and loss both at that load and across all ten loads. Signed mode
remained below 0.001\% loss across all loads at every depth.

\begin{table}[H]
\centering
\scriptsize
\setlength{\tabcolsep}{2pt}
\begin{tabular}{@{}rlrrr@{}}
\toprule
Depth & Mode & Median QPS & Loss at 20k & Loss, all loads \\
\midrule
2 & Baseline & 19,998.902 & 55 / 29,998,235 & 130 / 164,997,520 \\
2 & Unsigned & 19,998.068 & 37 / 29,997,548 & 141 / 164,996,990 \\
2 & Signed   & 19,695.803 & 196 / 28,231,110 & 386 / 163,155,575 \\
\midrule
3 & Baseline & 19,995.483 & 101 / 29,993,381 & 232 / 164,988,590 \\
3 & Unsigned & 19,995.706 & 132 / 29,991,123 & 294 / 164,986,692 \\
3 & Signed   & 19,545.222 & 439 / 29,132,169 & 1,095 / 163,704,716 \\
\midrule
5 & Baseline & 19,959.482 & 195 / 29,939,714 & 836 / 164,903,749 \\
5 & Unsigned & 19,955.340 & 201 / 29,934,516 & 657 / 164,895,498 \\
5 & Signed   & 19,423.159 & 499 / 28,979,101 & 1,595 / 163,372,922 \\
\bottomrule
\end{tabular}
\caption{Attestation depth-load matrix. Each depth--mode--load cell has five
300-second repetitions. Achieved QPS is the median at 20,000 offered QPS; loss
cells give lost queries over total queries at that load or across all ten loads.
The complete matrix comprises 450 runs and 135,000.268 measured seconds.}
\label{tab:attestation-depth-load}
\end{table}

\paragraph{Separate CPU/latency and heterogeneous experiments.}
A separate CPU/latency experiment ran at 2,000~QPS. At the measured
36.8\,\si{\micro\second} signing cost, the arithmetic service-rate estimate is
approximately 27,000 signatures/s per core; this estimate is distinct from the
observed end-to-end depth-load results above. At depths 2, 3, and 5, CPU cost
rose from 76.7 to 189.4, 101.8 to 202.1, and 122.4 to
224.4\,\si{\micro\second}/query, respectively. P95 latency rose from 0.247 to
0.463\,ms at depth two, 0.303 to 0.495\,ms at depth three, and 0.073 to 0.239\,ms
at depth five. The offered 2,000~QPS completed with zero measured loss.

In heterogeneous runs, verification rate was 1.0 and observed truncation and
TCP fallback were 0.0. Throughput deltas relative to Technique~2 were $-1.29\%$
for \texttt{bind\_bind\_bind}, $-1.36\%$ for
\texttt{bind\_knot\_bind}, $-3.35\%$ for
\texttt{bind\_unbound\_knot}, and $-5.56\%$ for
\texttt{bind\_unbound\_bind}; corresponding p95 latencies were 6.527, 7.807,
7.807, and 11.519\,ms. The full heterogeneous run reduced throughput by about
1.3\%. These controlled participating-chain results authenticate received
assertions but do not test assertion truth or hidden-hop omission.

\section{Measurement Accounting and Reproducibility}
\label{sec:appx:reproducibility}

\begin{table*}[t]
\centering
\scriptsize
\setlength{\tabcolsep}{3.5pt}
\begin{tabular}{@{}lp{3.3cm}p{2.3cm}rrrr@{}}
\toprule
Run & UTC collection window & Measurement-ID span & Atlas meas. & Requested & Scheduled & Probes with result \\
\midrule
2023 & 2023-01-05 02:45:53--2023-01-09 08:40:25 & 48540346--48579274 & 12,281 & 12,281 & 11,956 & 10,729 \\
2024 & 2024-08-19 18:20:29--2024-08-28 04:58:45 & 77472062--77715847 & 23,095 & 23,095 & 19,557 & 10,082 \\
2025A & 2025-05-29 15:14:19--2025-05-30 04:03:00 & 106351841--106460342 & 11,909 & 11,909 & 10,488 & 10,039 \\
2025B & 2025-10-01 19:47:32--2025-10-02 08:34:00 & 130689926--130790733 & 11,444 & 11,444 & 10,572 & 9,611 \\
\bottomrule
\end{tabular}
\caption{RIPE Atlas campaign manifest. Times are UTC. Measurement-ID values are
reported as the minimum--maximum span for each run; the ``Atlas meas.'' column
reports the campaign count rather than implying that every intervening integer
belongs to the run. Requested and scheduled counts are per run; ``Probes with result'' counts
distinct probes returning at least one result.}
\label{tab:campaign-reproducibility}
\end{table*}

\begin{table*}[t]
\centering
\footnotesize
\begin{tabular}{@{}lrrp{7.3cm}r@{}}
\toprule
Run & Successful & Failed / timeout & Excluded after success & Retained \\
\midrule
2023 & 24,460 & 538 & 465 missing frontend ASN & 23,995 \\
2024 & 10,087 & 6,376 & 147 missing frontend ASN; 41 missing egress ASN & 9,899 \\
2025A & 19,178 & 1,870 & 145 missing client ASN & 19,033 \\
2025B & 16,636 & 1,833 & 75 missing client ASN; 6 missing client and ingress ASN; 2 missing ingress ASN & 16,553 \\
\bottomrule
\end{tabular}
\caption{Campaign-level result and exclusion accounting. Within every row,
successful results minus the listed exclusions equals retained chains. Campaign
2025B supplies the 16,636 successful rows used as the raw 2025 analysis base;
16,553 remain in the ASN-complete subset after the campaign-level exclusions
shown here. The full-population client--frontend lower bound retains all 16,636
rows by counting the 83 ASN-incomplete rows as same-AS.}
\label{tab:campaign-outcomes}
\end{table*}

The two 2025 runs together account for 23,353 requested probe measurements,
21,060 scheduled probes, 35,814 successful results, 3,703 failed or timed-out results,
228 listed ASN exclusions, and 35,586 retained chains. Campaign~B is the source
of the paper's 16,636-row 2025 cross-sectional coverage base.

\begin{table}[H]
\centering
\footnotesize
\caption{IPs, ASes, and organizations in the 2025 dataset.}
\label{tab:unique_entities}
\begin{tabular}{lrrr}
\toprule
Entity & Clients & Ingress & Egress \\
\midrule
Unique IPs & 11,444 & 5,538 & 7,563 \\
Unique ASes & 3,805 & 2,294 & 1,565 \\
Unique organizations & 3,133 & 2,174 & 1,478 \\
\bottomrule
\end{tabular}
\end{table}

\begin{table*}[t]
\centering
\footnotesize
\setlength{\tabcolsep}{4pt}
\begin{tabular}{@{}lrrrrrr@{}}
\toprule
Run & \makecell{Both endpoint\\IPs} & \makecell{Different endpoint\\IPs} & Share & \makecell{Public-ingress\\observations} & \makecell{Different public\\endpoint IPs} & Share \\
\midrule
2025A & 19,178 & 18,143 & 94.60\% & 11,155 & 10,121 & 90.73\% \\
2025B & 16,636 & 15,812 & 95.05\% & 9,927 & 9,103 & 91.70\% \\
Combined & 35,814 & 33,955 & 94.81\% & 21,082 & 19,224 & 91.19\% \\
\bottomrule
\end{tabular}
\caption{Observed ingress--egress IP divergence in the 2025 campaigns. Each
row counts observations with both endpoint IPs; the public-ingress subset
excludes private ingress. IP inequality establishes different visible endpoint
addresses, not an additional logical resolver or the number of intermediate
hops. ``Combined'' concatenates Campaigns~A and~B.}
\label{tab:endpoint-ip-divergence}
\end{table*}

\begin{table*}[t]
\centering
\footnotesize
\setlength{\tabcolsep}{4pt}
\begin{tabular}{@{}lrrrrrr@{}}
\toprule
Run & \makecell{Comparable\\IPv4 pairs} & \makecell{Different\\\texttt{/24}} & Share & \makecell{Public-ingress\\IPv4 pairs} & \makecell{Public-ingress\\different \texttt{/24}} & Share \\
\midrule
2025A & 14,909 & 13,404 & 89.91\% & 8,080 & 6,576 & 81.39\% \\
2025B & 13,040 & 11,845 & 90.84\% & 7,273 & 6,078 & 83.57\% \\
Combined & 27,949 & 25,249 & 90.34\% & 15,353 & 12,654 & 82.42\% \\
\bottomrule
\end{tabular}
\caption{Ingress--egress IPv4 prefix divergence. A pair is comparable when
both endpoints are IPv4 and is divergent when the first 24 bits differ. The
public-ingress population is the primary endpoint-opacity population because
private ingress can differ from public egress solely through address
translation. Different \texttt{/24}s do not establish distinct operators or
an additional logical resolver.}
\label{tab:endpoint-prefix-divergence}
\end{table*}

\begin{table*}[t]
\centering
\footnotesize
\begin{tabular}{@{}lrrrr@{}}
\toprule
Combined \texttt{/24}-divergent population & ASN known & Different ASN & Organization known & Different organization \\
\midrule
All ingress & 25,243 & 10,592 (41.96\%) & 24,223 & 9,325 (38.50\%) \\
Public ingress & 12,654 & 3,372 (26.65\%) & 12,383 & 2,983 (24.09\%) \\
Non-public ingress & 12,589 & 7,220 (57.35\%) & 11,840 & 6,342 (53.56\%) \\
\bottomrule
\end{tabular}
\caption{Administrative characterization of the combined Campaign~A and~B
IPv4 \texttt{/24}-divergent observations. Percentages are conditional on
\texttt{/24} divergence and availability of the corresponding mapping; they
are not prevalence rates over all comparable IPv4 pairs. Different AS or
organization mappings establish administrative separation between the
observed ingress and egress endpoints, not an intermediate forwarding path.}
\label{tab:endpoint-divergence-administration}
\end{table*}

\begin{table*}[t]
\centering
\footnotesize
\begin{tabular}{p{0.22\textwidth}rrp{0.47\textwidth}}
\toprule
Stage & Base & Retained & Excluded / use \\
\midrule
2025B successful results & 16,636 & 16,636 & Raw campaign-B analytical coverage base \\
Campaign-level ASN completeness & 16,636 & 16,553 & 83 rows with missing client and/or ingress ASN \\
Full-population AS lower bound & 16,636 & 16,636 & 83 ASN-incomplete rows retained and counted as same-AS \\
2025 per-side observations & 16,636 & 16,636 & Coverage base \\
Public frontend geolocation & 16,636 & 9,923 & 6,713 private-ingress observations reported separately \\
Valid country triples & 16,636 & 16,563 & 73 without a retained complete triple \\
Plausibility-consistent subset & 16,636 & 16,141 & 495 chains contradicted by active checks; counted in-country in the full-population analysis \\
Anycast serving-site subset & 5,498 & 3,869 & 1,629 without usable serving-site geolocation \\
Non-anycast egress subset & 16,563 & 15,125 & 1,438 anycast-egress triples excluded \\
\bottomrule
\end{tabular}
\caption{Filtering and analysis-population accounting for the 2025B dataset.}
\label{tab:filter-accounting}
\end{table*}

\clearpage

\section{Open Validation Questions}
\label{sec:appx:open-items}

One question remains outside the scope of the completed evaluation and is not
used to support stronger claims: the operational distribution of ingress- and
intermediate-cache-hit path lengths beyond the controlled cache matrix. The
previously open cross-implementation conformance question is addressed by the
18-case published-vector experiment in
Table~\ref{tab:response-bound-conformance}, including selected-answer/status and
sibling substitution, cache freshness and reuse, partial participation,
stripping, key state, overflow, separate logs, and packet captures. The two
standalone codebases do not establish provenance from independent human teams
or authors.

\section{Supplementary Measurement Results}
\label{sec:appx:resultslongitudinal}

This section records the longitudinal material that is too detailed for the
main text. It includes the probe-weighted stable-panel summary used in the main
paper, older full-population campaign tables for 2023--2024, and the 2025
client--frontend AS-boundary summary. The older tables remain useful as
baseline context but do not control for probe-composition churn. All AS-level
tables in this subsection use the conservative same-AS convention: every private
ingress is assigned to the client AS. For the 2025 full-population client--frontend
result, the 83 ASN-incomplete observations are also counted as same-AS, so the
different-AS share is a lower bound.
The longitudinal 2025 row uses the deduplicated union of Campaigns~A and~B;
the 16,636-row cross-sectional analyses use Campaign~B alone.

\begin{table}[H]
	\centering
	\footnotesize
	\resizebox{\columnwidth}{!}{%
	\begin{tabular}{|l|c|c|c|c|c|}
		\hline
		Year & Classified Probes & $C{=}I{=}E$ & $C{=}I, I{\neq}E$ & $C{\neq}I, I{=}E$ & $C{\neq}I, I{\neq}E$ \\
		\hline
		2023 & 2,432 & 41.5\% & 23.1\% & 19.0\% & 16.5\% \\
		\hline
		2024 & 2,439 & 38.9\% & 20.8\% & 13.7\% & 26.6\% \\
		\hline
		2025 & 2,468 & 37.9\% & 20.0\% & 18.2\% & 23.9\% \\
		\hline
	\end{tabular}
	}
	\caption{Probe-weighted AS-level chain shares under the conservative same-AS convention for the strict stable panel of 2,468 probes present in the 2023, 2024, and 2025 year-level datasets with unchanged client AS and country. The 2025 dataset is the deduplicated union of Campaigns~A and~B. Within each year-level dataset, each classified probe contributes total weight 1, split across its classified resolver chains.}
	\label{tab:longitudinal-stable-panel-appendix}
\end{table}

\begin{table}[H]
	\centering
	\footnotesize
	\caption{AS and organizational relationship between clients (C), ingress (I), and egress (E) resolvers in the 2023 campaign under the conservative same-AS convention.}
	\label{tab:dnscategorization_priorstudy}
	\begin{tabular}{lccc}
		\toprule
		C \% & C/I ASes & I/E ASes & Organizational Relationship\\
		\midrule
		43\%  & $=$ & $=$ & All in the same organization\\
		31\% & $=$ & $\neq$ & \begin{tabular}{@{}l@{}}C = I = E: 4.4\% \\ C = I $\neq$ E: 95.6\%\end{tabular} \\
		15.4\% & $\neq$ & $=$ & \begin{tabular}{@{}l@{}}C = I = E: 1.1\% \\ C $\neq$ I = E: 98.9\%\end{tabular} \\
		10.1\% & $\neq$ & $\neq$ & \begin{tabular}{@{}l@{}}C = I = E: 1.8\% \\ C $\neq$ I $\neq$ E: 95\% \\ C = I $\neq$ E: 2.8\% \\ C $\neq$ I = E: 0.4\%\end{tabular} \\
		\bottomrule
	\end{tabular}
\end{table}

\begin{table}[H]
	\centering
	\footnotesize
	\caption{AS and organizational relationship between clients (C), ingress (I), and egress (E) resolvers in the 2024 campaign under the conservative same-AS convention.}
	\label{tab:dnscategorization2024}
	\begin{tabular}{lccc}
		\toprule
		C \% & C/I ASes & I/E ASes & Organizational Relationship\\
		\midrule
		37.2\%  & $=$ & $=$ & All in the same organization\\
		26.6\% & $=$ & $\neq$ & \begin{tabular}{@{}l@{}}C = I = E: 4.4\% \\ C = I $\neq$ E: 95.6\%\end{tabular} \\
		23.0\% & $\neq$ & $=$ & \begin{tabular}{@{}l@{}}C = I = E: 1.1\% \\ C $\neq$ I = E: 98.9\%\end{tabular} \\
		13.3\% & $\neq$ & $\neq$ & \begin{tabular}{@{}l@{}}C = I = E: 1.8\% \\ C $\neq$ I $\neq$ E: 95\% \\ C = I $\neq$ E: 2.8\% \\ C $\neq$ I = E: 0.4\%\end{tabular} \\
		\bottomrule
	\end{tabular}
\end{table}

\begin{table}[H]
	\centering
	\footnotesize
	\caption{Client--frontend AS-boundary result in Campaign~2025B under the conservative convention that assigns every private-ingress observation to the client AS and counts all 83 observations lacking a client and/or ingress ASN as same-AS. The 16,636 denominator is therefore the full successful-result population; 16,553 rows are ASN-complete.}
	\label{tab:dnscategorization2025}
	\begin{tabular}{lrrr}
		\toprule
		Comparison & Count & $N$ & Share\\
		\midrule
		Client ASN $\neq$ frontend ASN & 6,622 & 16,636 & 39.8\% \\
		\bottomrule
	\end{tabular}
\end{table}

\begin{table}[h!]
	\centering
	\footnotesize
		\caption{IPs and AS/organization analysis labels in the 2024 dataset. Private-ingress labels follow the conservative same-AS convention.}
	\label{tab:unique_entities_2024}
	\begin{tabular}{lrrr}
		\toprule
		Entity Type & Clients & Ingress & Egress \\
		\midrule
		Unique IPs           & 10,082 & 2,874 & 6,031 \\
		Unique ASes          & 3,258  & 2,236 & 1,757 \\
		Unique Organizations & 3,048  & 2,093 & 1,644 \\
		\bottomrule
	\end{tabular}
\end{table}

\section{Supplementary Protocol Evaluation}
\label{sec:appx:cloud-capacity}

This section expands the protocol-evaluation results from
Section~\ref{sec:tracingdnspath}. It provides the per-behavior cloud-capacity
breakdown at the common 20{,}000~QPS comparison point, the wire-level
depth-scaling summary for Technique~2, and the full throughput, resource, and
cache figures referenced in the main text.

\subsection{Strict Response-Path Correctness}

The final correctness matrix contains all 185 required runs, exactly 37 in each
of five repetitions. Both strict validators pass across all 23 scenario
conditions and all 32 topology conditions. Three pre-fix failures and two
interrupted attempts remain preserved separately rather than being counted in
the required-run matrix.

\begin{table*}[t]
\centering
\footnotesize
\setlength{\tabcolsep}{4pt}
\begin{tabular}{@{}p{0.25\textwidth}p{0.24\textwidth}p{0.43\textwidth}@{}}
\toprule
Condition & Population & Validated result \\
\midrule
Fully participating paths & 63,042 individual responses; 47,042 selected responses & Every path matched the linked resolver events exactly after excluding the intentionally non-participating P1 condition. \\
P1 partial participation & 1,000 responses & Every response explicitly reported missing or truncated disclosure; none reported false completeness. \\
Selected cache hits & 17,000 responses & Every current-response path was exact; historical-path reuse was 0/17,000. \\
Multiple upstream responses & 29,042 parent-observed responses; 13,042 selected responses & Every observed and selected response path was exact; sibling/cross-response merges were 0. \\
Technique~1 comparison & 7,000 diagnostic/ordinary-response pairs & 4,000 pairs disagreed (57.14\%), confirming that configured diagnostic topology and ordinary response paths are distinct objects. \\
\bottomrule
\end{tabular}
\caption{Strict controlled response-path correctness. ``Exact'' means agreement
with the linked client-query, upstream-response, and timestamped per-resolver
event evidence for the same controlled transaction. The table validates the
tested operational behavior, not path completeness under arbitrary
non-participation.}
\label{tab:response-path-correctness}
\end{table*}

The retained evidence contains 49,242 controlled client-query records, 65,242
upstream-response records, and 445,729 timestamped resolver events. All query
names were controlled: the validators found 0 uncontrolled qnames.

\subsection{Completed Local Performance Matrices}

Table~\ref{tab:completed-local-throughput} reports only workloads whose cells
are final. Warm cache additionally completed 100/100 repetitions and passed its
strict validator. The cold-cache and conditional-forwarding rows are scoped to
their respective completed 20-cell matrices.

\begin{table*}[t]
\centering
\footnotesize
\begin{tabular}{@{}lcrrrr@{}}
\toprule
Workload & Completed cells & Baseline & Tracing disabled & Unsigned & Signed \\
\midrule
Warm cache & 20/20 & 19,999.633 & 19,999.303 & 19,998.139 & 19,993.933 \\
Cold cache & 20/20 & 19,999.865 & 19,999.982 & 19,998.474 & 19,999.899 \\
Conditional forwarding & 20/20 & 17,609.977 & 17,633.096 & 17,589.189 & 17,633.980 \\
\bottomrule
\end{tabular}
\caption{Median completed QPS at a 20,000-QPS offered load for the completed
local workload matrices. Each row is a separately completed workload; the
table makes no aggregate claim across unfinished workload families.}
\label{tab:completed-local-throughput}
\end{table*}

\begin{table*}[t]
\centering
\footnotesize
\resizebox{\textwidth}{!}{%
\begin{tabular}{@{}lrrrrrrr@{}}
\toprule
Mode & Completed QPS & p50 (ms) & p95 (ms) & p99 (ms) & Chain CPU (\si{\micro\second/query}) & Peak memory (MB) & Bytes/query \\
\midrule
Baseline & 19,999.633 & 0.012 & 0.017 & 0.163 & 13.124 & 243.58 & 127 \\
Tracing disabled & 19,999.303 & 0.012 & 0.018 & 0.163 & 12.987 & 246.12 & 127 \\
Unsigned & 19,998.139 & 0.012 & 0.018 & 0.167 & 13.358 & 251.64 & 163 \\
Signed & 19,993.933 & 0.069 & 0.127 & 0.423 & 76.135 & 247.44 & 309 \\
\bottomrule
\end{tabular}%
}
\caption{Final warm-cache performance at a 20,000-QPS offered load (20/20
cells, 100/100 repetitions, strict validation pass). CPU is the summed resolver-
chain cost. Signed client verification had a median cost of
226.01\,\si{\micro\second}.}
\label{tab:warm-performance-complete}
\end{table*}

\begin{table*}[t]
\centering
\footnotesize
\begin{tabular}{llrrrrr}
\toprule
Behavior & Mode & Completed QPS & p95 (ms) & Loss & CPU (\si{\micro\second/query}) & Mem (MB) \\
\midrule
Domestic & Vanilla & 18,084.6 [18,027.0--18,857.5] & 13.311 [7.807--13.567] & 0.000\% & 216.182 & 97.77 \\
Domestic & Tracing Off & 18,240.9 [17,827.7--18,257.4] & 12.287 [11.519--15.103] & 0.000\% & 238.551 & 98.33 \\
Domestic & Technique~1 & 19,999.9 [19,999.9--20,000.0] & 0.543 [0.543--0.575] & 0.000\% & 55.730 & 65.91 \\
Domestic & Technique~2 & 18,290.6 [18,008.3--18,459.9] & 12.543 [10.751--13.567] & 0.000\% & 206.842 & 98.84 \\
Egress Out, London & Vanilla & 12,562.5 [12,334.2--13,265.4] & 44.031 [44.031--44.031] & 0.000\% & 313.060 & 98.11 \\
Egress Out, London & Tracing Off & 12,954.8 [12,598.9--13,515.2] & 44.031 [44.031--44.031] & 0.000\% & 246.291 & 94.13 \\
Egress Out, London & Technique~1 & 20,000.0 [20,000.0--20,000.0] & 0.327 [0.311--0.391] & 0.000\% & 23.890 & 67.53 \\
Egress Out, London & Technique~2 & 12,837.7 [12,703.1--13,206.1] & 44.031 [44.031--44.031] & 0.000\% & 351.514 & 97.26 \\
Ingress Out, Egress Same & Vanilla & 9,292.1 [9,256.2--9,371.6] & 14.591 [14.335--15.103] & 0.000\% & 358.625 & 94.38 \\
Ingress Out, Egress Same & Tracing Off & 9,116.2 [9,104.7--9,287.6] & 15.359 [14.335--15.359] & 0.000\% & 377.865 & 93.28 \\
Ingress Out, Egress Same & Technique~1 & 10,149.9 [10,087.1--10,200.8] & 10.495 [10.239--10.751] & 0.000\% & 91.719 & 66.78 \\
Ingress Out, Egress Same & Technique~2 & 9,164.1 [9,098.5--9,209.0] & 14.591 [14.079--16.127] & 0.000\% & 389.218 & 95.93 \\
Ingress Out, Egress Third & Vanilla & 5,075.0 [4,976.4--5,212.7] & 60.415 [60.415--60.415] & 0.000\% & 453.789 & 92.17 \\
Ingress Out, Egress Third & Tracing Off & 5,008.2 [4,959.6--5,096.1] & 60.415 [60.415--60.415] & 0.000\% & 452.412 & 90.93 \\
Ingress Out, Egress Third & Technique~1 & 9,996.5 [9,984.5--10,044.4] & 10.751 [10.751--10.751] & 0.000\% & 114.469 & 66.01 \\
Ingress Out, Egress Third & Technique~2 & 4,834.7 [4,718.4--4,915.3] & 62.463 [62.463--62.463] & 0.000\% & 531.305 & 92.36 \\
Bounce Back & Vanilla & 5,459.8 [5,382.9--5,567.1] & 51.199 [51.199--52.223] & 0.000\% & 447.450 & 90.93 \\
Bounce Back & Tracing Off & 5,382.1 [5,205.4--5,404.3] & 53.247 [53.247--53.247] & 0.000\% & 427.962 & 91.12 \\
Bounce Back & Technique~1 & 10,105.8 [9,998.4--10,143.6] & 10.751 [10.751--11.007] & 0.000\% & 81.357 & 65.42 \\
Bounce Back & Technique~2 & 5,168.4 [5,103.9--5,241.5] & 55.295 [54.271--55.295] & 0.000\% & 468.703 & 89.97 \\
\bottomrule
\end{tabular}
\caption{Cloud-capacity results at the common 20{,}000~QPS comparison point. Values are medians across five repetitions; brackets show interquartile ranges.}
\label{tab:cloud-20k-breakdown}
\end{table*}

\begin{figure*}[t]
\begin{minipage}[t]{0.45\textwidth}
\centering
\footnotesize
\captionof{table}{Technique~2 depth-scaling wire-level metrics at 20{,}000~QPS. Path recovery is 100\% at every depth with zero truncation or TCP fallback.}
\label{tab:depth-scaling-wire-metrics}
\vspace{0.5em}
\begin{tabular}{rrrrr}
\toprule
Depth & Pos.\ rate & Path len.\ & Resp.\ (B) & TC / TCP \\
\midrule
2  & 1.000 & 2.0  & 91.9  & 0 / 0 \\
3  & 1.000 & 3.0  & 96.9  & 0 / 0 \\
5  & 1.000 & 5.0  & 106.9 & 0 / 0 \\
8  & 1.000 & 8.0  & 121.9 & 0 / 0 \\
10 & 1.000 & 10.0 & 132.9 & 0 / 0 \\
15 & 1.000 & 15.0 & 157.9 & 0 / 0 \\
\bottomrule
\end{tabular}
\end{minipage}
\hfill
\begin{minipage}[t]{0.50\textwidth}
\centering
\includegraphics[width=0.85\textwidth]{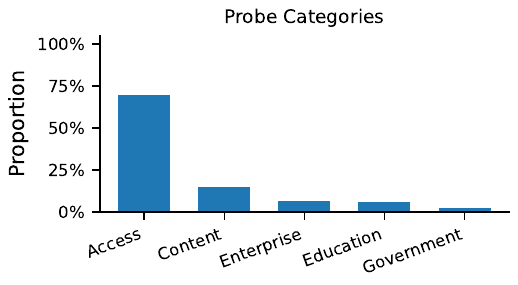}
\captionof{figure}{Probe distribution by ASN type, dominated by access networks.}
\label{fig:asntype}
\end{minipage}
\end{figure*}

\begin{table*}[t]
\centering
\footnotesize
\caption{Technique~1 \texttt{TYPE65400} diagnostic latencies across the cloud topologies (200 samples per topology). These dedicated diagnostic-query timings are not ordinary-query performance.}
\label{tab:t1-cloud-decomposition}
\begin{tabular}{lrrrrrrr}
\toprule
Topology & Success & Ingr. p50 & Mid. p50 & Egr. p50 & Auth. p50 & Total p50 & Total p95 \\
\midrule
Domestic & 200/200 & 0.624 & 0.362 & 0.428 & 0.456 & 2.117 & 2.380 \\
Egr.\ Out & 200/200 & 0.392 & 0.766 & 10.021 & 10.170 & 21.574 & 22.252 \\
Ingr.\ Out/Same & 200/200 & 10.280 & 10.220 & 10.226 & 10.178 & 40.896 & 41.939 \\
Ingr.\ Out/Third & 200/200 & 10.260 & 10.285 & 20.189 & 20.111 & 61.174 & 62.041 \\
Bounce Back & 200/200 & 10.486 & 0.492 & 0.304 & 0.231 & 11.913 & 12.575 \\
\bottomrule
\end{tabular}
\end{table*}

\begin{table}[t]
\centering
\footnotesize
\caption{End-to-end Technique~1 diagnostic time in the Docker depth-scaling experiment (200 samples per depth).}
\label{tab:t1-depth-scaling}
\begin{tabular}{lrrrrrr}
\toprule
Metric & 2 & 3 & 5 & 8 & 10 & 15 \\
\midrule
p50 (ms) & 0.401 & 0.573 & 0.837 & 1.318 & 1.858 & 4.015 \\
p95 (ms) & 0.928 & 1.299 & 1.649 & 2.730 & 3.334 & 5.158 \\
\bottomrule
\end{tabular}
\end{table}

The diagnostic latency fit was 0.2654\,ms/hop at p50 and 0.3227\,ms/hop at
p95. Successful walks recovered the configured topology and issued
depth$+1$ queries because the walk ended with an authoritative lookup. A
configured hop exposed by this diagnostic need not process an ordinary query,
especially when a resolver answers from cache.

\begin{figure*}[t]
\centering
\begin{subfigure}[t]{\textwidth}
\centering
\includegraphics[width=\textwidth]{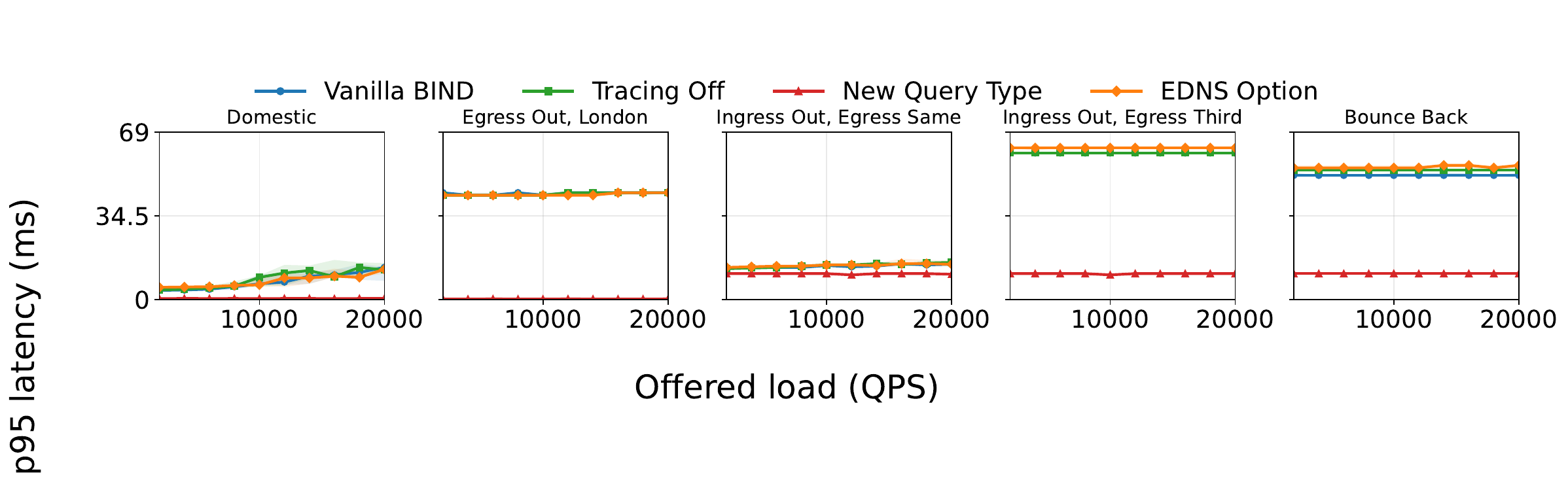}
\caption{p95 latency. Technique~1's lower latency reflects different query semantics, not a comparable improvement.}
\label{fig:appx-capacity-latency}
\end{subfigure}\\[0.4em]
\begin{subfigure}[t]{\textwidth}
\centering
\includegraphics[width=\textwidth]{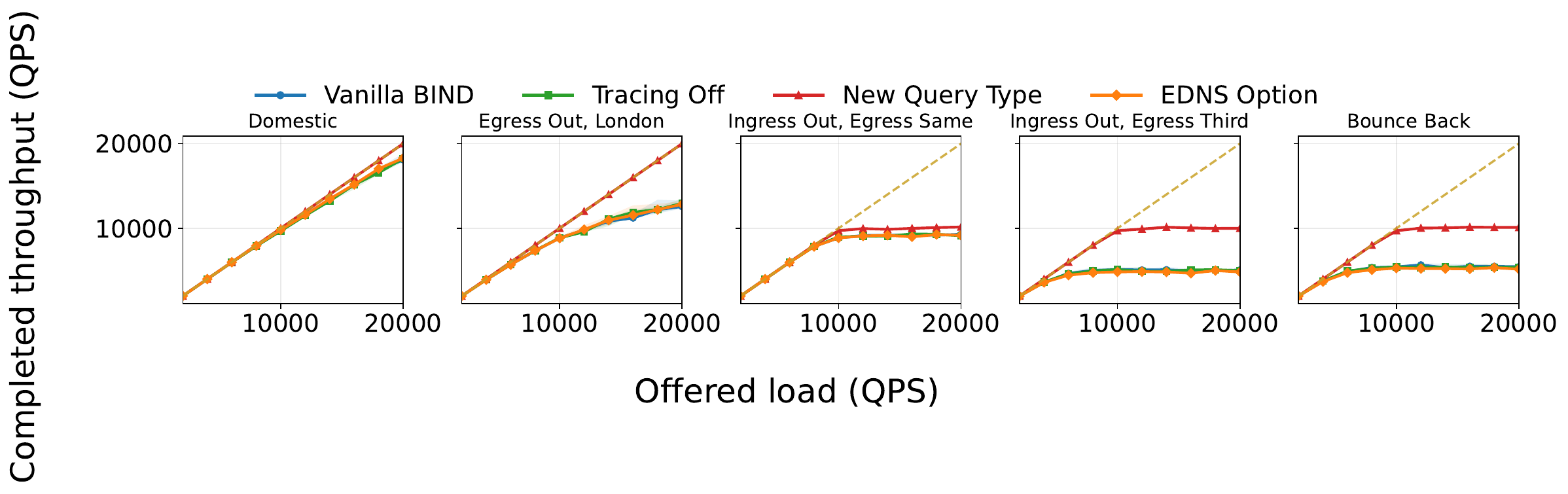}
\caption{Completed throughput. Technique~1's higher throughput is expected (different query semantics).}
\label{fig:appx-capacity-throughput}
\end{subfigure}\\[0.4em]
\begin{subfigure}[t]{\textwidth}
\centering
\includegraphics[width=\textwidth]{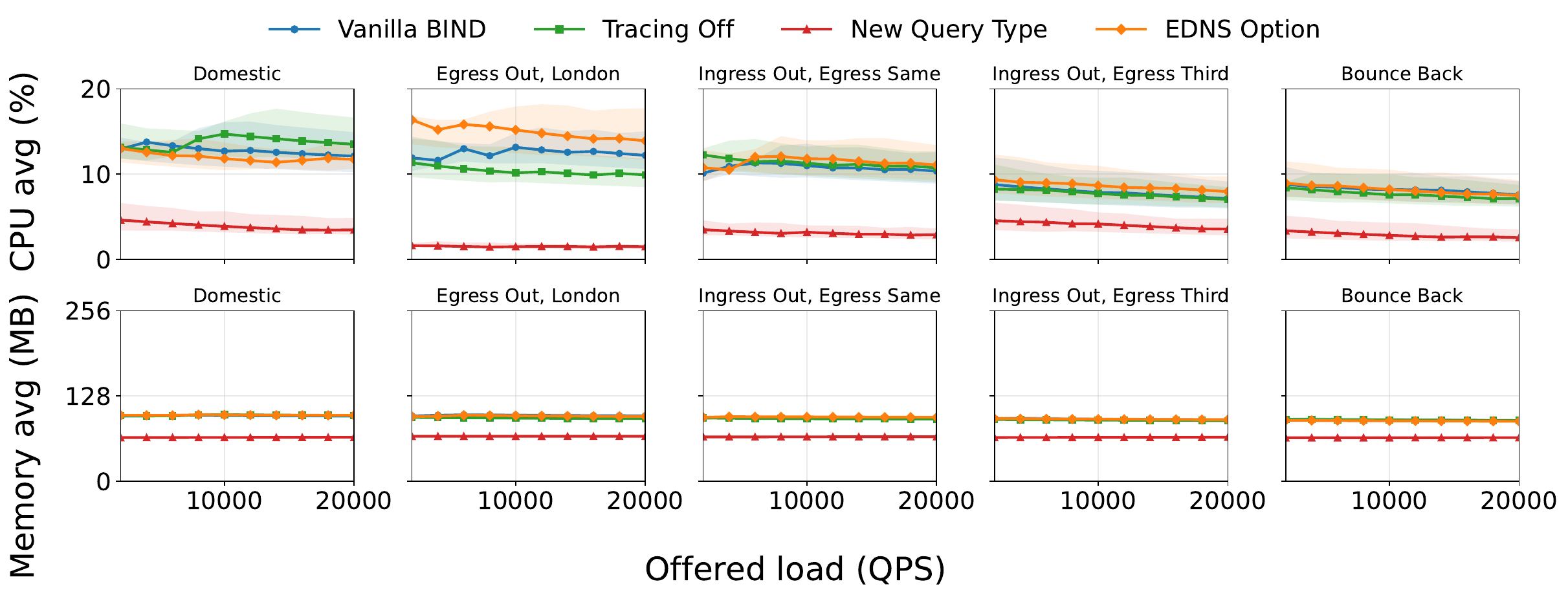}
\caption{Resolver-side CPU and memory. Technique~1 is not directly comparable (lightweight query).}
\label{fig:appx-capacity-resources}
\end{subfigure}
\caption{Cloud-capacity load-sweep results across the five observed path behaviors and four resolver modes over the 2{,}000--20{,}000~QPS load sweep. Panels~(a)--(b) show client-side metrics; panel~(c) shows resolver-side resource usage.}
\label{fig:appx-capacity-curves}
\end{figure*}

\begin{figure*}[t]
\centering
\begin{subfigure}[t]{0.48\textwidth}
\centering
\includegraphics[width=\textwidth]{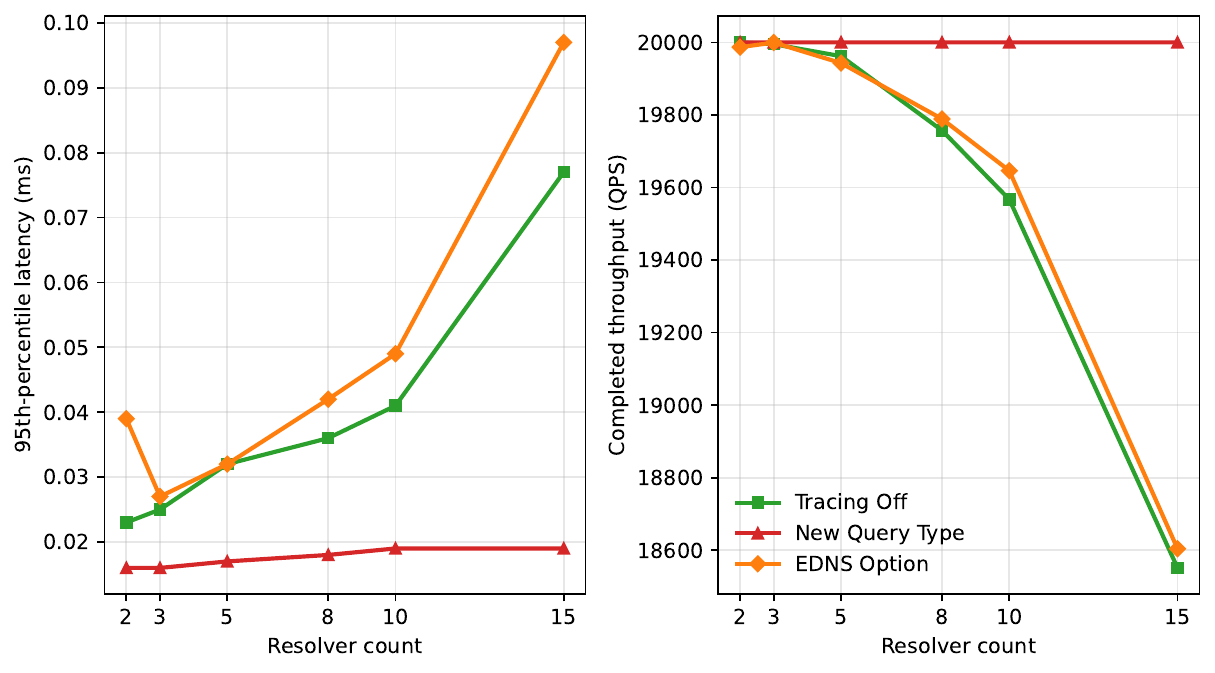}
\caption{Depth scaling at 20{,}000~QPS over resolver chains of length 2--15.}
\label{fig:appx-depth-scaling}
\end{subfigure}
\hfill
\begin{subfigure}[t]{0.48\textwidth}
\centering
\includegraphics[width=\textwidth]{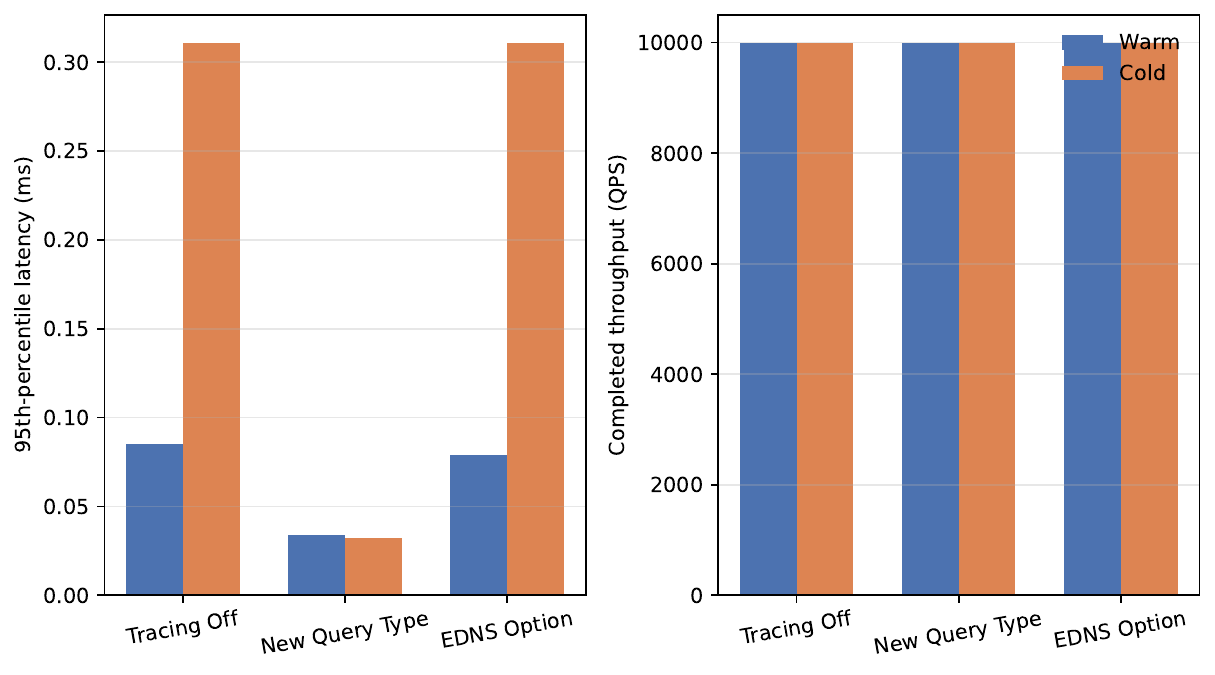}
\caption{Warm-cache and cold-cache behavior at 10{,}000~QPS.}
\label{fig:appx-cache-behavior}
\end{subfigure}
\caption{Depth-scaling and cache experiments. Technique~2 remains throughput-stable as depth increases; neither tracing mechanism shows a meaningful penalty under warm- or cold-cache conditions.}
\label{fig:appx-depth-cache}
\end{figure*}

\begin{figure*}[t]
\centering
\includegraphics[width=\textwidth]{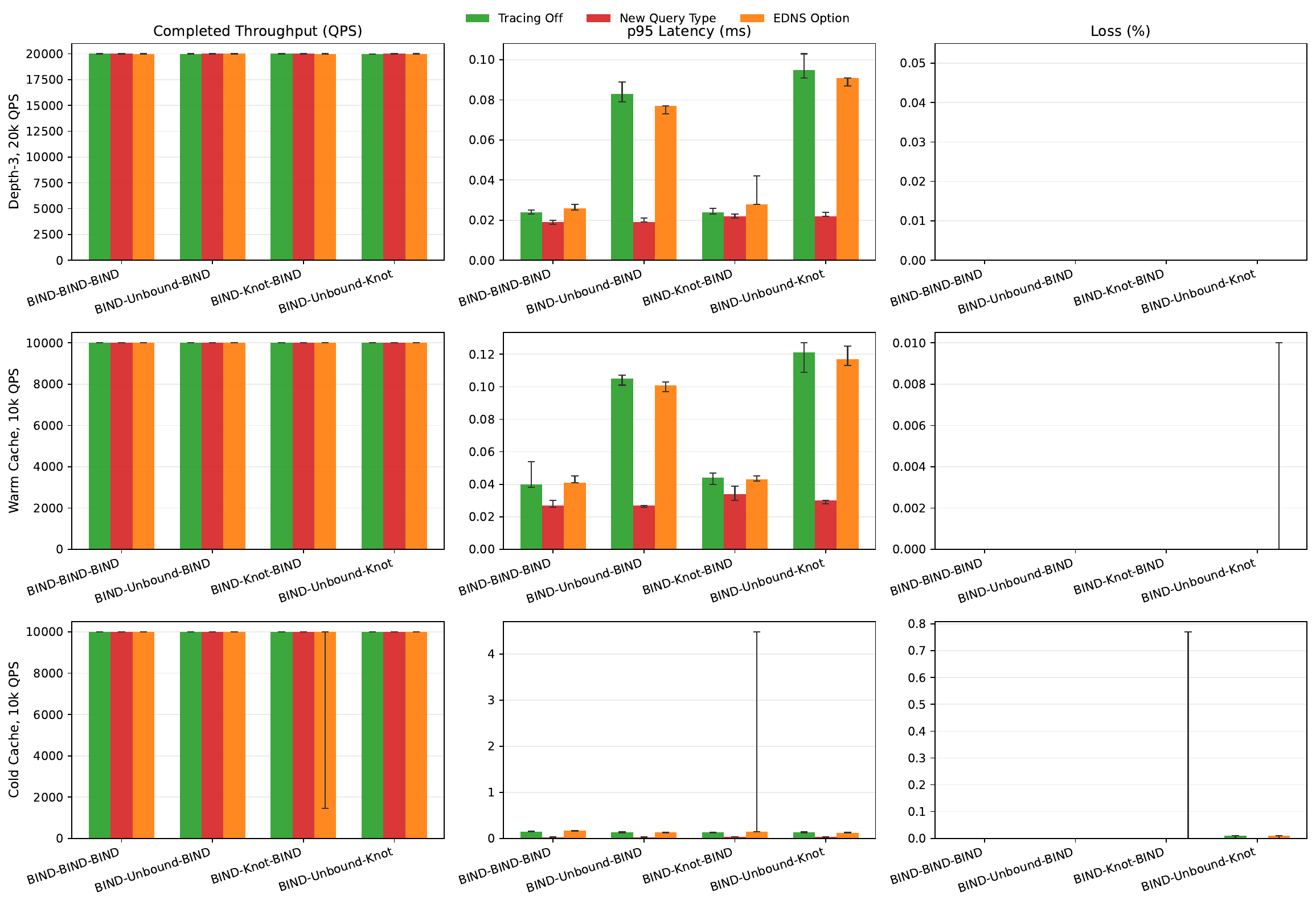}
\caption{Heterogeneous depth-3 and cache-performance overview. The figure compares the homogeneous BIND control against mixed \emph{BIND $\rightarrow$ Unbound $\rightarrow$ BIND}, \emph{BIND $\rightarrow$ Knot $\rightarrow$ BIND}, and \emph{BIND $\rightarrow$ Unbound $\rightarrow$ Knot} chains under the same local offered-load settings used in the main text.}
\Description{A multi-panel overview figure summarizing throughput and latency for heterogeneous depth-3 resolver chains and their warm-cache and cold-cache variants.}
\label{fig:appx-heterogeneous-performance}
\end{figure*}

\section{Supplementary Geolocation Analysis}
\label{sec:appx:geo}

This section documents the geolocation pipeline behind the country-level
results and collects the supporting figures that would otherwise interrupt the
main narrative. We first summarize the passive and active validation workflow,
then provide additional country-level visualizations and the full frontend
country-selection table used to choose the main-text countries.

\begin{table}[t]
\centering
\footnotesize
\caption{Passive geolocation signals used for resolver-country inference.}
\label{tab:geolocation-signals}
\begin{tabular}{lrr}
\toprule
Signal source & Country signals & Agreement vs.\ IPinfo \\
\midrule
TXT / resolver identity & 3,973 & 13.4\% (531 / 3,972) \\
Aleph PTR & 3,944 & 92.5\% (3,417 / 3,695) \\
IPinfo & 13,098 & -- \\
\bottomrule
\end{tabular}
\end{table}

Resolver geolocation combines passive and active measurements for publicly routable addresses. Private ingress is handled separately: an RFC~1918/private-use address establishes the near-side hop but supplies neither an independently geolocatable address nor a public BGP origin ASN. All-observed country summaries place private ingress in the client country, a conservative convention analogous to the same-AS assignment.

\textbf{Passive Measurements:} IPInfo supplies country-level geolocation data~\cite{ipinfo:io}, while the Aleph extracts location hints from reverse DNS hostnames~\cite{aqua:aleph}. For public anycast ingress resolvers, operator-exposed identity strings (NSID or CHAOS TXT records such as \texttt{id.server}) provide the primary site signal because IP-registration databases frequently map anycast addresses to corporate headquarters rather than the serving PoP. Cloudflare confirmed in private correspondence with the authors that \texttt{id.server} generally identifies the ingress colo; during overload, internal forwarding can cause it to continue naming the client-facing site while a different downstream resolver supplies the recursive egress.

\textbf{Active Measurements:} RIPE Atlas probes test every egress resolver whose country attribution comes from IPInfo and every ingress resolver for which IPInfo is the deciding signal. The validation collects RTT and traceroute measurements, compares implied distance against the claimed country under a fiber-speed bound of 200,000~km/s, inspects the last responsive traceroute hop, and measures the share of domestic hops. A country attribution enters the plausibility-consistent subset only when these measurements do not contradict it. When they do, the full-population analysis conservatively assigns the observation to the probe country rather than counting it as cross-border, while the stricter subset excludes it. This rule produces residual cases such as \textit{Ingress Out} in the main-text structure figure: the ingress geolocates abroad, but no plausibility-consistent egress country remains available.

Reconciling passive signals with the active checks yields 16,141 frontend/egress country chains whose inferred countries remain plausible, the stricter robustness-check subset. The other cases were actively tested rather than left unvalidated: when an IPinfo location was contradicted, the full-population analysis counted the observation as in-country. The 16,636 per-side geolocation-coverage base used in Table~\ref{tab:geolocation_source_exclusive_coverage_2025} and the main-text country-level mismatch claims therefore apply this conservative convention; Table~\ref{tab:filter-accounting} gives the complete accounting.

\subsection{Additional Geolocation Results}
\label{sec:appx:addl-geo}

This subsection provides additional detail behind the geolocation findings in
Section~\ref{sec:Findings}. The summary statistics below and the figures that
follow are grouped here to preserve space in the main text while still showing
the broader EU-scope, ingress-mismatch, and egress-mismatch patterns. For
comparison, the full 2025 dataset contains 16,563 valid
probe--frontend--egress country triples,
including the chains later excluded from the main-text non-anycast egress
comparison: probe--frontend mismatch occurs in 1,430 cases (8.63\%),
probe--egress mismatch in 17.06\% of cases, frontend--egress mismatch in
18.20\%, and all three countries differ in 0.42\%.

\begin{figure*}[t]
	\centering
	\begin{subfigure}[t]{0.54\textwidth}
		\centering
		\includegraphics[width=\textwidth]{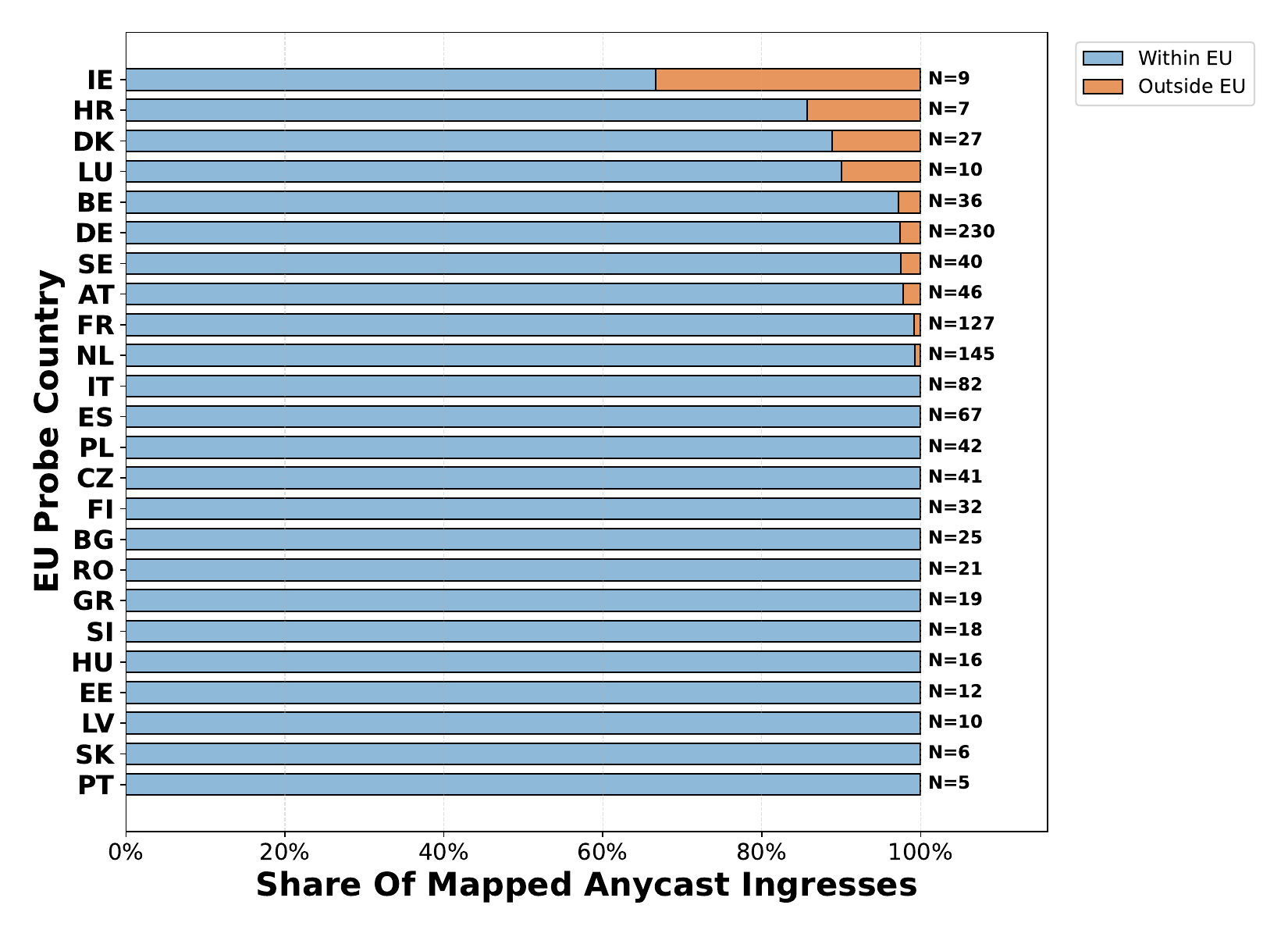}
		\caption{EU-scope anycast ingress \texttt{N}/\texttt{S}/\texttt{M} rates.}
		\label{fig:appx_anycast_ingress_eu_scope}
	\end{subfigure}
	\hfill
	\begin{subfigure}[t]{0.40\textwidth}
		\centering
		\includegraphics[width=\textwidth]{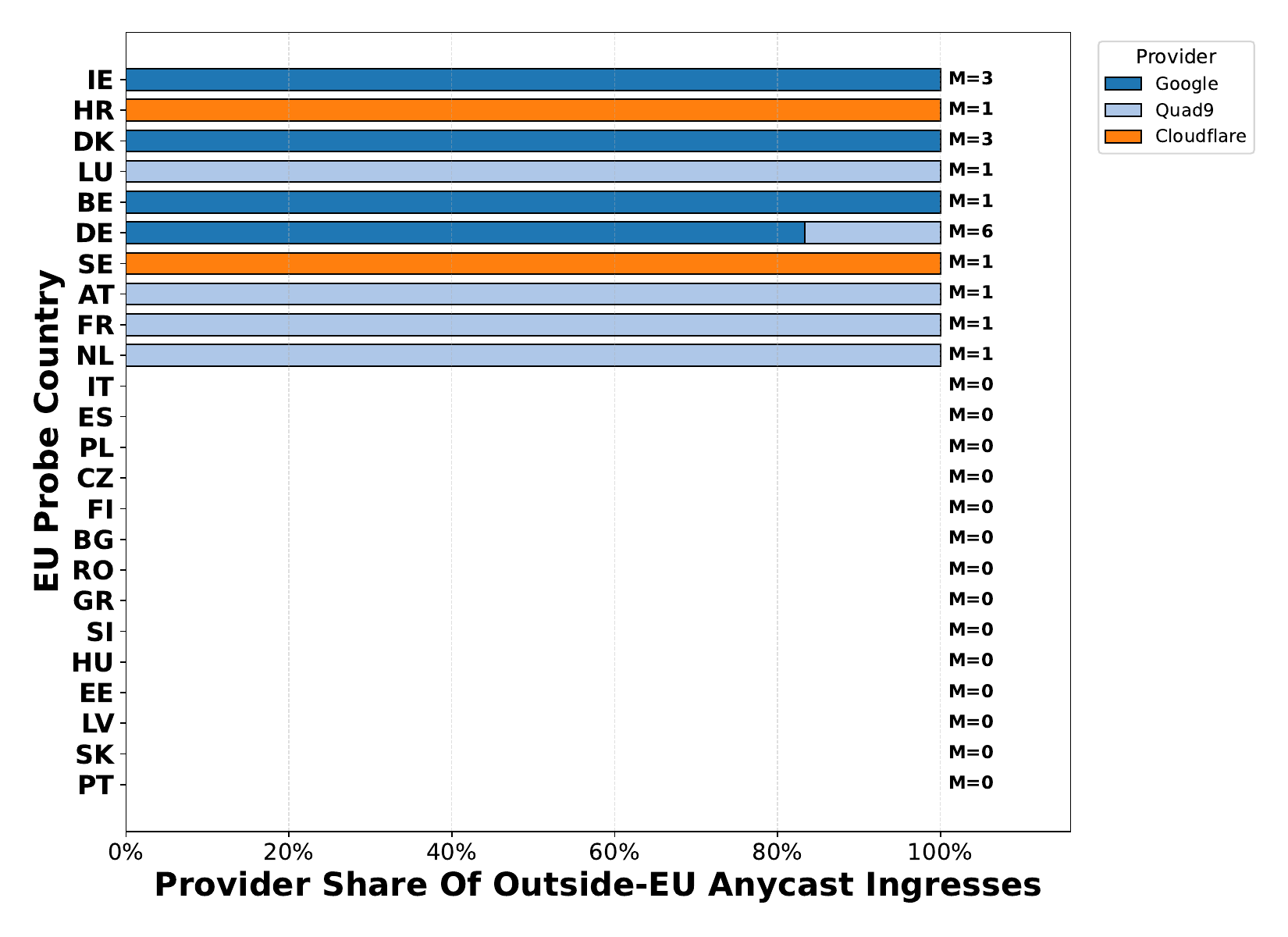}
		\caption{EU-scope provider mix of out-of-country anycast ingress cases.}
		\label{fig:appx_anycast_ingress_eu_provider_mix}
	\end{subfigure}
	\caption{Supporting EU-scope anycast ingress visualizations. In panel~(a), \texttt{N} is the out-of-country ingress share, \texttt{S} is the subset whose egress stays in the same foreign country as ingress, and \texttt{M} is the subset whose egress moves to a third country. All three bars use the same denominator, namely all anycast probe--resolver pairs for each probe country; unresolved replies remain in the denominator and are counted as in-country. Panel~(b) decomposes the out-of-country anycast ingress cases for the countries shown in panel~(a).}
	\Description{A two-panel figure for the EU-scope anycast analysis. The left panel is a three-bar country-level chart using the same N, S, and M semantics as the main anycast-ingress figure. The right panel shows provider mix for the corresponding out-of-country anycast ingress cases.}
	\label{fig:appx_anycast_ingress_eu}
\end{figure*}

\begin{figure*}[t]
	\centering
	\begin{subfigure}[t]{0.47\textwidth}
		\centering
		\includegraphics[width=\textwidth]{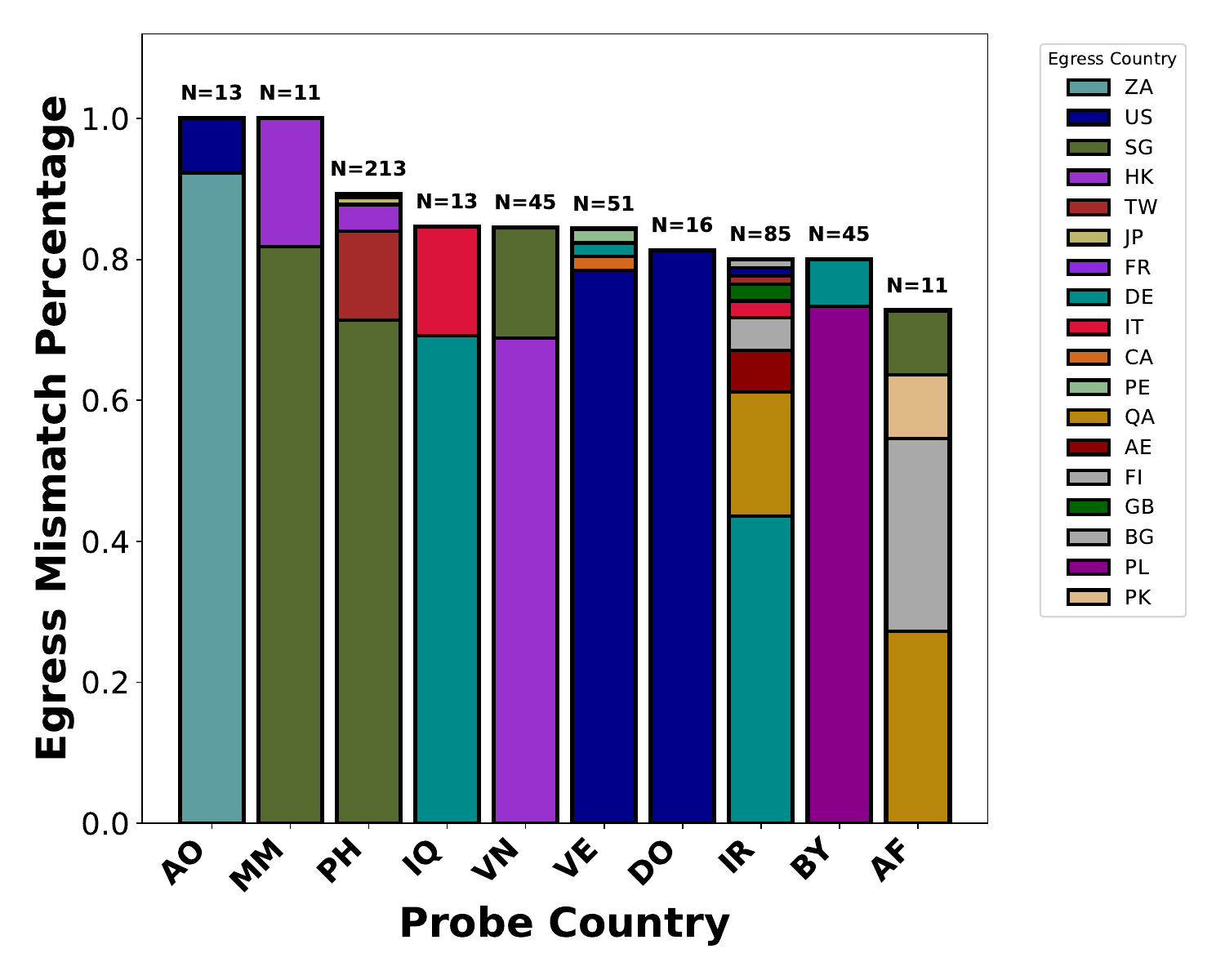}
		\caption{Egress mismatch bar plot.}
		\label{fig:appx_egress_mismatch_bar}
	\end{subfigure}
	\hfill
	\begin{subfigure}[t]{0.49\textwidth}
		\centering
		\includegraphics[width=\textwidth]{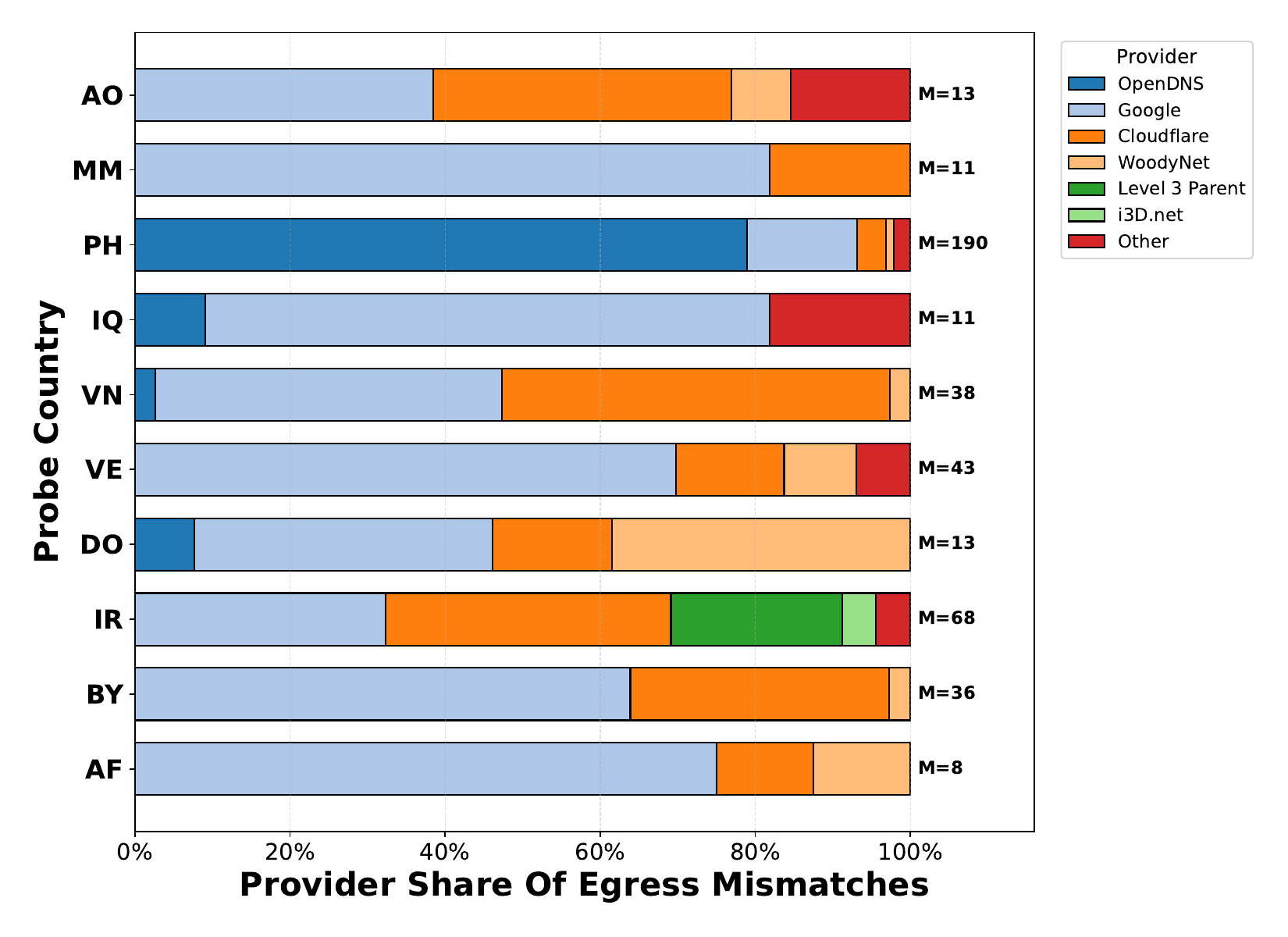}
		\caption{Egress mismatch provider mix.}
		\label{fig:appx_egress_mismatch_provider_mix}
	\end{subfigure}\\[0.8em]
	\begin{subfigure}[t]{0.78\textwidth}
		\centering
		\includegraphics[width=\textwidth]{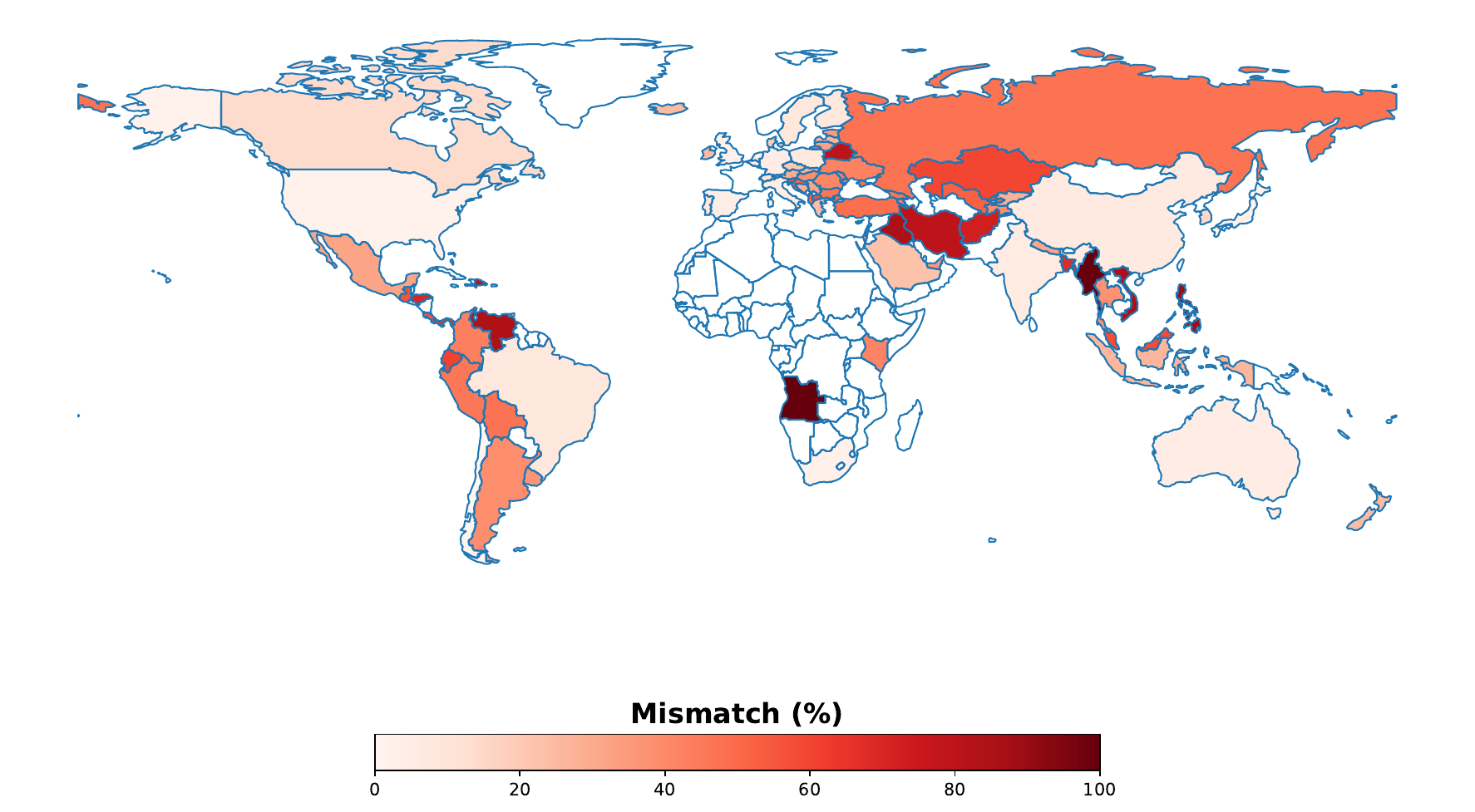}
		\caption{Egress mismatch heatmap.}
		\label{fig:appx_egress_mismatch_heatmap}
	\end{subfigure}
	\caption{Supporting egress mismatch visualizations. Countries such as Angola (13/13), Myanmar (11/11), the Philippines (190/213), Iraq (11/13), Vietnam (38/45), and Venezuela (43/51) show high out-of-country egress rates. Panel~(b) shows the provider mix; panel~(c) maps mismatch rates geographically.}
	\Description{A three-panel figure with an egress mismatch bar chart, a provider-mix chart, and a world heatmap showing per-country egress mismatch rates.}
	\label{fig:appx_egress_mismatch}
\end{figure*}

\subsection{Frontend Country Selection Table}
\label{sec:appx:frontend-country-selection}

This final table reports the complete per-country frontend-selection metrics for
all 171 observed countries. It serves as the full appendix counterpart to the
top-country views shown in the main text and allows readers to inspect which
countries were excluded by the support thresholds.

\section{Ethics}
\label{sec:appx:ethics}

This section summarizes the ethical constraints that shaped both the RIPE Atlas
measurement campaign and the protocol-evaluation workload.

The Atlas campaign queries domains under controlled authoritative service and
uses returned IP, ASN, and country metadata in aggregate. It does not observe or
intercept third-party user traffic, and private ingress is not assigned an
operator identity. Atlas use follows the platform's Acceptable Use Policy and
Terms of Service.

The performance workload derives from a ten-minute university DNS trace
collected on January~13, 2025, from 1:00--1:10~p.m. CST (19:00--19:10~UTC) by
the university IT department during routine network operations. The authors
received only an anonymized version of the trace, in which client identifiers
and queried names had been removed or anonymized prior to release. The
organization that owns and operates the network approved secondary research use
and release in transformed form; we omit the organization's identity to
preserve double-blind review. The transformed trace was released with the
organization's permission for research reproduction.

The evaluation ran in a Docker testbed whose local authoritative service was
configured to answer every replay name. Published results remain aggregated at
country or AS level, with no per-probe identifiers in the released dataset.

\clearpage
\begin{table*}[p]
\centering
\begingroup
\caption{Per-country frontend selection metrics. $A$\,=\,Anycast Frontend, $G$\,=\,All Resolver Chains.
Countries above the rule in the first column are \emph{eligible} for the main-text figures ($N_A \ge 25$, $N_G \ge 100$, $P_T \ge 15$ unique probes);
all countries are ranked by $XB_\mu=(XB_A+XB_G)/2$.}\label{tab:frontend_country_selection_full}
\centering\footnotesize\setlength{\tabcolsep}{1.25pt}\renewcommand{\arraystretch}{0.80}
\newcommand{\tblhdr}{%
  \hline
  Rk & CC & $N_A$ & $N_G$ & $P_T$ & $XB_A$ & $XB_G$ & $XB_\mu$ \\
  \hline}
\begin{minipage}[t]{0.326\textwidth}\centering
\begin{tabular}{rlrrrrrr}
\tblhdr
1 & PH & 26 & 213 & 113 & 92.3 & 89.2 & 90.8 \\
2 & RU & 246 & 665 & 401 & 78.1 & 50.7 & 64.4 \\
3 & UA & 64 & 227 & 131 & 78.1 & 43.6 & 60.9 \\
4 & BG & 33 & 108 & 62 & 72.7 & 47.2 & 60.0 \\
5 & RO & 38 & 121 & 66 & 68.4 & 41.3 & 54.9 \\
6 & ID & 27 & 131 & 69 & 63.0 & 29.8 & 46.4 \\
7 & DK & 36 & 138 & 84 & 61.1 & 25.4 & 43.2 \\
8 & AT & 83 & 272 & 151 & 51.8 & 31.2 & 41.5 \\
9 & CZ & 60 & 384 & 215 & 60.0 & 21.6 & 40.8 \\
10 & GR & 32 & 113 & 62 & 50.0 & 23.9 & 37.0 \\
11 & FR & 210 & 1282 & 711 & 52.9 & 19.4 & 36.1 \\
12 & BE & 55 & 244 & 128 & 50.9 & 16.4 & 33.7 \\
13 & FI & 44 & 198 & 115 & 20.5 & 20.2 & 20.3 \\
14 & CA & 91 & 397 & 231 & 20.9 & 16.4 & 18.6 \\
15 & CH & 80 & 395 & 233 & 20.0 & 11.4 & 15.7 \\
16 & SE & 72 & 252 & 156 & 18.1 & 9.9 & 14.0 \\
17 & AU & 88 & 306 & 189 & 10.2 & 11.4 & 10.8 \\
18 & SG & 62 & 167 & 103 & 4.8 & 16.8 & 10.8 \\
19 & BR & 27 & 120 & 75 & 11.1 & 10.0 & 10.6 \\
20 & PL & 60 & 227 & 136 & 11.7 & 8.8 & 10.2 \\
21 & DE & 385 & 2237 & 1255 & 12.5 & 7.0 & 9.7 \\
22 & ZA & 64 & 166 & 92 & 4.7 & 6.0 & 5.4 \\
23 & GB & 189 & 737 & 421 & 3.2 & 6.0 & 4.6 \\
24 & IN & 46 & 145 & 96 & 2.2 & 6.9 & 4.5 \\
25 & NL & 235 & 783 & 467 & 3.8 & 5.2 & 4.5 \\
26 & ES & 94 & 307 & 180 & 2.1 & 6.8 & 4.5 \\
27 & IT & 105 & 418 & 245 & 2.9 & 5.7 & 4.3 \\
28 & US & 602 & 2343 & 1401 & 2.5 & 3.5 & 3.0 \\
29 & JP & 58 & 269 & 171 & 0.0 & 5.6 & 2.8 \\
\hline
30 & MM & 7 & 11 & 6 & 100.0 & 100.0 & 100.0 \\
31 & AO & 2 & 13 & 7 & 100.0 & 100.0 & 100.0 \\
32 & VI & 6 & 7 & 3 & 100.0 & 100.0 & 100.0 \\
33 & MT & 3 & 8 & 7 & 100.0 & 100.0 & 100.0 \\
34 & BJ & 2 & 5 & 3 & 100.0 & 100.0 & 100.0 \\
35 & GH & 3 & 4 & 2 & 100.0 & 100.0 & 100.0 \\
36 & CD & 2 & 4 & 3 & 100.0 & 100.0 & 100.0 \\
37 & CM & 3 & 3 & 2 & 100.0 & 100.0 & 100.0 \\
38 & FM & 2 & 4 & 2 & 100.0 & 100.0 & 100.0 \\
39 & GG & 3 & 3 & 1 & 100.0 & 100.0 & 100.0 \\
40 & GQ & 3 & 3 & 1 & 100.0 & 100.0 & 100.0 \\
41 & LA & 2 & 3 & 1 & 100.0 & 100.0 & 100.0 \\
42 & ZM & 2 & 3 & 2 & 100.0 & 100.0 & 100.0 \\
43 & CI & 1 & 3 & 1 & 100.0 & 100.0 & 100.0 \\
44 & DJ & 2 & 2 & 2 & 100.0 & 100.0 & 100.0 \\
45 & KY & 2 & 2 & 1 & 100.0 & 100.0 & 100.0 \\
46 & PW & 2 & 2 & 1 & 100.0 & 100.0 & 100.0 \\
47 & TV & 2 & 2 & 1 & 100.0 & 100.0 & 100.0 \\
48 & KI & 1 & 2 & 2 & 100.0 & 100.0 & 100.0 \\
49 & SS & 1 & 2 & 1 & 100.0 & 100.0 & 100.0 \\
50 & BW & 1 & 1 & 1 & 100.0 & 100.0 & 100.0 \\
51 & QA & 1 & 1 & 1 & 100.0 & 100.0 & 100.0 \\
52 & TT & 1 & 1 & 1 & 100.0 & 100.0 & 100.0 \\
53 & AF & 5 & 11 & 7 & 100.0 & 90.9 & 95.5 \\
54 & IQ & 2 & 13 & 5 & 100.0 & 84.6 & 92.3 \\
55 & VN & 18 & 45 & 25 & 100.0 & 84.4 & 92.2 \\
56 & VE & 22 & 51 & 27 & 100.0 & 84.3 & 92.2 \\
57 & TD & 5 & 6 & 3 & 100.0 & 83.3 & 91.7 \\
\hline
\end{tabular}
\end{minipage}%
\begin{minipage}[t]{0.326\textwidth}\centering
\begin{tabular}{rlrrrrrr}
\tblhdr
58 & NG & 2 & 6 & 3 & 100.0 & 83.3 & 91.7 \\
59 & BY & 12 & 45 & 26 & 100.0 & 82.2 & 91.1 \\
60 & DO & 7 & 16 & 10 & 100.0 & 81.2 & 90.6 \\
61 & SN & 2 & 5 & 2 & 100.0 & 80.0 & 90.0 \\
62 & MK & 6 & 14 & 8 & 100.0 & 71.4 & 85.7 \\
63 & RE & 2 & 51 & 31 & 100.0 & 68.6 & 84.3 \\
64 & ZW & 1 & 6 & 4 & 100.0 & 66.7 & 83.3 \\
65 & PA & 3 & 23 & 17 & 100.0 & 65.2 & 82.6 \\
66 & KE & 3 & 19 & 10 & 100.0 & 63.2 & 81.6 \\
67 & TZ & 4 & 8 & 5 & 100.0 & 62.5 & 81.2 \\
68 & MV & 1 & 8 & 7 & 100.0 & 62.5 & 81.2 \\
69 & GU & 4 & 13 & 7 & 100.0 & 61.5 & 80.8 \\
70 & EC & 2 & 28 & 16 & 100.0 & 60.7 & 80.4 \\
71 & IM & 2 & 5 & 3 & 100.0 & 60.0 & 80.0 \\
72 & SI & 27 & 72 & 41 & 100.0 & 56.9 & 78.5 \\
73 & GT & 4 & 16 & 7 & 100.0 & 56.2 & 78.1 \\
74 & CR & 3 & 18 & 12 & 100.0 & 55.6 & 77.8 \\
75 & PY & 2 & 9 & 7 & 100.0 & 55.6 & 77.8 \\
76 & AL & 6 & 40 & 21 & 100.0 & 52.5 & 76.2 \\
77 & UZ & 9 & 21 & 10 & 100.0 & 52.4 & 76.2 \\
78 & MA & 3 & 6 & 2 & 100.0 & 50.0 & 75.0 \\
79 & NA & 1 & 6 & 3 & 100.0 & 50.0 & 75.0 \\
80 & KH & 2 & 4 & 2 & 100.0 & 50.0 & 75.0 \\
81 & PE & 12 & 24 & 13 & 91.7 & 54.2 & 72.9 \\
82 & HN & 4 & 10 & 5 & 75.0 & 70.0 & 72.5 \\
83 & MD & 11 & 36 & 19 & 100.0 & 44.4 & 72.2 \\
84 & HR & 12 & 42 & 26 & 91.7 & 52.4 & 72.0 \\
85 & TJ & 3 & 23 & 13 & 100.0 & 43.5 & 71.7 \\
86 & BO & 10 & 32 & 18 & 90.0 & 53.1 & 71.6 \\
87 & KR & 1 & 26 & 18 & 100.0 & 42.3 & 71.2 \\
88 & KZ & 23 & 57 & 34 & 82.6 & 59.6 & 71.1 \\
89 & MY & 18 & 50 & 27 & 83.3 & 58.0 & 70.7 \\
90 & MU & 2 & 5 & 4 & 100.0 & 40.0 & 70.0 \\
91 & LU & 13 & 61 & 33 & 84.6 & 54.1 & 69.4 \\
92 & UY & 5 & 22 & 12 & 100.0 & 36.4 & 68.2 \\
93 & AZ & 6 & 6 & 4 & 66.7 & 66.7 & 66.7 \\
94 & NP & 4 & 22 & 14 & 100.0 & 31.8 & 65.9 \\
95 & BD & 13 & 22 & 12 & 61.5 & 68.2 & 64.9 \\
96 & BA & 5 & 24 & 11 & 100.0 & 29.2 & 64.6 \\
97 & IS & 9 & 16 & 10 & 77.8 & 50.0 & 63.9 \\
98 & RS & 21 & 70 & 39 & 90.5 & 32.9 & 61.7 \\
99 & CO & 14 & 39 & 22 & 78.6 & 43.6 & 61.1 \\
100 & AR & 18 & 62 & 38 & 83.3 & 38.7 & 61.0 \\
101 & EE & 22 & 55 & 30 & 81.8 & 36.4 & 59.1 \\
102 & GE & 6 & 15 & 10 & 66.7 & 46.7 & 56.7 \\
103 & MN & 1 & 8 & 5 & 100.0 & 12.5 & 56.2 \\
104 & LT & 8 & 48 & 31 & 75.0 & 35.4 & 55.2 \\
105 & TR & 21 & 64 & 44 & 61.9 & 48.4 & 55.2 \\
106 & HU & 24 & 61 & 36 & 70.8 & 39.3 & 55.1 \\
107 & KG & 5 & 32 & 16 & 80.0 & 28.1 & 54.1 \\
108 & CN & 4 & 132 & 62 & 100.0 & 7.6 & 53.8 \\
109 & CY & 3 & 22 & 12 & 66.7 & 40.9 & 53.8 \\
110 & TH & 15 & 45 & 29 & 53.3 & 46.7 & 50.0 \\
111 & NC & 2 & 8 & 5 & 50.0 & 50.0 & 50.0 \\
112 & ME & 0 & 6 & 5 & 0.0 & 100.0 & 50.0 \\
113 & MZ & 0 & 4 & 2 & 0.0 & 100.0 & 50.0 \\
114 & PR & 0 & 3 & 1 & 0.0 & 100.0 & 50.0 \\
\hline
\end{tabular}
\end{minipage}%
\begin{minipage}[t]{0.326\textwidth}\centering
\begin{tabular}{rlrrrrrr}
\tblhdr
115 & BT & 0 & 2 & 1 & 0.0 & 100.0 & 50.0 \\
116 & CG & 0 & 2 & 2 & 0.0 & 100.0 & 50.0 \\
117 & EG & 0 & 2 & 1 & 0.0 & 100.0 & 50.0 \\
118 & GP & 0 & 2 & 1 & 0.0 & 100.0 & 50.0 \\
119 & HT & 0 & 2 & 1 & 0.0 & 100.0 & 50.0 \\
120 & LS & 0 & 2 & 1 & 0.0 & 100.0 & 50.0 \\
121 & MP & 0 & 2 & 1 & 0.0 & 100.0 & 50.0 \\
122 & PG & 0 & 2 & 2 & 0.0 & 100.0 & 50.0 \\
123 & PS & 0 & 2 & 1 & 0.0 & 100.0 & 50.0 \\
124 & TO & 0 & 2 & 1 & 0.0 & 100.0 & 50.0 \\
125 & AX & 0 & 1 & 1 & 0.0 & 100.0 & 50.0 \\
126 & DZ & 0 & 1 & 1 & 0.0 & 100.0 & 50.0 \\
127 & KN & 0 & 1 & 1 & 0.0 & 100.0 & 50.0 \\
128 & MH & 0 & 1 & 1 & 0.0 & 100.0 & 50.0 \\
129 & LV & 14 & 46 & 30 & 64.3 & 30.4 & 47.4 \\
130 & IE & 15 & 95 & 60 & 66.7 & 26.3 & 46.5 \\
131 & AE & 16 & 28 & 19 & 56.2 & 35.7 & 46.0 \\
132 & SA & 3 & 22 & 13 & 66.7 & 22.7 & 44.7 \\
133 & AM & 7 & 28 & 16 & 42.9 & 42.9 & 42.9 \\
134 & IL & 9 & 31 & 18 & 55.6 & 29.0 & 42.3 \\
135 & KW & 2 & 6 & 3 & 50.0 & 33.3 & 41.7 \\
136 & NO & 19 & 123 & 69 & 57.9 & 25.2 & 41.5 \\
137 & IR & 0 & 86 & 64 & 0.0 & 80.2 & 40.1 \\
138 & NZ & 13 & 125 & 75 & 53.8 & 24.8 & 39.3 \\
139 & PT & 15 & 279 & 130 & 66.7 & 9.0 & 37.8 \\
140 & TL & 0 & 4 & 2 & 0.0 & 75.0 & 37.5 \\
141 & MX & 15 & 59 & 34 & 40.0 & 33.9 & 37.0 \\
142 & BF & 0 & 6 & 4 & 0.0 & 66.7 & 33.3 \\
143 & SK & 8 & 66 & 36 & 37.5 & 28.8 & 33.1 \\
144 & FJ & 0 & 5 & 2 & 0.0 & 60.0 & 30.0 \\
145 & TN & 0 & 9 & 4 & 0.0 & 55.6 & 27.8 \\
146 & NI & 0 & 6 & 4 & 0.0 & 50.0 & 25.0 \\
147 & LC & 0 & 4 & 3 & 0.0 & 50.0 & 25.0 \\
148 & MW & 0 & 4 & 2 & 0.0 & 50.0 & 25.0 \\
149 & CV & 0 & 2 & 2 & 0.0 & 50.0 & 25.0 \\
150 & GF & 0 & 5 & 2 & 0.0 & 40.0 & 20.0 \\
151 & LB & 0 & 3 & 2 & 0.0 & 33.3 & 16.7 \\
152 & HK & 23 & 63 & 36 & 8.7 & 19.0 & 13.9 \\
153 & AD & 0 & 4 & 3 & 0.0 & 25.0 & 12.5 \\
154 & JO & 0 & 4 & 2 & 0.0 & 25.0 & 12.5 \\
155 & LK & 0 & 6 & 5 & 0.0 & 16.7 & 8.3 \\
156 & PK & 1 & 9 & 5 & 0.0 & 11.1 & 5.6 \\
157 & PF & 0 & 9 & 5 & 0.0 & 11.1 & 5.6 \\
158 & TW & 44 & 84 & 50 & 0.0 & 6.0 & 3.0 \\
159 & CL & 23 & 111 & 63 & 0.0 & 0.9 & 0.5 \\
160 & LI & 0 & 7 & 4 & 0.0 & 0.0 & 0.0 \\
161 & UG & 0 & 7 & 3 & 0.0 & 0.0 & 0.0 \\
162 & BH & 0 & 5 & 3 & 0.0 & 0.0 & 0.0 \\
163 & OM & 0 & 4 & 2 & 0.0 & 0.0 & 0.0 \\
164 & BB & 0 & 2 & 1 & 0.0 & 0.0 & 0.0 \\
165 & CK & 0 & 2 & 1 & 0.0 & 0.0 & 0.0 \\
166 & SM & 0 & 2 & 1 & 0.0 & 0.0 & 0.0 \\
167 & SV & 0 & 2 & 1 & 0.0 & 0.0 & 0.0 \\
168 & VA & 0 & 2 & 1 & 0.0 & 0.0 & 0.0 \\
169 & BI & 0 & 1 & 1 & 0.0 & 0.0 & 0.0 \\
170 & BZ & 0 & 1 & 1 & 0.0 & 0.0 & 0.0 \\
171 & MG & 0 & 1 & 1 & 0.0 & 0.0 & 0.0 \\
\hline
\end{tabular}
\end{minipage}
\endgroup

\end{table*}

\end{document}